\documentclass[12pt]{article}

\usepackage{amsmath,amssymb,amsfonts,bbm,subfigure}
\usepackage{epsfig,cancel,cite}

\numberwithin{equation}{section}

\newlength{\xtrawidth}
\newlength{\xtraheight}
\newcommand{\beq}{\begin{equation}}
\newcommand{\eeq}{\end{equation}}
\newcommand{\ba}{\begin{array}}
\newcommand{\ea}{\end{array}}
\newcommand{\bea}{\begin{eqnarray}}
\newcommand{\eea}{\end{eqnarray}}
\newcommand{\bean}{\begin{eqnarray*}}
\newcommand{\eean}{\end{eqnarray*}}
\newcommand{\eref}[1]{(\ref{#1})}

\newcommand{\nn}{\nonumber}
\newcommand{\comment}[1]{}

\newcommand{\cO}{{\cal O}}

\newcommand{\cF}{{\cal F}}

\newcommand{\cL}{{\cal L}}

\def\fnote#1#2{\begingroup\def\thefootnote{#1}\footnote{#2}
     \addtocounter{footnote}{-1}\endgroup}

\newtheorem{theorem}{\sf THEOREM}

\begin{document}

\title{{\LARGE
Transitions in the Web of Heterotic Vacua
}}
\author{
Lara B. Anderson${}^{1}$,
James Gray${}^{2,3}$,
 and Burt A. Ovrut${}^{1}$
}
\date{}
\maketitle
\begin{center}
{\small
${}^1${\it Department of Physics, University of
      Pennsylvania,\\ 209 South 33rd Street, Philadelphia, PA 19104-6395, U.S.A.}\\[0.2cm]
${}^2${\it Arnold-Sommerfeld-Center for Theoretical Physics, \\
Department f\"ur Physik, Ludwig-Maximilians-Universit\"at M\"unchen,\\
Theresienstra\ss e 37, 8033 M\"unchen, Germany}\\[0.2cm]
${}^3${\it Max-Planck-Institut f\"ur Physik -- Theorie,\\ 
F\"ohringer Ring 6, 80805 M\"unchen, Germany}\\[0.2cm]

\fnote{}{andlara@physics.upenn.edu}
\fnote{}{James.Gray@physik.uni-muenchen.de}
\fnote{}{ovrut@elcapitan.hep.upenn.edu}
}
\end{center}

\abstract

We analyze transitions between heterotic vacua with distinct gauge bundles using two complementary methods -- the effective four-dimensional field theory and the corresponding geometry. From the viewpoint of effective field theory, such transitions occur between flat directions of the potential energy associated with heterotic stability walls. Geometrically, this branch structure corresponds to smooth deformations of the gauge bundle coupled to the chamber structure of K\"ahler moduli space. We demonstrate how such transitions can change important properties of the effective theory, including the gauge symmetry and the massless spectrum. Geometrically, this study is divided into deformations of the vector bundle which preserve the rank of the gauge bundle and those which change the rank. In the latter case, our results provide explicit solutions to a class of Li-Yau type deformation problems. Finally, we use the framework of stability walls and their effective theory to study Donaldson-Thomas invariants on Calabi-Yau threefolds.

\newpage

\tableofcontents

\section{Introduction}\label{intro}
The choice of an $N=1$ supersymmetric heterotic vacuum \cite{Candelas:1985en,Witten:1985bz,GSW,Lukas:1997fg,Lukas:1998yy,Lukas:1998tt,Lukas:1998hk} is, in fact, a choice of geometry; specifically, a Calabi-Yau threefold, $X$, with a holomorphic, slope-stable vector bundle, $V$, defined over it.  In this paper, we explore a  long-standing question in heterotic geometry --  given $X$ and $V$, what geometries are closely connected in the ``neighborhood" of this compactification? From the point of view of effective field theory, this neighborhood is defined by fluctuations around the vacuum. By giving small vacuum expectation values (vevs) to massless particles, the flat directions of the potential define a local neighborhood that can be perturbatively explored. Geometrically, this is associated with changing the geometry of both the manifold and the gauge field configuration in the internal dimensions. In this paper, we describe how such changes can correspond to smooth transitions between different vector bundles.

Understanding such changes in geometry can yield important insights into string phenomenology  \cite{Donagi:2004ia,Donagi:2004qk,Donagi:2004su,Donagi:2003tb,burt4,burt5,burt6,burt7}. 
For example, following flat directions in the potential away from a given geometric vacuum can change the phenomenological features of the low-energy theory -- including the gauge symmetry and the massless spectrum.
Examples include moving a heterotic orbifold model away from the orbifold point by blowing up singularities (see\cite{Blaszczyk:2010db,Nibbelink:2009sp} for recent examples), and finding a flat direction which removes vector-like exotics from the massless spectrum of a smooth compactification \cite{Greene:1986bm,Greene:1986ar,Greene:1986jb,Candelas:2008wb,Braun:2009qy}.

There is a substantial advantage to understanding the geometry, rather than just the local vacuum space, associated with flat directions in the four-dimensional
theory. 
The effective theory of heterotic string and M-theory is constructed
as a perturbative expansion in the matter fields. Hence, one has a good
description of the physics only in a small portion of
moduli space, where the vevs of fields are relatively small.
There is no guarantee that a moduli
stabilization mechanism, if one could be found, would place the system
in this region. Furthermore, it can be hard to decide when a
flat directions might be ``obstructed''; that is, when the flat direction
is lifted by higher-order terms in the effective Lagrangian. In any
case, if the choice of a stringy vacuum corresponds to a
particular geometry, it is natural to ask what other ``nearby''
geometries are connected to it by flat directions? And if certain flat
directions lead to more phenomenologically desirable theories,
why not simply start with the more suitable geometry in the
first place? The answer, of course, is that it is frequently difficult
to determine the explicit form of these nearby geometries.

Mathematically, perturbations of the field theory vacuum corresponds to small deformations of the initial geometry; that is, to small topology preserving deformations of the manifold and the holomorphic vector bundle. The presence of flat directions can be easily observed in the effective theory, but what vector bundles these directions correspond to (and whether they really exist) is, in general, a difficult problem in deformation theory. In this paper, we study one aspect of this by providing concrete geometric descriptions of deformation neighborhoods (that is,  flat directions in the effective theory) associated with heterotic compactifications exhibiting K\"ahler cone substructure and stability walls \cite{Sharpe:1998zu, Anderson:2009sw,Anderson:2009nt,Anderson:2010tc}. We find that it is possible to completely determine not only the vacuum structure of the effective field theory, but to fully describe the geometry of a local deformation as well. This 
yields new insights into the full vacuum structure of heterotic theories and the moduli space of stable bundles, including the computation of Donaldson-Thomas invariants \cite{DT,Thomas,ellingsrud,Qin,Li_Qin,Li_Qin2,Friedman_Qin,Nekrasov1,Nekrasov2}.

In Section \ref{stab_wall_review}, we give a brief review of K\"ahler cone substructure and stability walls, from both a geometric and field theoretic standpoint \cite{Sharpe:1998zu,Anderson:2009sw,Anderson:2009nt,Anderson:2010tc}. A vector bundle need not be slope-stable everywhere in K\"ahler moduli space. Instead, $V$ may be stable only in certain ``chambers" of K\"ahler moduli space, separated from regions where it is unstable by co-dimension one boundaries called ``stability walls". As we review below, at a stability wall a bundle can only preserve supersymmetry by decomposing into a poly-stable sum of sub-sheaves. This decomposition leads to new anomalous $U(1)$ symmetries \cite{Lukas:1999nh,Blumenhagen:2005ga,Blumenhagen:2006wj,Kuriyama:2008pv} in the effective theory. The associated D-terms, along with the holomorphicity of $V$,  determine the supersymmetric vacuum structure. 

In Section \ref{rank_pres_sec}, we explore flat directions of the effective theory associated with the anomalous $U(1)$ symmetries. Unlike previous discussions in the heterotic literature, these flat directions correspond to non-isomorphic, slope-stable bundles in different chambers of K\"ahler moduli space. For bundles with stability walls, we study the full geometric deformation theory of these ``rank-preserving" transitions \cite{hartshorne_deformation,Huybrechts}, and explicitly construct the relevant vector bundles. As discussed  in Subsection \ref{branch_goodness}, the possible branches in each chamber of K\"ahler moduli space are characterized by the structure of S-equivalence classes on the stability walls \cite{ellingsrud,Qin,Li_Qin,Sharpe:1998zu}. We provide examples in which the low-energy physics  dramatically changes between different vacuum branches. In addition, we illustrate the correspondence between obstructions to these flat directions (including higher-order contributions to the superpotential) and obstructions to deformations of the corresponding vector bundles.

In Section \ref{rank_change_sec}, we extend our results to include so-called ``rank-changing" transitions \cite{GSW,Huybrechts_tangent, Li:2004hx}, which occur along flat directions that break the four-dimensional symmetry, $G$, to a subgroup $H \subset G$. Geometrically, these correspond to deforming not the original rank $n$ bundle $V$ but, rather, the direct sum $V\oplus \cO_{X}^{\oplus m}$. For a fixed K\"ahler modulus for which $V$ is slope-stable, the deformations of this direct sum were definitively studied by Li and Yau in \cite{Li:2004hx}, who proved that a properly stable deformation of $V\oplus \cO_{X}^{\oplus m}$ exists under certain conditions. However, despite this existence proof, it is often hard to explicitly construct the deformed rank $n+m$ bundle, $V_{n+m}$. In Subsection \ref{li_yau_stab}, we show that when $V$ has a stability wall, it is possible to use the presence of the stability wall -- specifically, the decomposition of $V$ into sub-sheaves on the wall -- to infer the form of the deformed Li-Yau solution everywhere within a stable chamber of K\"ahler moduli space. In Subsection \ref{gen_rank_change}, we extend our consideration of rank-changing deformations to different chambers of K\"ahler moduli space. That is, we consider branches of the effective theory that satisfy D- and F-flatness that can be found by giving vevs to $G$-charged matter. These correspond to deformations of the bundle $V$ which change not only the structure group, but also the slope-stable regions of K\"ahler moduli space.

Finally, in Section \ref{DT_section} we investigate stability walls as a tool for computing Donaldson-Thomas invariants. On a threefold $X$, a Donaldson-Thomas (DT) invariant is a geometric invariant, $\lambda_{\omega}(X,c)$, whose absolute value counts the number of holomorphic vector bundles $V$ with fixed total Chern class $c=(rank,c_1,c_2, c_3)$ that are stable for a given K\"ahler form $\omega$ \cite{Thomas}. That is, a DT-invariant counts the number of heterotic vacua corresponding to a given set of topological data and a K\"ahler form. In Subsection \ref{DT_su2}, we demonstrate that stability walls can provide a powerful tool to compute DT-invariants within a chamber of K\"ahler moduli space \cite{Qin,Li_Qin,Li_Qin2}. Moreover, we relate wall-crossing formula for DT-invariants to the vacuum branch counting of the effective field theory described throughout this paper, and provide an explicit example. In the final Subsection \ref{DT_recursion}, we demonstrate that stability walls can also be used to recursively compute DT-invariants for higher-rank vector bundles. To date, very few examples of DT-invariants have been computed for either compact Calabi-Yau threefolds or for vector bundles of rank $\geq 2$. In Subsection \ref{DT_recursion}, we use K\"ahler cone substructure to determine the DT-invariants of a rank $3$ bundle in terms of similar invariants for specific rank $2$ and rank $1$ sub-sheaves. In a specific rank $3$ example, we succeed in computing a wall-crossing formula for the DT-invariant.

In the Appendix, we provide a proof that the deformed vector bundles used thorough this paper, derived as the geometries corresponding to flat directions in the effective theory, are uniquely defined.

\section{Review of Stability Walls and K\"ahler Sub-Structure}\label{stab_wall_review}
In this paper, 
we investigate holomorphic vector bundles $V$ over Calabi-Yau (CY) threefolds $X$ that give rise to $N=1$ supersymmetric vacua in four-dimensions and, in particular, analyze the rich branch structure associated with them. To begin, we briefly review the relationship between slope-stability of holomorphic bundles and supersymmetric vacua in heterotic theory.

The conditions for supersymmetry  \cite{GSW} require that the background $E_{8}$ gauge field configuration
satisfies the Hermitian Yang-Mills equations
\beq\label{HYM}
g^{a\bar{b}}F_{a\bar{b}}=0~,~~~F_{ab}=F_{\bar{a}\bar{b}}=0 \ .
\eeq
Here $F$ is the two-form field strength of the gauge field and $g_{a\bar{b}}$ is the CY metric. By the Donaldson-Uhlenbeck-Yau theorem \cite{duy}, holomorphic vector bundles admitting a connection satisfying \eref{HYM} are in one-to-one correspondence with {\it slope poly-stable} vector bundles. The quantity $\mu$, called the slope, is defined for any coherent sheaf $\cF$ as
\beq
\label{slope}
\mu(\cF)=\frac{1}{rk(\cF)}\int_X c_1(\cF) \wedge \omega \wedge \omega=\frac{1}{rk(\cF)}d_{rst}c_1(\cF)^r t^s t^t \ ,
\eeq
where $\omega$ is the K\"ahler form (expanded in a basis of harmonic $(1,1)$ forms, $\omega=t^r \omega_r$) and $d_{rst}$ are the triple intersection numbers. A vector bundle is called {\it stable} 
if for all sub-sheaves $\cF \subset V$ with $0 < rk(\cF) < rk(V)$,
\beq
\mu(\cF) < \mu(V) \ .
\label{slope2}
\eeq
$V$ is {\it poly-stable} if it can be written as a direct sum of stable bundles, all with the same slope; that is,
\begin{equation}
\label{burt1}
V=\bigoplus_i V_i ~~~\text{with}~~~ \mu(V_i)=\mu(V)
\end{equation}
for all $i$ with each $V_i$ satisfying \eref{slope2}. Note that all stable bundles are trivially poly-stable. When $\mu(\cF)> \mu(V)$ for some sub-sheaf $\cF$, the bundle $V$ is called {\it unstable}. 
Finally, if $V$ is {\it not} poly-stable but
\begin{equation}
\label{burtbird}
\mu(\cF)= \mu(V)
\end{equation}
for some sub-sheaf $\cF \subset V$, then the bundle is called {\it semi-stable}.
Since the existence of an Hermitian-Yang-Mills connection, that is, one satisfying \eref{HYM}, is in one-to-one correspondence with poly-stable bundles, it is clear that both unstable and semi-stable bundles fail to give supersymmetric vacua.  

It follows from the definition of slope in \eqref{slope} that whether or not \eref{HYM} has a solution can 
depend on the choice of K\"ahler moduli. In addition, at each point in a supersymmetric region of K\"ahler moduli space the solution to \eref{HYM} has a number of arbitrary integration constants, the vector bundle moduli. These are elements of $H^1(X,V\otimes V^{*})$. In the following sections, we will see that it is the combined K\"ahler/vector bundle moduli space in which we need to carry out our analysis. In the presence of a given vector bundle, K\"ahler moduli space divides into chambers where supersymmetric vacua are possible and chambers where they are not; that is, where the bundle is stable and unstable respectively. A {\it stability wall} is a co-dimension one boundary between any two such chambers. On a stability wall, the bundle can split into a poly-stable direct sum, although for generic bundle moduli, it is semi-stable.
Such a division of the K\"ahler cone is referred to in the mathematics literature as ``K\"ahler cone sub-structure''.

In \cite{Sharpe:1998zu} and in recent work \cite{Anderson:2009sw,Anderson:2009nt,Anderson:2010tc}, this sub-structure was explored from the point of view of effective field theory. In this description, as the fields are varied into the supersymmetry
breaking region of moduli space, a new potential -- generated by D-terms associated with Green-Schwarz anomalous U(1) gauge factors \cite{Blumenhagen:2006wj,Blumenhagen:2005ga,Lukas:1999nh,Kuriyama:2008pv}-- appears for the scalar fields of the four-dimensional
effective theory. The slope stability properties of the bundle are described in terms of this potential.
As we review below, the form of this potential can be understood by noting that, on a stability wall between stable/unstable regions of the K\"ahler cone, a particular sub-sheaf of $V$ becomes important.

A sub-sheaf $\cF \subset V$ {\it de-stabilizes} $V$ in the regions of K\"ahler moduli space for which $\mu(\cF) \geq \mu(V)$.  In this case, one can always define $V$ as an ``extension'' of $\cF$ via the short exact sequence
\beq\label{extension}
0 \to \cF \to V \to V/\cF \to 0 \ .
\eeq
Such an extension sequence is called ``split'' (or trivial) if $V = \cF \oplus V/\cF$. The non-triviality of this extension is measured by the Ext group, 
\beq\label{ext_ness}
Ext^1(V/\cF, \cF)=H^1(X, \cF\otimes(V/\cF)^{*}) \ .
\eeq
That is, $0 \in H^1(X, \cF\otimes(V/\cF)^{*})$ corresponds to the bundle $V = \cF \oplus V/\cF$. This cohomology group is a subset of the vector bundle moduli space. Note that $V$ is not necessary stable anywhere in moduli space, even at the split point. 

Let us now assume that $V$ {\it is} slope-stable somewhere in K\"ahler moduli space. Then there must exist a stability wall, defined by $\mu(\cF)=\mu(V)$, between the stable and unstable regions. 
By the Donaldson-Uhlenbeck-Yau theorem, the only way for $V$ to preserve supersymmetry on the wall is for it to be poly-stable. That is, its bundle moduli must be chosen so that $V$ can be written as a direct sum of two stable pieces with the same slope. In terms of the extension sequence \eref{extension}, this means that $V$ will only preserve supersymmetry on the stability wall if the extension splits. We can gain insight into this by writing down the low energy effective theory for $V = \cF \oplus V/\cF$, and considering small fluctuations in the vacuum space away from this split locus.

The low energy gauge symmetry of the heterotic theory associated with an {\it indecomposable} (non-split) bundle $V$ with structure group $H=SU(n)$
is given by the commutant $G$ of $H$ inside $E_8$. In this paper, we will be interested in $n=2,3,4,5$ for which the low-energy gauge groups are $G=E_{7}, E_{6}, SO(10)$ and $SU(5)$. We will refer to zero-modes that transform non-trivially under $G$ as ``matter'' multiplets. The remaining fields are the geometric and vector bundle moduli. Generically, these two types of fields will be denoted by $f$ and $\phi$ respectively. Associated with the gauge group $G$ will be a D-term of the form
\begin{equation}
\label{bulb1}
D^{Ga}=  \sum g_{i \bar{j}} f^{i} T^{a} f^{\bar{j}} \ , \quad a=1,\dots, {\rm dim}G 
\end{equation}
where the sum is over all matter multiplets, $T^{a}$ are the Lie algebra generators of $G$ and the K\"ahler metric $g_{i \bar{j}}$ has positive definite eigenvalues. Here, and for every D-term in this paper, we suppress the gauge coupling parameter since it does not effect our conclusions.

If the bundle $V$ is chosen to be a {\it direct sum} of two pieces, as it is at a stability wall, its commutant inside $E_8$ will be enhanced by at least one extra  $U(1)$ factor from $G$ to $G \times U(1)$. This $U(1)$ is anomalous in the Green-Schwarz sense \cite{Lukas:1999nh,Blumenhagen:2005ga}. Near the stability wall, a subset of both the matter multiplets and the moduli in the effective field theory can become charged under this $U(1)$. While continuing to refer to all matter multiplets and {\it uncharged} moduli as $f$ and $\phi$ respectively, we now generically denote the $U(1)$ {\it charged} bundle moduli as $C$.
The D-term associated with this enhanced $U(1)$ can be written as a sum of a Fayet-Illiopolous (FI) term, the $U(1)$ charged matter fields $f^{i}$ (with charge $q^{i}$), and the $U(1)$ charged bundle moduli $C^L$ (with charge $Q^L$) as \cite{Anderson:2009nt}
\beq\label{dterm}
D^{U(1)}=\frac{3}{16}\frac{\epsilon_{S}\epsilon_{R}^{2}}{\kappa_4^{2}}\frac{\mu(\cF)}{{\cal V}} - \sum Q^L G_{L\bar{M}} C^L C^{\bar M}  - \sum q^i {\cal{G}}_{i\bar{j}} f^i f^{\bar j} \ .
\eeq
Here $\kappa_{4}^{2}$ is the four-dimensional Planck constant, $\epsilon_{S},\epsilon_{R}$ are constants related to the perturbative expansion parameters of heterotic M-theory \cite{Lukas:1997fg,Lukas:1998tt,Lukas:1998yy,Lukas:1998hk}, ${\cal V}$ is the volume of the Calabi-Yau threefold and $\mu(\cF)$ is the slope of the de-stabilizing sub-sheaf $\cF$. The K\"ahler metrics $G_{L\bar{M}}$ and ${\cal{G}}_{i\bar{j}}$ all have positive definite eigenvalues. Note that these metrics can be different for each type of $C$ and $f$ field. Nonetheless, this does not effect our analysis and, hence, when multiple 
K\"ahler metrics appear, we will not distinguish between them. Importantly, note that the FI term is a {\it function of the K\"ahler moduli} through both the Calabi-Yau volume and the slope.

The simplest possible example is when $V$ is an $SU(2)$ bundle destabilized by a line bundle $\cF=\cL$ with locally-free quotient $V/\cF=\cL^{*}$. In this case,
the $SU(2)$ bundle must decompose on a stability wall, where $\mu(\cL)=0$, to $V \to \cL \oplus \cL^{*}$ with structure group $S[U(1) \times U(1)]$. Locally, $S[U(1)\times U(1)] \simeq  U(1)$ and so, near the stability wall, the spectrum of the low-energy $E_7 \times U(1)$ theory is obtained from the decomposition
\beq
{\bf 248} \to ({\bf 56})_{+1/2}\oplus({\bf 56})_{-1/2}  \oplus ({\bf 1})_{-1}\oplus ({\bf 1})_{+1}
\eeq
of the adjoint representation of $E_{8}$ under $E_7 \times U(1)$.
The zero-modes of this decomposition are given by the bundle cohomology groups associated with the line bundle $\cL$. These are presented in Table \ref{SU2_table}. 

To illustrate the moduli dependence of slope stability from the point of view of effective field theory, consider the simple case in which $\cL$ is chosen so that only one sign of $U(1)$ charge is present in the low-energy spectrum. For example, take the line bundle $\cL=\cO_{X}(-1,1)$ on the Calabi-Yau threefold 
\begin{equation}
\label{burt2}
\left[ \ba{c |c }
\mathbb{P}^1 & 2 \\
\mathbb{P}^3 & 4
\ea \right]^{2,86} \ ,
\end{equation}
where the superscripts denote the Hodge numbers $h^{1,1}$ and $h^{2,1}$, respectively. Then \eref{extension} becomes
\beq\label{E7_eg}
0 \to \cO_{X}(-1,1) \to V \to \cO_{X}(1,-1) \to 0 \ .
\eeq
The space of $SU(2)$ extension bundles $V$ is given by
\begin{equation}
\label{hope1}
Ext^{1}({\cal{L}}^{*},{\cal{L}})=H^1(X, \cL^2)=H^1(X, \cO_{X}(-2,2))
\end{equation}
of dimension $10$. For the CY threefold \eqref{burt2}, the triple intersection numbers are
\beq
\label{someday2}
d_{122}=d_{212}=d_{221}=4,~~d_{222}=2,~~\text{all others zero}~.
\eeq
Using these intersection numbers and \eref{slope}, we find that the quantities relevant to the FI term in \eref{dterm} are
\beq
\label{hope2}
\mu(\cF)=\mu(\cO_{X}(-1,1))=-2(t^2)^2+8t^1t^2~,~~{\cal V}=2t^1(t^2)^2+\frac{1}{3}(t^2)^3 \ .
\eeq
Hence, on the line in K\"ahler moduli space given by $t^2/t^1=4$ for which $\mu(\cO_{X}(-1,1))=0$, the line bundle $\cO_{X}(-1,1)$ de-stabilizes $V$. That is, $V$ is stable for all K\"ahler moduli with $t^{2}/t^{1}>4$ and unstable for all $t^{2}/t^{1}<4$. The locus $t^2/t^1=4$ defines the stability wall. In order to preserve supersymmetry on the wall itself, $V$ must be poly-stable and, hence, decompose into the direct sum 
\begin{equation}
\label{hope3}
V \rightarrow \cO_{X}(-1,1)\oplus \cO_{X}(1,-1) \ .
\end{equation}

Writing the low energy theory associated with this bundle, we find the spectrum given in the last column of Table \ref{SU2_table}. Note that it consists only of bundle moduli\footnote{For simplicity, when a $4$-dimensional field, $C$, is counted by a cohomology $H^1(X, U)$, for some bundle $U$, we denote this relationship informally as $C \in H^1(X, U)$. 
} $C_{2}^{L}  \in H^1(X,{\cal L}^{2})$ with $U(1)$ charge $-1$. Clearly, the $E_{7}$ D-terms of the form \eqref{bulb1} trivially vanish.
It follows from $\eqref{dterm}$ and $\eqref{hope2}$ that the $U(1)$ D-term is of the form
\beq\label{dterm_e7}
D^{U(1)} \sim \frac{4t^1-t^2}{t^{2}(t^1 + \frac{1}{6}t^2)} +G_{L\bar{M}} C_{2}^{L} C_{2}^{\bar M} \ .
\eeq
\begin{table}[t]
\begin{center}
\begin{tabular}{|c|c|c||c|}
  \hline
  Representation & Field Name & Cohomology & Multiplicity \\ \hline
  ${\bf 1}_{+1}$ &  $C_1$ & $H^1(X,{{\cal L}^{{*}}}^{2})$ & $10$ \\ \hline
  ${\bf 1}_{-1}$ &  $C_2$ & $H^1(X,{\cal L}^{2})$ & $0$ \\ \hline
  ${\bf 56}_{+1/2}$ & $f_1$ &  $H^1(X,{\cal L}^{*})$ & $0$ \\ \hline
   ${\bf 56}_{-1/2}$ & $f_2$ &  $H^1(X,{\cal L})$ & $0$ \\ \hline
\end{tabular}
\caption{{\it The representations, fields and the associated cohomology groups for the $E_7 \times U(1)$ stability wall theory given in Section \ref{stab_wall_review}. The multiplicities for the rank $2$ vector bundle $V$ defined in \eref{E7_eg} are given in the fourth column.}}
\label{SU2_table}
\end{center}
\end{table}
What happens when the K\"ahler moduli are chosen so that $\mu(\cO_{X}(-1,1))<0$ and, hence, the FI term is negative? In this case, a linear combination of the charged $C_{2}^L$ fields can acquire a non-vanishing vev which cancels the FI term in vacuum. This preserves  D-flatness and, hence, the supersymmetry of the theory. This is a description, in field-theoretic language, of the fact that in such a region of moduli space $V$ in \eref{E7_eg} is slope-stable. 
Since $C_{2}^L \in H^1(X,{\cal L}^{2})$, such a field acquiring a non-zero vev is equivalent to choosing a non-trivial element of the extension group $Ext^1(\cL^{*}, \cL)=H^1(X, \cL^2)$. That is, we are defining an indecomposable $SU(2)$ bundle $V$ of the form \eref{E7_eg}. 
Furthermore, as a $C_2^{L}$ field acquires a vev, the mass of the anomalous $U(1)$ vector multiplet increases as we move deeper into the stable region of K\"ahler moduli space. At some distance from the stability wall, this reaches the compactification scale and must be integrated out of the effective theory \cite{Anderson:2009nt}. Thus, the purely $E_7$ gauge group one would expect from the indecomposable $SU(2)$ bundle in \eref{E7_eg} is obtained. Finally, we note that as a charged bundle modulus $C_2^{L}$ acquires a vev, its fluctuations also obtain a mass from the $D^{U(1)}$ contribution to the potential. As a result, one expects the number of vector bundle moduli in the stable region to be the {\it number of $C_2$-fields $-1$}. Direct computation yields
\beq\label{moduli_count}
h^1(X,V\otimes V^{*})=h^1(X, \cL^2)-1=10-1=9 \ ,
\eeq
as required. A general argument that guarantees the matching of $h^1(X,V\otimes V^*)$ with massless mode counting in the supersymmetric region is given in the Appendix of \cite{Anderson:2009nt}.

\begin{figure}
\centering
\mbox{\subfigure{\includegraphics[width=3in]{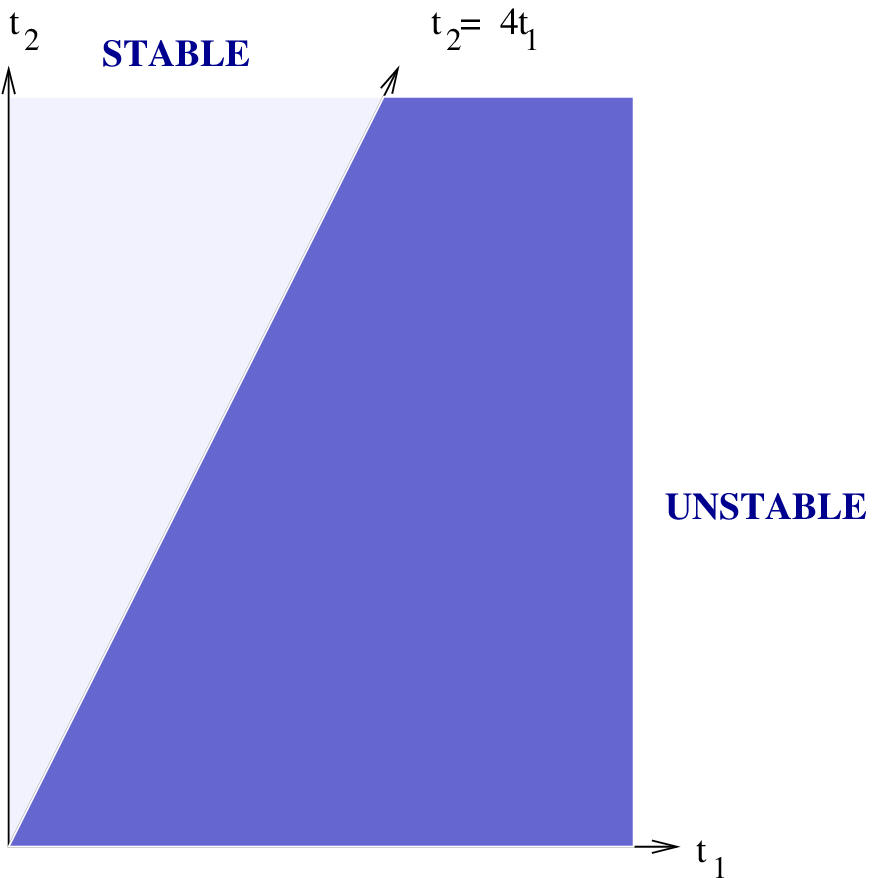}}\quad
\subfigure{\includegraphics[width=3in]{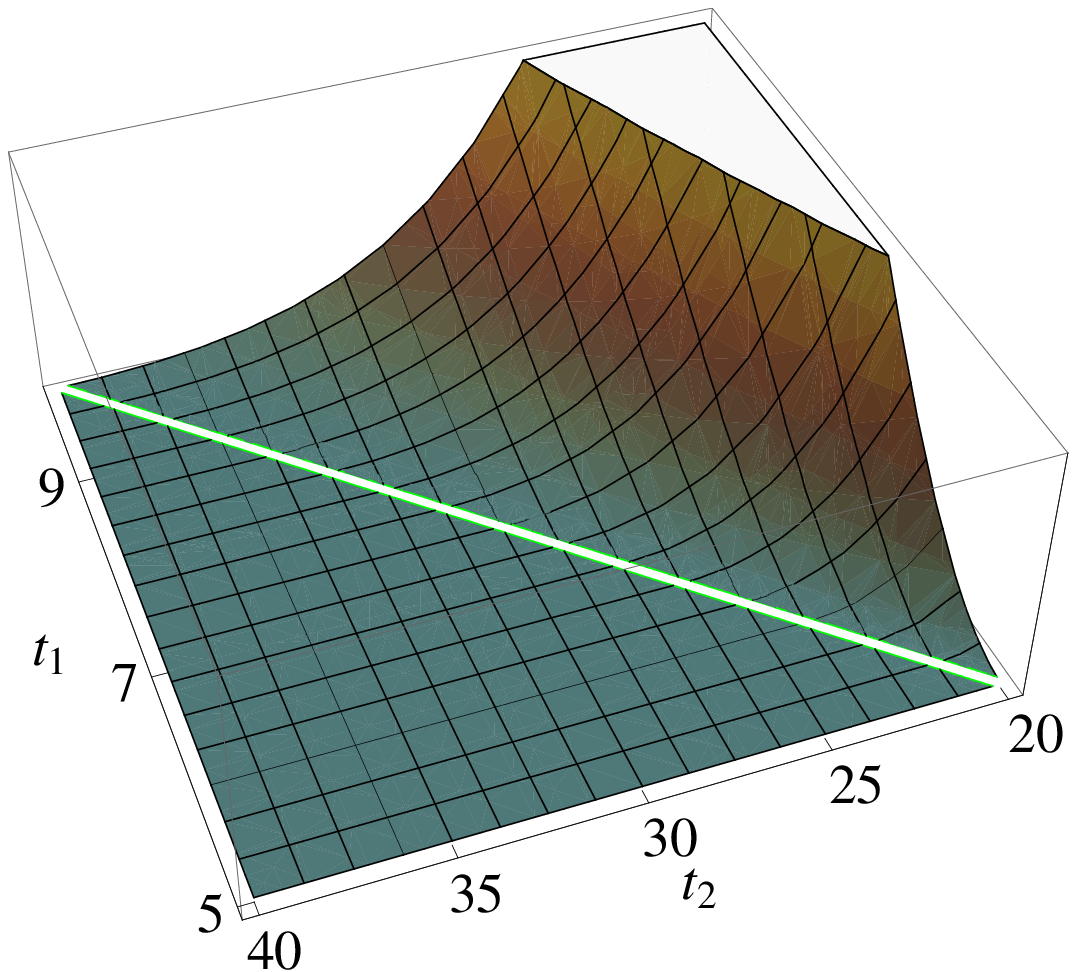} }}
\caption{\emph {An example of a stability wall associated with the bundle \eref{E7_eg} in Section \ref{stab_wall_review}. In the two chambers of the K\"ahler cone (shown at left) the bundle $V$, is respectively, stable/unstable. The non-trivial potential (shown at right) forces the system back into the supersymmetric region of moduli space.}} \label{fig_stable}
\end{figure}

Similarly, in the region of moduli space where $\mu(\cO_{X}(-1,1))>0$, that is, where $V$ is unstable, the field theory correctly describes the geometry. In this region, we find that since there are only negatively charged $C_{2}$ fields, there is nothing that can cancel the positive FI term. As a result, $D^{U(1)} \neq 0$ and supersymmetry is spontaneously broken in this region. The potential energy forces the vacuum back towards the supersymmetric region of moduli space. The chamber structure of the K\"ahler cone associated with \eref{E7_eg} and the non-trivial D-term potential are shown in Figure \ref{fig_stable}.

As demonstrated above, the algebraic geometry of slope-stability and K\"ahler cone substructure can be simply understood from the point of view of effective field theory as the properties of an anomalous $U(1)$ D-term. In this language, whether or not supersymmetric vacua exist in a given region of moduli space can be decided by considering the effective theory near the stability wall and, in particular, the $U(1)$-charged moduli content of the theory. When charged moduli exist with the appropriate sign to cancel the FI term, $D^{U(1)}=0$ and the vacuum is supersymmetric. However, if no such moduli are available, the potential is non-zero and the bundle is unstable.
What happens when we have more than one sign of charged moduli? Naively, this seems strange since the bundle under consideration is clearly unstable when $\mu(F)>0$, but the positively charged matter indicate that D-flatness (that is, stability) can be maintained everywhere in K\"ahler moduli space. A geometric analysis resolves this confusion. As we will see in the next section, the different $U(1)$ charged moduli will correspond to branch structure in the theory. In fact, stability walls and the simple effective field theory description of them can provide insight into transitions between distinct geometries. The study of these transitions will be the subject of the remainder of this paper.

\section{Rank-Preserving Transitions and S-Equivalence Classes}\label{rank_pres_sec}
As discussed in the previous section, the slope-stability properties of a bundle can be accurately described in effective field theory by an enhanced $U(1)$ symmetry at a stability wall and the D-term associated with it. Specifically, whether or not a bundle is stable in a given region of moduli space is decided by the sign of the FI term and the $U(1)$ charges of the matter fields and bundle moduli. In the simple example of an $SU(2)$ bundle with a stability wall, the $E_7 \times U(1)$ ``wall-theory" spectrum contained no matter fields and only $C_2^{L}$ charged bundle moduli. The vevs of these fields can cancel the FI term, setting the $U(1)$ D-term to zero in the region $t^2/t^1>4$ where the slope of the de-stabilizing sub-sheaf (and, hence, the FI term) are negative.
However, the absence of any bundle moduli with positive charge makes it clear that D-flatness cannot be preserved in the region of K\"ahler moduli space with $t^2/t^1<4$.

What happens if bundle moduli with more than one sign are present? To answer this, again consider the general case of an $SU(n)$ bundle $V$ de-stabilized by a single sub-bundle $\cF$. 
Then
\beq\label{gen_V}
0 \to \cF \to V \to {\cal K} \to 0 \ ,
\eeq
where $V/\cF={\cal K}$. Assume, for simplicity, that $\cF$ and ${\cal K}$ are both stable bundles. At the stability wall defined by $\mu(\cF)=\mu(V)=0$, $V$ will break into a sum of two pieces
\beq
V=\cF \oplus {\cal K}~~\text{where}~~ n_1=rk(\cF),~~n_2=rk({\cal K})~~\text{and}~~n_1\oplus n_2=n \ .
\eeq
At this locus, the structure group changes from $SU(n)$ to $S[U(n_1) \times U(n_2)] \simeq SU(n_1) \times SU(n_2) \times U(1)$. This  induces an enhancement of the low energy gauge group from $G$ (the commutant of $SU(n)$ in $E_{8}$) to $G \times U(1)$. 

Now suppose that there are {\it two types} of charged bundle moduli present, namely, the $C_1$ and $C_2$ associated with the cohomology groups
\beq\label{c_charges}
C_1 \in H^1(X, {\cal K}\otimes \cF^{*} )~~~\text{and}~~~C_2 \in H^1(X, \cF  \otimes{\cal K}^{*}) \ .
\eeq
Generically these two types of fields have opposite $U(1)$ charge, which we denote by $q$ and $-q$ respectively. Here, $q$ is a positive number whose magnitude depends on the rank $n$ of the structure group. Furthermore, in this section we will take the vevs of all $G$ charged matter multiplets to vanish.\footnote{Note that there are several circumstances for which this naturally occurs: 1) there are {\it no} matter multiplets in the spectrum, as in our previous example, and 2) there {\it are} matter multiplets present, but their vevs are set to zero to assure $G$ D-flatness and/or F-flatness. Our analysis will also apply more generally to any vacua where we simply focus our attention on the locus of vanishing matter vevs.} In this case, the $G$ D-term is trivially zero and the $U(1)$-D-term takes the form
\beq\label{dterm_doubleC}
D^{U(1)} \sim \frac{\mu(\cF)}{{\cal V}} -qG_{L\bar{M}} C_{1}^{L} C_{1}^{\bar M}+qG_{L\bar{M}} C_{2}^{L} C_{2}^{\bar M} \ .
\eeq
Note that, in addition to this D-term, one must also consider the moduli contribution to the superpotential. This is given by 
\bea\label{w_high}
W \sim \lambda_0 (C_1 C_2) ^2 \ldots\ ,
\eea
where the indices on both fields and couplings are suppressed. Here and elsewhere in this paper, higher dimension contributions to $W$ that do not change the conclusions are ignored. 

To begin our analysis of the vacuum structure, consider the region of K\"ahler moduli space for which $V$ is slope-stable; that is, when $\mu(\cF)<0$. As before, one can cancel the FI term and, hence, the entire $U(1)$ D-term \eqref{dterm_doubleC} via a vev 
$\langle C_2\rangle \neq 0$. This cancellation does not, by itself,  preclude the possibility that $C_{1}$ also has a non-vanishing vev. However, we now show generically that in the presence of superpotential terms such as \eref{w_high}, one must take $\langle C_1 \rangle=0$
in the slope-stable region of $V$ .
To see this, note that
in addition to the vanishing of the $U(1)$ D-term, the supersymmetric, Minkowski vacua we are seeking must satisfy the equations
\bea \nonumber
\partial_{C_1} W &=& \lambda_{0} C_2 (C_1 C_2) =0 \ , \\  \label{Weqns}
\partial_{C_2} W &=& \lambda_{0} C_1 (C_1 C_2) =0 \ , \\ \nonumber W &=&
\lambda_{0} (C_1 C_2)^{2} \ .  \eea 
That is, they must be F-flat with vanishing cosmological constant. With $\mu(\cF)<0$ in \eref{dterm_doubleC}, one might suppose that, to preserve supersymmetry, the fields $C_1$ and
$C_2$ could {\it both} get vevs so that the last two terms in $D^{U(1)}$ cancel the FI
term. However, substituting these two non-zero vevs into equations \eqref{Weqns}, it is clear that no such solution is generically possible. Thus, to move into the stable region of $V$ and obtain a
Minkowski vacuum, the only clearly unobstructed choice available is to take all $\langle C_{1} \rangle=0$ and choose non-vanishing $C_{2}$ vevs to cancel the FI term in \eqref{dterm_doubleC}. Now consider the region where $V$ is unstable; that is, when $\mu(\cF)>0$. Here, can set the D-term to zero by choosing $\langle C_1 \rangle \neq 0$. In this case, to obtain vanishing F-terms and $W$ it is necessary to take $\langle C_2 \rangle=0$. In general then, F-flatness and $W=0$ require that the vevs of $C_{1}$ and $C_{2}$ cannot be simultaneously non-trivial. 

It is clear that when both charged moduli in \eref{c_charges} are present in the spectrum, one can cancel the FI term in the region in K\"ahler moduli space for which $\mu(\cF)<0$ via a vev $\langle C_2 \rangle \neq 0$, as well as when $\mu(\cF)>0$ with $\langle C_1 \rangle \neq 0$. Recall however, that a direct analysis of bundle \eref{gen_V} shows that $V$ is manifestly unstable in the region of moduli space where $\mu(\cF)>0$. How then do the geometry and field theory match?
The answer is that D- and F-flatness actually define two ``branches" of the vacuum space, separated by the stability wall. These branches, namely $\langle C_1 \rangle \neq 0$, $\langle C_2 \rangle =0$ and 
$\langle C_1 \rangle = 0$, $\langle C_2 \rangle \neq 0$, respectively correspond to {\it two different vector bundles}. The bundle described by re-mixing the decomposed sum $\cF \oplus {\cal K}$ via a non-trivial element $C_1 \in H^1(X, {\cal K}\otimes \cF^{*})$ is {\it not isomorphic} to the bundle $V$ in \eref{gen_V}. By crossing the stability wall and canceling the FI term in \eref{dterm_doubleC} with the moduli in $H^1(X, {\cal K}\otimes \cF^{*})$, we are defining a new geometry. Specifically,
the $Ext^1$ groups determining the $C$-moduli correspond to the moduli spaces of two different extension bundles \cite{AG,AG2, Braun:2005ux}, 
\begin{eqnarray} \label{branch_bundles}
&& 0 \to \cF \to V \to {\cal K} \to 0~~~\Leftrightarrow~~~ \textnormal{Ext}^1({\cal K},{\cal F})  \\ \label{branch_bundles2}
&& 0 \to {\cal K} \to {\tilde V} \to \cF \to 0~~~\Leftrightarrow~~~ \textnormal{Ext}^1({\cal F},{\cal K})  
 \end{eqnarray}
respectively.
The sequences, \eref{branch_bundles} and \eref{branch_bundles2}, are referred to as an extension sequence and its ``reverse'' extension\footnote{Note that in the literature \eref{branch_bundles2} is frequently referred to as the ``dual" extension to \eref{branch_bundles}. Here, we use different terminology to avoid confusion with the dual, $V^*$, of a bundle $V$.}. Note that these two types of bundle deformations -- corresponding to turning on vevs for charged bundle moduli in $Ext^1({\cal F},{\cal K})$ or $Ext^1({\cal K},{\cal F})$ respectively -- preserve the rank of the structure group and, hence, the gauge symmetry of the four-dimensional effective theory. Except at the zero of their 
$Ext^1$ groups, where each splits into the direct sum ${\cal{F}} \oplus {\cal{K}}$, $V$ and $\tilde{V}$ are not isomorphic. 

\begin{figure}
\centering
\mbox{\subfigure{\includegraphics[width=3in]{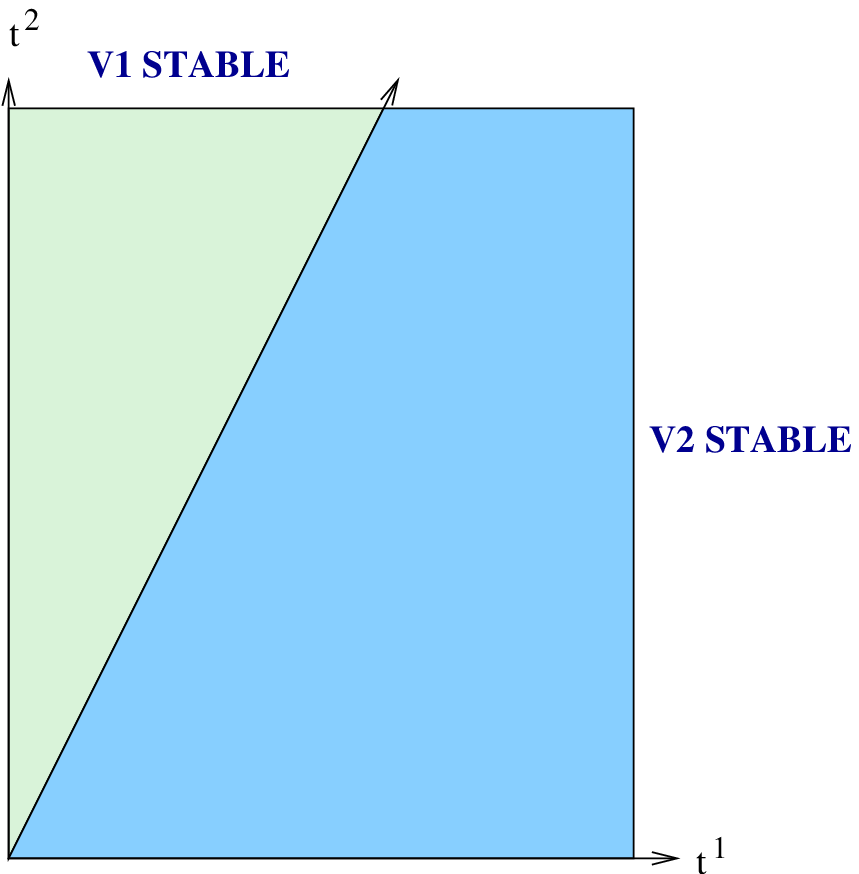}}\quad
\subfigure{\includegraphics[width=3in]{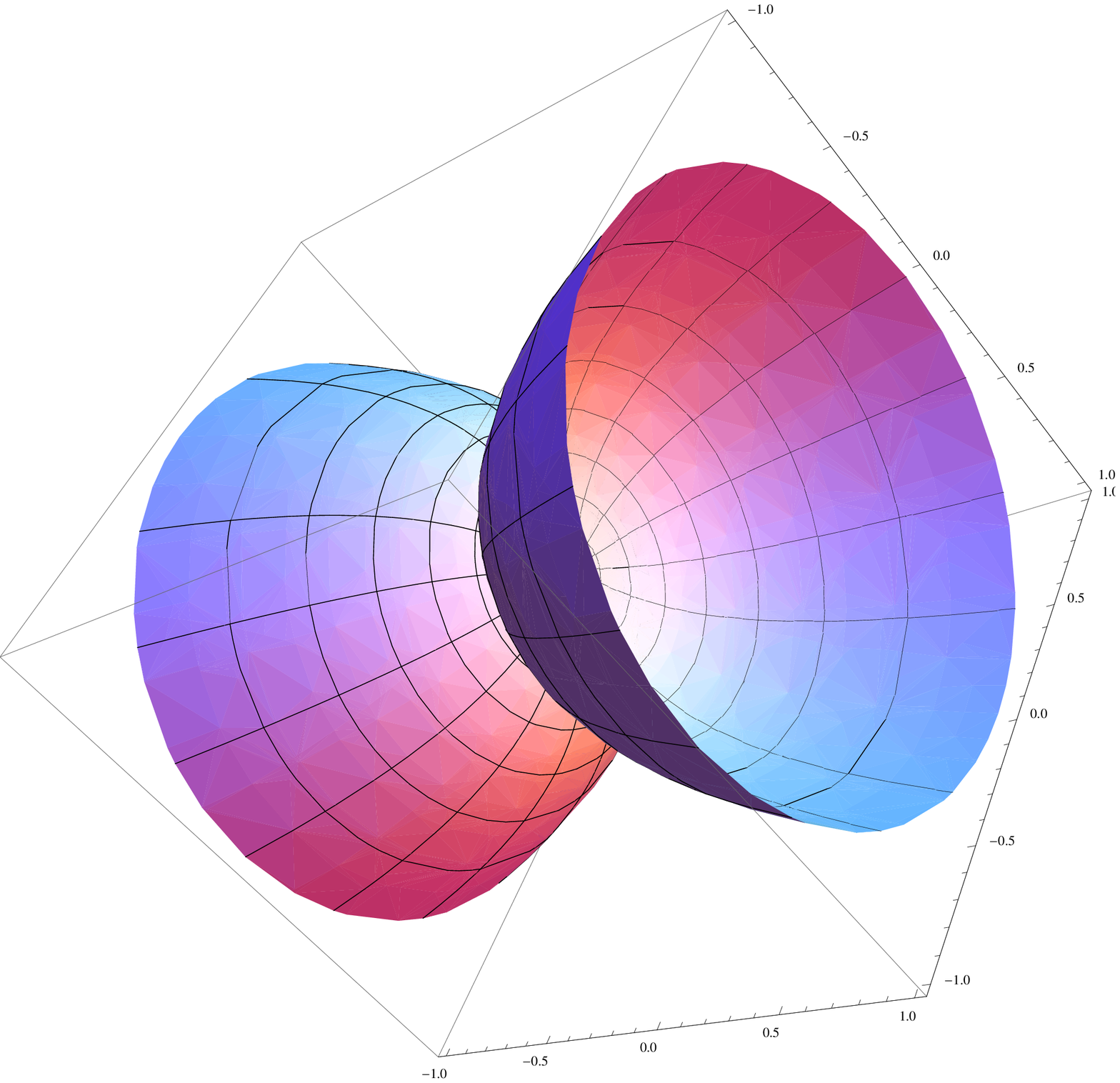} }}
\caption{\emph {An illustration of a stability wall connecting two bundles, $V_{1}$ and $V_{2}$. In the two chambers of the K\"ahler cone (shown at left), $V_1$ and $V_2$, respectively, are stable. The moduli spaces of the two stable bundles (shown at right) are connected at only at a point, corresponding to the shared zero of the two $Ext^1$ groups in \eref{branch_bundles2}.}} \label{fig12}
\end{figure}

To see this, note that 
for an $SU(n)$ structure group $c_{1}(V)=0$ and, hence, from \eqref{gen_V} that 
\begin{equation}
\label{nw1}
c_{1}({\cal{F}})= -c_{1}({\cal{K}}) \ .
\end{equation}
It follows from the definition of slope in \eqref{slope} that, away from the stability wall, $\mu({\cal{F}})$ and $\mu{(\cal{K}})$ are both non-vanishing with opposite sign. We then see from sequences \eqref{branch_bundles} and \eqref{branch_bundles2} that 
\begin{equation}
\label{what2}
V~ {\rm stable} ~\Leftrightarrow~ \mu({\cal{F}})<0 ~\Leftrightarrow~ \mu({\cal{K}})>0 ~\Leftrightarrow~ {\tilde{V}}~ {\rm unstable} \ .
\end{equation}
That is, at a point in K\"ahler moduli space if one extension bundle is stable its reverse bundle is unstable and vice versa. Hence, $V$ and ${\tilde{V}}$ are never isomorphic. It follows that
the stable moduli spaces of $V$ and $\tilde{V}$ touch at only one point -- namely the shared zero of the two extension groups $Ext^1({\cal K},\cF)$ and $Ext^1(\cF,{\cal K})$. These moduli spaces are represented schematically\footnote{A detailed discussion of birational transitions between such moduli spaces (similar to flips) can be found in \cite{ellingsrud}.} in Figure \ref{fig12}. That is, generically, we can supersymmetrically define only one of $V$ and $\tilde{V}$ for a give chamber in K\"ahler moduli space.
Why one cannot give non-vanishing vevs to $C_1$ and $C_2$ simultaneously can be explained geometrically by the correspondence between higher-order superpotential terms, such as \eref{w_high}, and constraints on the  the vector bundle arising from holomorphicity -- specifically, an obstruction to the Kodaira-Spencer class in $H^1(X,V\otimes V^*)$. We will discuss such obstructions in future work \cite{us_soon}.

We conclude that, at the stability wall, if the relevant cohomologies
are all non-zero, then there are always at least two ways of
transforming the split $S[U(n_1) \times U(n_2)]$ bundle into an
$SU(n)$ bundle. One can move smoothly between them along $D$- and
$F$-flat directions in K\"ahler and vector bundle moduli space.  We
refer to a transition from a bundle of the form \eref{branch_bundles}
to that in \eref{branch_bundles2}, and vice versa, as a ``rank
preserving transition".
\subsection{An Example}\label{so10_eg}
In this section, we consider a simple example of a rank preserving transition. The effective field theory contains two branches, corresponding to two distinct geometries of the form shown in \eref{branch_bundles} and \eref{branch_bundles2}. This example will clearly illustrate that while the topology of the bundle is preserved in these transitions, many properties of the the low energy theory, including the massless spectrum, can change dramatically.

For this example, we use the same complete intersection Calabi-Yau (CICY) threefold as in the previous section; that is, the degree $\{2,4\}$ hypersurface in $\mathbb{P}^1\times \mathbb{P}^3$. Over this threefold, consider the split rank $4$ bundle
\beq\label{so10_V}
V= {\cal Q}\oplus {\cal Q}^{*} \ ,
\eeq
where ${\cal Q}$ is itself a stable rank $2$ bundle defined by
\beq\label{Qdef}
0 \to {\cal Q} \to \cO_{X}(1,0) \oplus \cO_{X}(0,1)^{\oplus 2} \to \cO_{X}(2,1) \to 0
\eeq
with $c_1({\cal Q})=(-1,1)$. $V$ has the reduced structure group $S[U(2) \times U(2)]\simeq 
SU(2) \times SU(2) \times U(1)$ and is poly-stable along the co-dimension one line in K\"ahler moduli space defined by $\mu({\cal Q})=\mu({\cal Q}^{*})=0$. This corresponds to a stability wall at $t^2/t^1=4$, as in our previous example. The associated effective field theory here has enhanced $SO(10) \times U(1)$ symmetry (the commutant of the structure group in $E_{8}$). To determine which indecomposable, stable $SU(4)$ bundles can be obtained by deforming the split bundle, one must examine the $SO(10)\times U(1)$ particle content of the stability wall theory. This is given in Table \ref{table1}. By inspection, we find that there are two types of $U(1)$-charged bundle moduli, 
\beq\label{c_choices}
C_1:~h^1(X,{\cal Q} \otimes {\cal Q})=27~~~{\rm and}~~~C_2:~h^1(X,{\cal Q}^{*} \otimes {\cal Q}^{*})=1 \ ,
\eeq
with charges $+2$ and $-2$ respectively. 
Giving a vev to $C_1$ or $C_2$ defines two branches,
\bea \label{office1}
& & \langle C_1 \rangle\neq 0:~~~\quad 0 \to {\cal Q} \to V_1 \to {\cal Q}^{*} \to 0  \\ \label{office2}
& & \langle C_2 \rangle \neq 0:~~~\quad 0 \to {\cal Q}^{*} \to V_2 \to {\cal Q} \to 0  ~~\ ,
\eea

\begin{table}[t]
\begin{center}
\begin{tabular}{|c|c|c||c|}
  \hline
  Representation & Field Name & Cohomology & Multiplicity \\ \hline
  ${\bf 1}_0$ &  $\phi_1$ & $H^1(X,{\cal Q} \otimes {\cal Q}^{*})$ & 8 \\ \hline
  ${\bf 1}_{+2}$ &  $C_1$ & $H^1(X,{\cal Q} \otimes {\cal Q})$ & 27 \\ \hline
  ${\bf 1}_{-2}$ &  $C_2$ & $H^1(X,{\cal Q}^{*} \otimes {\cal Q}^{*})$ & 1 \\ \hline
  ${\bf 16}_{+1}$ & $f_1$ &  $H^1(X,{\cal Q})$ & 2 \\ \hline
  $\overline{{\bf 16}}_{+1}$ & $\tilde{f}_2$ &  $H^1(X,{\cal Q^{*}})$ & 2 \\ \hline
  ${\bf 10}_{+2}$ & ${h}_1$ &  $H^1(X,\wedge^{2}{\cal Q})$ & 0 \\ \hline
  ${\bf 10}_{-2}$ & ${h}_2$ &  $H^1(X,\wedge^{2}{\cal Q}^{*})$ & 0 \\ \hline
\end{tabular}
\caption{{\it The representations, fields and the associated cohomology groups for the $SO(10)\times U(1)$ stability wall theory given in Subsection \ref{so10_eg}. The multiplicities for the specific decomposable rank $4$ vector bundle $V={\cal Q}\oplus {\cal Q}^{*}$ defined in \eref{so10_V} are presented in the fourth column. The field labels are chosen to match Appendix A of \cite{Anderson:2010tc}. }}
\label{table1}
\end{center}
\end{table}
\noindent in the vacuum space. Let us analyze each branch from a both a geometrical and effective field theory  point of view. 
\vspace{5pt}

\noindent {\bf Branch 1:} In this branch, we consider the extension sequence
\beq\label{V1def}
0 \to {\cal Q} \to V_1 \to {\cal Q}^{*} \to 0  ~~ \Leftrightarrow ~~ Ext^{1}({\cal Q}^{*},{\cal Q})=H^{1}(X,{\cal Q} \otimes {\cal Q}) \ .
\eeq
We see from this and \eref{c_choices} that giving a vev to at least one of the $27$ $C_1$-fields defines an indecomposable $SU(4)$ bundle $V_{1}$. Furthermore, this bundle will be stable when $\mu({\cal{Q}})<0$. This occurs in the region of K\"ahler moduli space with $t^2/t^1<4$. From the long exact sequence in cohomology associated with \eref{V1def}, we find that
\beq
\label{scr1}
h^1(X,V_1)=h^1(X, {\cal Q})=2~,~h^1(X,V_{1}^{*})=h^2(X, {\cal Q}^{*})=2
\eeq
and 
\beq
\label{good_count0}
h^1(X,V_1 \otimes V_{1}^{{*}})= 34 \ .
\eeq
The latter result is simply 
\beq\label{good_count}
h^1(X,V_1 \otimes V_{1}^{{*}})= h^1(X, {\cal Q}\otimes {\cal Q})-1 +h^1(X,  {\cal Q}\otimes {\cal Q}^{*})=27-1+8=34 \ .
\eeq
It follows from \eqref{scr1}, \eqref{good_count0} and Table \ref{tableVcomp} that there are 2 ${\bf {16}}$ and 2 ${\bf {\overline{16}}}$ matter multiplets and 34 vector bundle moduli in the zero-mode spectrum of bundle $V_{1}$. How does one interpret \eqref{good_count} in terms of the effective field theory? To define $V_1$, we are using 
$\langle C_1 \rangle$ to cancel the FI term in \eref{dterm_doubleC}. As a result, we expect one $C_1$ superfield to become massive  through the super-Higgs mechanism as one moves from the stability wall into the $V_1$-branch. In addition, the single $C_2$ field gains a mass through the $(C_1C_2)^2$ term in the superpotential. As a result, after integrating out the massive fields one would expect the number of remaining massless bundle moduli to be the {\it number of $C_1$-fields $-1$}~+~{\it number of $\phi_1$-fields}. This is precisely the counting given in \eref{good_count}.
\begin{table}[t]
\begin{center}
\begin{tabular}{|c|c|c|}
  \hline
  Representation & Cohomology of $V_1$ & Cohomology of $V_2$ \\ \hline
  ${\bf 1}$ &  $h^1(X,V_1 \otimes V_{1}^{*})=34$& $h^1(X,V_2 \otimes V_{2}^{*})=8$\\ \hline
  ${\bf 16}$ & $h^1(X,V_1)=2$&  $h^1(X,V_2)=0$ \\ \hline
  $\overline{{\bf 16}}$ & $h^1(X,V_{1}^{*})=2$ &  $h^1(X,V_{2}^{*})=0$ \\ \hline
  ${\bf 10}$ & $h^1(X,\wedge^2 V_1)=0$ &  $h^1(X,\wedge^{2}V_{2})=0$ \\ \hline
\end{tabular}
\caption{{\it The representations and associated cohomology groups for the two $SO(10)$ theories corresponding to the branch structure described in Subsection \ref{so10_eg}. The particle multiplicities for each of the rank $4$ vector bundles, $V_1$ and $V_2$, defined in \eref{V1def} and \eref{V2def} are given in the second and third columns.}}
\label{tableVcomp}
\end{center}
\end{table}

\vspace{5pt}

\noindent {\bf Branch 2:}
In this branch, we consider the reverse extension sequence
\beq\label{V2def}
0 \to {\cal Q}^{*} \to V_2 \to {\cal Q} \to 0 ~~ \Leftrightarrow ~~ Ext^{1}({\cal Q},{\cal Q}^{*})=H^{1}(X,{\cal Q}^{*} \otimes {\cal Q}^{*}) \ .
\eeq
From this and \eqref{c_choices}, we see that giving a vev to the $C_2$-field defines an indecomposable $SU(4)$ bundle $V_{2}$. This bundle will be stable when $\mu({\cal{Q}}^{*})<0$. This occurs in the region of K\"ahler moduli space with $t^2/t^1>4$. 
What is the low energy spectrum on this side of the stability wall? From the long exact sequence in cohomology associated with \eref{V2def}, we find that
\beq
0 \to H^1(X,V_2) \to H^1(X, {\cal Q}) \stackrel{h}{\to} H^2(X, {\cal Q}^{*}) \to H^2(X, V_2) \to 0 \ .
\eeq
Furthermore, the co-boundary map $h$ is an element of $H^1(X, {\cal Q}^{*}\otimes {\cal Q}^{*})$. Since this space is one-dimensional, then $h=\langle C_2 \rangle$ and the map is an isomorphism. As a result,
\beq
\label{scr2}
h^1(X, V_2)=h^1(X, V_{2}^{*})=0
\eeq
and 
\beq\label{good_mod}
h^1(X,V_2 \otimes V_{2}^{{*}})=8 \ .
\eeq
It follows from \eqref{scr2}, \eqref{good_mod} and Table \ref{tableVcomp} that there are {\it no} ${\bf {16}}$ or ${\bf {\overline{16}}}$ matter multiplets and $8$ vector bundle moduli in the zero-mode spectrum of bundle $V_{2}$. In terms of the effective theory, we expect the result \eqref{good_mod} since the single $C_2$ field becomes massive through the super-Higgs effect and the $27$ $C_1$ fields get mass through the $(C_{1}C_{2})^2$ term in the superpotential, analogously to Branch 1. The $8$ uncharged $\phi_1$ fields are all that remain of the bundle moduli, as in \eref{good_mod}.

\vspace{5pt}

Having analyzed both supersymmetric branches in moduli space, it is of interest to ask how the low-energy field theory describes the transition between the two? More specifically, from the point of view of the effective theory, how does one reconcile the change in spectrum between the two branches? To understand this, consider the following contributions to the superpotential near the stability wall,
\beq
W \sim f_1\tilde{f}_2C_2 + \ldots +(C_1C_2)^2 + \ldots \ ,
\eeq
where $f_1$ and $\tilde{f}_2$ are the ${\bf 16}$ and $ {\bf {\overline{16}}}$ fields respectively in 
Table \ref{table1}. We have suppressed the coupling parameters and indices for simplicity. The first term in $W$ produces F-term contributions to the potential energy of the form
\beq\label{f_term}
|{\tilde f}_2|^2|C_2|^{2}+|f_1|^2|C_2|^2 \ .
\eeq
For the $V_1$-branch of the vacuum, in which $\langle C_2 \rangle=0$, these F-terms are trivial and the $2$ ${\bf 16}$ and $2$ ${\bf {\overline{16}}}$ fields remain massless. However, in the branch associated with $V_2$, where $\langle C_2 \rangle \neq 0$, it follows from \eref{f_term} that both  $f_1$ and ${\tilde f}_2$ acquire mass. This is consistent with the results in Table \ref{tableVcomp}. While the topology is preserved throughout the smooth deformation of $V_1$ into $V_2$, this example illustrates how the low energy spectrum can change significantly. 

\subsection{Branch Structure and S-Equivalence Classes}\label{branch_goodness}
The branch structure possible near a stability wall can be considerably more complex than the examples presented above. Depending on the properties of the de-stabilizing sub-sheaves of a bundle, there can be many more than two branches to the effective theory at a stability wall. In general, a rank $n$ bundle can decompose into a direct sum of $m \leq n$ sheaves on the stability wall; that is,
\beq\label{wall_decomp}
V \to \bigoplus_{i}^{m} V_i~,~~\mu(V_i)=\mu(V)=0 \ .
\eeq
If there are $m$ terms in the decomposition of $V$, the low-energy gauge group $G$ will be enhanced by $m-1$ anomalous $U(1)$ factors to $G\times U(1)^{m-1}$. Each of these $U(1)$ symmetries will contribute its own D-term, with a K\"ahler moduli-dependent FI parameter, to the effective potential.

How many branches in the moduli space do we expect? As a first step, note that generically there will 
be $m^2-m$ different types of $U(1)$-charged bundle moduli. These are described by the cohomology groups of the form
\beq\label{ext_types}
Ext^1(V_j, V_i)=H^1(X, V_i \otimes V_{j}^{*})~,~~i \neq j \ .
\eeq
Hence, considering D-flatness only, to determine a branch it is clear that one must choose $m-1$ moduli from the $m^2-m$ types in \eref{ext_types} to acquire vevs to cancel the $m-1$ FI terms. However, we must satisfy {\it both D-flatness and F-flatness}. To count the actual number of branches, the superpotential and its associated F-terms must be taken into account, as well as possibly equivalent directions. This counting can rapidly become complicated and dependent on the particular $U(1)$ charges of the fields associated with \eref{ext_types}. Several examples of constraints from F-terms will be presented below.

To organize the study of such branch structure, we will make use of mathematical statements concerning the form that the direct sum \eref{wall_decomp} can take and use them to classify possible bundles arising at a stability wall. To begin, note that on the stability wall itself a bundle $V$ is {\it semi-stable} for generic values of its moduli. It is only for special choices of the moduli (that is,  for $\left<C_i \right> \to 0$ as discussed in the previous section) that $V$ becomes poly-stable and splits into the direct sum in \eref{wall_decomp}. To understand the chamber structure of K\"ahler moduli space away from the stability wall (and, hence, the stable bundles which live in each chamber), we will use results about the wall itself and the classification of semi-stable bundles. In particular, the family of distinct bundles (describing different vacuum branches) that we hope to classify share a stability wall. While the structure of their sub-sheaves may be very different, the decomposition of each bundle on the wall where it becomes semi-stable is in fact, universal. The following theorem makes this idea explicit and explains how unstable sheaves may be described in terms of semi-stable sheaves, and semi-stable sheaves in terms of stable sheaves~\cite{Huybrechts}.

\begin{theorem}[Harder-Narasimhan]
  \label{HS_fil}
  Given a holomorphic bundle $V$ over a closed K\"ahler manifold
  $X$ (with K\"ahler form $\omega$), there is a filtration (called the
  Harder-Narasimhan filtration) by sub-sheaves
  \begin{equation}\label{hs_unstable}
    0=\cF_0 \subset \cF_1 \subset \ldots \cF_{m}=V
  \end{equation}  
  such that $\cF_i/\cF_{i-1}$ are semi-stable sheaves for $i=1,\ldots m$ and the slope of the quotients are ordered
  \begin{equation}
    \mu(\cF_1) > \mu(\cF_2/\cF_1)> \ldots \mu(\cF_{m}/\cF_{m-1}) \ .
  \end{equation}
  If $V$ is semi-stable, then there is a filtration by sub-sheaves (called the Jordan-H\"{o}lder filtration)
  \begin{equation}\label{filt}
    0=\cF_0 \subset \cF_1 \subset \ldots \cF_{m}=V
  \end{equation}  
  such that the quotients $\cF_i/\cF_{i-1}$ are all stable sheaves and have slope $\mu(\cF_{i}/\cF_{i-1})=\mu(V)$. In addition
  \begin{equation}\label{graded_sum}
    Gr(V)=\cF_1 \oplus \cF_{2}/\cF_{1}\oplus \ldots \cF_{m}/\cF_{m-1}
  \end{equation}
  is uniquely determined up to isomorphism (and is called the `graded sum').
\end{theorem}

One consequence of Theorem \ref{HS_fil} is a description of the moduli
space of semi-stable sheaves. Two semi-stable bundles $V_1$ and
$V_2$ are called \emph{S-equivalent} if
$Gr(V_1)=Gr(V_2)$. The concept of S-equivalence arises when
trying to define the notion of a moduli space of sheaves. Stable
bundles correspond to unique points in their moduli space, while a
moduli space of semi-stable sheaves can only be made Hausdorff if each
point corresponds to an S-equivalence class \cite{Huybrechts}. Note
that each S-equivalence class contains a unique poly-stable
representative, namely, the graded sum \eqref{graded_sum}.
In the following, we will see that this is relevant to stability wall branch structure  because, at the stability wall, the direct sum decomposition in \eref{wall_decomp} is the unique graded sum \eref{graded_sum} associated with each of the possible branches.  As a result, all stable bundles that can be constructed {\it away from a stability wall} sweep out an S-equivalence class {\it on the wall}.

As discussed above, the possible branches of supersymmetric vacua are determined by the $Ext^1$ groups which mix together various pieces of the direct sum \eref{wall_decomp}. The charged bundle moduli in \eref{ext_types}, that is, those which can appear in some $U(1)$ D-term, are described by
\beq\label{graded_ext}
Ext^1(\cF_i/\cF_{i-1},\cF_j/\cF_{j-1})
\eeq
for $i<j \leq m$ associated with terms $\cF_i/\cF_{i-1}$ in the graded sum $Gr(V)$ of \eref{graded_sum}.  It is straightforward to verify that any extension bundle built this way will be semi-stable {\it at the stability wall} and have a wall-decomposition $Gr(V)$ of form \eref{graded_sum}. This will be demonstrated with an explicit example in the following subsection. 

The central observation is that, at a stability wall, the mathematical and physical notions of a moduli space coincide. The effective field theory is entirely described by the decomposed form of the bundle on the stability wall; that is, the unique poly-stable description of a semi-stable bundle. 
On the other hand, the S-equivalence class is uniquely determined by the graded sum \eref{graded_sum}.\footnote{This description of semi-stable (and unstable) bundles in terms of S-equivalence classes has been of use \cite{Anderson:2010ke} in classifying the behavior of holomorphic bundles under numerical implementations of the generalized Donaldson Algorithm \cite{burt2,burt3}. 
In that classification, the gauge connection on an arbitrary bundle is described by its decomposition into a graded sum of the form \eref{graded_sum}.} Since the wall decomposition and the graded sum are the same object, it follows that the physical description is identical to the mathematical notion of an S-equivalence class.
The web of connected, rank-preserving vacua arising from a stability wall is simply related to an  S-equivalence class, a point in the moduli space of semi-stable sheaves. We will see this explicitly in the following example.

 \subsection{An Example of More Complicated Branch Structure}\label{more_comp}
To illustrate the more complicated types of branch structure possible,
we now consider a bundle whose poly-stable decomposition at a stability wall consists of multiple factors. In particular, we examine an $SU(3)$ bundle which decomposes into a sum of three line bundles, 
\beq \label{3piece} 
V \rightarrow  L_1 \oplus L_2 \oplus L_3 \ ,
\eeq
with structure group $S[U(1)\times U(1) \times U(1)] \simeq   U(1) \times U(1)$ at the stability wall. In the four-dimensional theory, the symmetry will be enhanced from $E_{6}$ to $E_6 \times U(1) \times U(1)$.
\begin{figure}[!ht]
  \centerline{\epsfxsize=3in\epsfbox{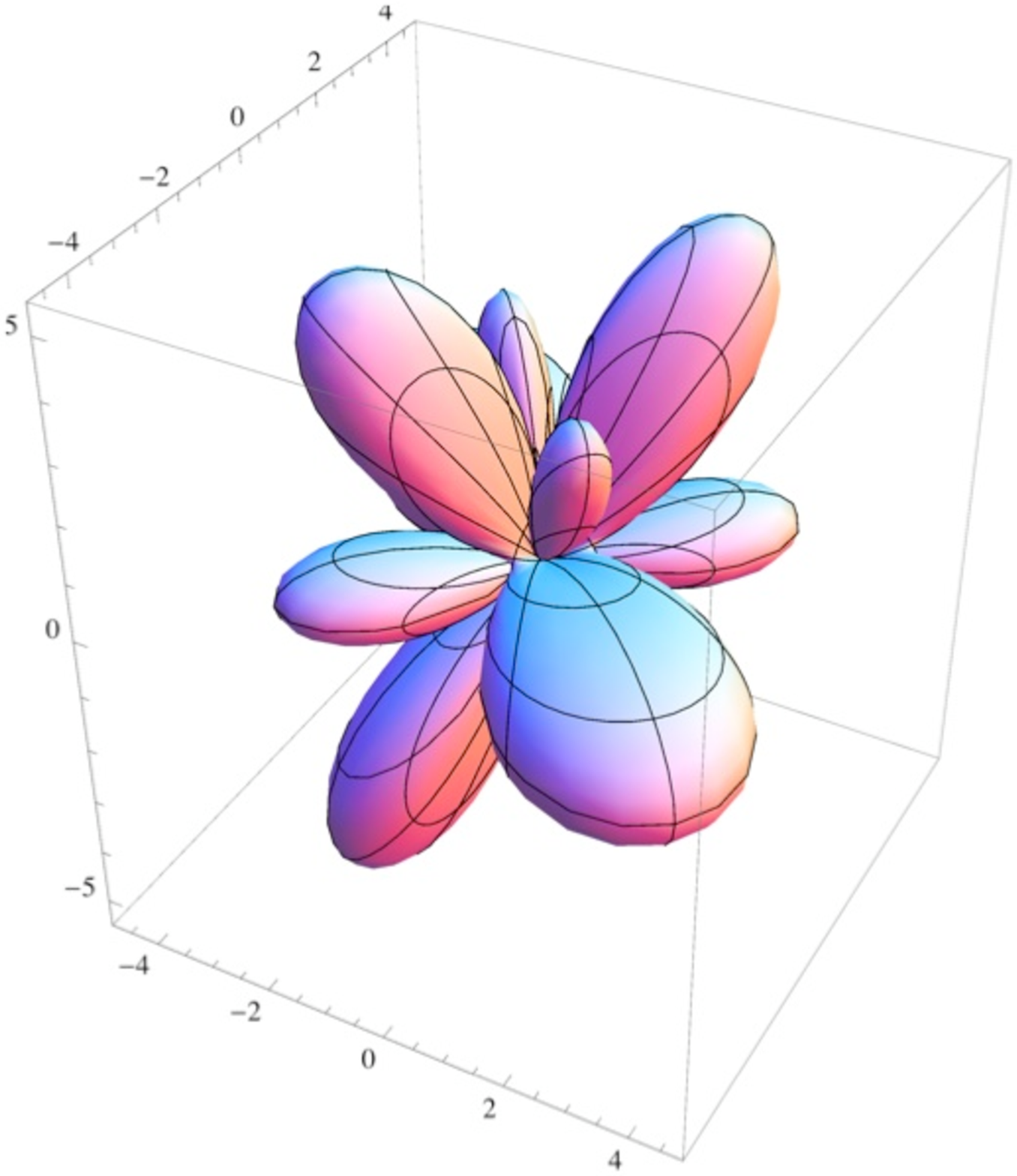}}
  \caption{\emph {An example of a stability wall connecting multiple bundles $V_{i}$. The local moduli spaces of the stable extension bundles, shown above, are connected at only at a point, corresponding to the shared zero of the $Ext^1$ groups in \eref{ext_types} and \eref{graded_ext}.}} \label{fig13} \end{figure}
As a result, on the stability wall, the multiplets in the four-dimensional theory will carry two additional $U(1)$ charges. The ${\bf 248}$ of $E_8$ decomposes as
 \bea \label{2Ddecomp} 
 E_8 &\supset& E_6 \times U(1) \times U(1) \ ,
\\ \nonumber {\bf 248} &=& {\bf 1}_{0,0} + {\bf 1}_{\frac{1}{2},3} +
{\bf 1}_{-\frac{1}{2},3} + {\bf 1}_{\frac{1}{2},-3} + {\bf
  1}_{-\frac{1}{2},-3} + {\bf 1}_{1,0}+ {\bf 1}_{-1,0} + {\bf
  78}_{0,0} \\ \nonumber && + {\bf 27}_{0,-2} + {\bf
  27}_{\frac{1}{2},1} + {\bf 27}_{-\frac{1}{2},1} + \overline{{\bf
    27}}_{0,2} + \overline{{\bf 27}}_{\frac{1}{2},-1} + \overline{{\bf
    27}}_{-\frac{1}{2},-1} \ ,
    \eea 
where the bold face number is the dimension of the $E_6$ representation and the subscripts are the
two $U(1)$ charges, $q_1,q_2$. The multiplicity of each such multiplet is given by the dimensions of  bundle-valued cohomology groups associated with \eqref{3piece}. The representations, field names and the cohomology groups for a generic sum of three line bundles are listed in the first three columns of Table \ref{tabledoubled}. 

Without loss of generality, one can write the two anomalous D-terms as functions of the slopes of any {\it two} of the line bundles in \eqref{3piece}. We choose $L_{1}$ and $L_{2}$ for specificity. The D-terms are then of the {\it generic} form
\bea\label{two_ds}
D_{1}^{U(1)} \sim \frac{\mu(L_1)}{{\cal V}} -q_1 G_{IJ}C^{I}{\bar C}^{J} \ , \quad
D_{2}^{U(1)} \sim \frac{\mu(L_2)}{{\cal V}} -q_2 G_{IJ}C^{I}{\bar C}^{J} 
\eea
where the charged bundle moduli and their two $U(1)$ charges $q_{1}$ and $q_{2}$ are given in \eref{2Ddecomp} and Table \ref{tabledoubled}. Although both slopes vanish on the stability wall, the assumption that the associated line bundles $L_{1}$ and $L_{2}$ destabilize $V$ implies that their slopes become negative in the interior of some stable chamber. Note that there are now {\it six} different types of charged bundle moduli, that is, $C$-fields in the $Ext^1$ groups  \eref{ext_types}.
Hence, we could consider all $15$ pair-wise combinations of the six $C$-fields which would satisfy D-flatness and attempt to determine the separate branches. However, as we now show, this number is generically reduced to only $6$ branches once full D- and F-flatness is taken into account.

In addition to the two D-terms in \eref{two_ds}, one must consider the superpotential. Focusing only on the $E_{6}$ singlets, this can be written as
\bea
\label{xmas1}
W= \lambda_{1} C_1 \tilde{C}_2 \tilde{C_3} +  \lambda_{2} \tilde{C}_1 C_2 C_3 +\lambda_{3} (C_1 \tilde{C}_1)^2+\lambda_{4} (C_2 \tilde{C}_2)^2 +\lambda_{5} (C_3
\tilde{C}_3)^2 + \ldots  \;.
\eea
For simplicity, we suppress indices and include terms
only to the dimension required for our analysis. In any stable
region, the four-dimensional effective theory has a supersymmetric
vacuum with vanishing cosmological constant. Therefore, as we vary the
K\"ahler moduli away from the stability wall into the $\mu (L_{1})
<0$, $\mu (L_{2})<0$ region, in addition to the vanishing of the two
D-terms we require
\begin{equation}
\label{fterms_all}
\partial_{C_{i}}W=\partial_{ {\tilde{C}}_{j}}W= W=0 \ , \quad  i,j=1,2,3 \ .
\end{equation}
Terms of the form $(C_i \tilde{C}_i)^2$, $i=1,2,3$ in
\eqref{xmas1} ensure that,  for each index $i$, either $C_i$ or $\tilde{C}_i$, but not both, can have a non-zero vev. The terms of the form $C_1 \tilde{C}_2 \tilde{C}_3$
and $\tilde{C}_1 C_2 C_3$ in \eqref{xmas1} generically ensure that only one of
$C_1$, $\tilde{C}_2$ and $\tilde{C}_3$, and only one of $\tilde{C}_1$,
$C_2$ and $C_3$, obtains a non-zero vev.  
Combining these results with the requirement that $D_{1}^{U(1)}=D_{2}^{U(1)}=0$, we find that for generic regions of moduli space there are $\it six$ supersymmetric branches associated with this stability wall in the K\"ahler cone -- each branch specified by a pair of non-vanishing $C$ fields. It is worth noting that for {\it special values} of the complex structure moduli or special choices of $C$-field vevs, these F-term obstructions may not be present and more branches may be accessible to the effective theory. We first consider the form of the six generically present branches before discussing any additional branches that might  appear at special loci in moduli space.

In terms of sequences, the different possible $C$ field vevs
correspond to the different ways of building a bundle $V$
from the three constituent line bundles $L_1$, $L_2$ and $L_3$. As a specific example,
take the case where
\beq
\label{bad1}
\langle \tilde{C}_2 \rangle \neq0~~\ , ~~\langle C_3 \rangle \neq 0
\eeq with all other $\langle C_i \rangle=0$. That is, we will use the
fields $\tilde{C}_2$ and $C_3$ to cancel the FI terms in \eref{two_ds}
and set the D-terms to zero. Geometrically, this corresponds to
defining a stable indecomposable rank $3$ bundle via the two sequences
\bea \label{firstone} & &0 \to L_1 \to {\cal{W}} \to L_3 \to 0 \
,\\ \label{secondone} & &0 \to L_2 \to V \to {\cal{W}} \to 0 \ .  \eea
The space of possible extensions associated with the first sequence is
given by $Ext^1(L_3,L_1) \cong H^1 (X,L_1 \otimes
L_3^{*})$. Therefore, this extension is non-trivial, that is,
${\cal{W}} \neq L_1 \oplus L_3$, if and only if one is at a non-zero
element of this cohomology group. We see from Table \ref{tabledoubled}
that this corresponds, in field theory language, to $< \tilde{C}_2>
\neq 0$ .  Similarly, sequence \eqref{secondone} is a non-trivial
extension if and only if one is at a non-trivial element in
$Ext^1({\cal{W}},L_2) \cong H^1(X,L_2 \otimes {\cal{W}}^{*})$. To see
how the cohomology $H^1(X,L_2 \otimes {\cal{W}}^{*})$ is related to
$\langle C_3 \rangle$, consider the dual sequence to
\eqref{firstone}. Tensoring with $L_{2}$, we find
\bea \label{tensored} 0 \to L_2 \otimes L_3^{*} \to L_2 \otimes
{\cal{W}}^{*} \to L_2 \otimes L_1^{*} \to 0 \;.  \eea The long exact
sequence associated with \eqref{tensored} is
\beq\label{tensored_cohom}
0 \to H^1(X,L_2\otimes L_3^{*}) \to H^1(X,L_2 \otimes {\cal{W}}^{*}) \to H^1(X,L_2\otimes L_{1}^{*}) \stackrel{\delta}{\to} \ldots \ . 
\eeq
From this, it follows that $H^1(X,L_2 \otimes {\cal{W}}^{*})$ takes the form
\beq
 H^1(X,L_2 \otimes {\cal{W}}^{*}) =H^1(X,L_2\otimes L_3^{*}) +Ker(\delta) \ .
\eeq
In the chosen vacuum, we note that all vevs for the $C_1$ fields vanish. 
It then follows from Table \ref{tabledoubled} that this branch is confined to the zero-element
of $H^1(X,L_2 \otimes L_1^{*})$ and, hence, to the zero in $Ker(\delta) \in H^1(X,L_2 \otimes L_1^{*})$. As a result, the non-triviality of the extension sequence \eref{secondone} is controlled precisely by an extension class in $H^1(X,L_2 \otimes L_3^{*})$, the cohomology associated with the fields $C_3$. That is, any non-zero element of $H^1(X,L_2 \otimes L_3^{*})$ defines a unique non-zero element of $H^1(X,L_2 \otimes {\cal{W}}^{*})$. Hence, the deviation of the bundle $V$ away from its split point in sequence \eqref{secondone} is controlled by the $\langle C_3 \rangle
\neq 0$ condition in the field theory, as expected. 
Putting everything together, we conclude that $V$ in \eref{firstone} and \eqref{secondone} is indeed the bundle corresponding to the branch of the vacuum space where
$\langle \tilde{C}_2 \rangle \neq 0,\langle C_3 \rangle \neq 0$ and all other $C$ vevs vanish. That being said, we want to point out that the pair of sequences,  \eref{firstone} and \eqref{secondone}, leading to this result are not unique. There is at least one different set of sequences that also corresponds to this branch. However, one can show that the bundle described is isomorphic to $V$ in \eref{secondone}. See the Appendix for a detailed discussion. A  similar analysis can be performed for any other allowed branch.

\begin{table}[h]
\begin{center}
\begin{tabular}{|c|c|c||c|}
  \hline
  Representation & Field Name & Cohomology & Multiplicity \\ \hline
  ${\bf 1}_{\frac{1}{2},3}$ & $C_1$  & $H^1(X,L_1^{*} \otimes L_2)$& 8\\ \hline 
  ${\bf 1}_{-\frac{1}{2},-3}$ & $\tilde{C}_1$ &  $H^1(X,L_1 \otimes L_2^{*})$& 8\\ \hline  
  ${\bf 1}_{-\frac{1}{2},3}$ & $C_2$ &  $H^1(X, L_1^{*} \otimes L_3)$  & 8\\ \hline
  ${\bf 1}_{\frac{1}{2},-3}$ & $\tilde{C}_2$ & $H^1(X,L_1 \otimes L_3^{*})$ & 8\\ \hline 
  ${\bf 1}_{1,0}$ & $C_3$  & $H^1(X,L_2 \otimes L_3^{*})$ & 8\\ \hline
  ${\bf 1}_{-1,0}$ & $\tilde{C}_3$  & $H^1(X,L_2^{*} \otimes L_3)$ & 8\\ \hline  
  ${\bf 27}_{0,-2}$ & $f_1$  & $H^1(X,L_1)$ & 0\\ \hline 
  ${\bf 27}_{\frac{1}{2},1}$ & $f_2$ & $H^1(X,L_2)$ & 0\\ \hline 
  ${\bf 27}_{-\frac{1}{2},1}$ & $f_3$  &  $H^1(X,L_3)$& 0\\ \hline
  $\overline{{\bf 27}}_{0,2}$ &  $\tilde{f}_1$  &  $H^1(X,L_1^{*})$& 4 \\ \hline 
  $\overline{{\bf 27}}_{-\frac{1}{2},-1}$ & $\tilde{f}_2$ &  $H^1(X,L_2^{*})$&4 \\ \hline 
  $\overline{{\bf 27}}_{\frac{1}{2},-1}$ & $\tilde{f}_3$  & $H^1(X,L_3^{*})$ & 4\\ \hline 
  \end{tabular}
\caption{The representations, fields and cohomology groups of a generic $E_6 \times U(1) \times U(1)$ theory associated with a poly-stable bundle $V= L_1 \oplus L_2 \oplus L_3 $ on the stability wall.
The multiplicities for the explicit bundle defined by \eqref{wow} are given in the fourth column.}
\label{tabledoubled}
\end{center}
\end{table}

To illustrate more concretely the possible branch structure, we turn now to a specific manifold and rank $3$ vector bundle.
\subsection*{A Specific Example}
As an example of a stability wall of the type discussed
in this section, consider the bundle
\bea \label{2dext}
0 \to {\cal O}_{X}(-1,-1,2,0) \oplus {\cal O}_{X}(2,-1,-1,0) \to V \to {\cal O}_{X}(-1,2,-1,0) \to 0
\eea
defined on the CICY threefold
\begin{eqnarray}
\label{eg3cy3}
\left[ \ba{c |c }
\mathbb{P}^1 & 2 \\
\mathbb{P}^1 & 2 \\
\mathbb{P}^1 & 2 \\
\mathbb{P}^1 & 2 \\
\ea \right]^{4,68} \ .
\end{eqnarray}
Equation \eqref{2dext} describes $V$ as an extension of direct sums of
line bundles. This bundle has a stability wall defined by a co-dimension $1$ hyperplane
in its four-dimensional K\"ahler cone. On this locus, the bundle
splits as 
\bea  \label{wow}
 V = {\cal O}_{X}(-1,-1,2,0) \oplus {\cal O}_{X}(2,-1,-1,0) \oplus {\cal
  O}_{X}(-1,2,-1,0)\;.
  \eea 
Hence, we can identify $L_1$, $L_2$ and $L_3$ of the
previous discussion as ${\cal O}_{X}(-1,-1,2,0)$, ${\cal O}_{X}(2,-1,-1,0)$ and ${\cal
  O}_{X}(-1,2,-1,0)$ respectively.
Within this explicit example, one can calculate the multiplicity of each
multiplet described in \eqref{2Ddecomp}. These are presented in the fourth
column of Table \ref{tabledoubled}. This bundle provides an example of a maximal and highly symmetric $U(1)$-charged moduli content. Note that all six of the possible $C_i$ fields in Table \ref{tabledoubled} are present. In addition, there are $12$ generations of ${\overline{27}}$ multiplets. The $E_6$ D-term associated with these $\tilde{f}_i$ fields will set their vevs to zero in vacuum. Hence, one need only be concerned with the charged bundle moduli.

Let us analyze the possible branch structure. According to the general discussion above, the six types of $C$-moduli listed in Table \ref{tabledoubled} can be pair-wise combined in \eref{two_ds} to give rise to $\binom{6}{2}=15$ (potentially equivalent) D-flat directions. However, we must also address the F-terms in \eref{fterms_all} as well. Generically, this produces only $6$ possible D- and F-flat branches. These six branches are listed in Table \ref{tablebranches}, along with the bundles they correspond to. 
As mentioned above, for {\it each fixed set} of $C$-field vevs there are actually several equivalent sets of extension sequences that can be used to define the bundle in that branch. In 
Table \ref{tablebranches} (and elsewhere in the text) we present only one set of sequences without loss of generality. The necessary isomorphism results are presented in the Appendix.

\begin{table}[h]
\begin{center}
\begin{tabular}{|c|c|c|c|}
  \hline
Branch & Field vevs & Bundle & Ext \\ \hline
  $1$ & $\langle C_1 \rangle,\langle C_2 \rangle \neq 0$ &  $0 \to L_2 \to {\cal W}_1 \to L_1 \to 0$ & $H^1(X,L_2 \otimes L_1^{*})$ \\ 
 &  &  $0 \to L_3 \to V_1 \to {\cal W}_1 \to 0$ & $H^1(X,L_3 \otimes L_1^{*})$\\ \hline
$2$ & $\langle C_1 \rangle,\langle C_3 \rangle \neq 0$ &  $0 \to L_2 \to {\cal W}_1 \to L_1 \to 0$ & $H^1(X,L_2 \otimes L_1^{*})$ \\ 
 &  &  $0 \to {\cal W}_1 \to V_2 \to L_3  \to 0$& $H^1(X,L_2 \otimes L_3^{*})$\\ \hline
 $3$ & $\langle \tilde{C}_2 \rangle,\langle \tilde{C}_1 \rangle \neq 0$ &  $0 \to L_1 \to {\cal W}_2 \to L_3 \to 0$ & $H^1(X,L_1 \otimes L_3^{*})$ \\ 
 &  &  $0 \to {\cal W}_2 \to V_3 \to L_2  \to 0$ & $H^1(X,L_1 \otimes L_2^{*})$\\ \hline
 $4$ & $\langle \tilde{C}_2 \rangle,\langle C_3 \rangle \neq 0$ &  $0 \to L_1 \to {\cal W}_2 \to L_3 \to 0$& $H^1(X,L_1 \otimes L_3^{*})$ \\ 
 &  &  $0 \to L_2 \to V_4 \to {\cal W}_2 \to 0$ & $H^1(X,L_2 \otimes L_3^{*})$\\ \hline
 $5$ & $\langle \tilde{C}_3 \rangle,\langle \tilde{C}_1 \rangle \neq 0$ &  $0 \to L_3 \to {\cal W}_3 \to L_2 \to 0$ & $H^1(X,L_3 \otimes L_2^{*})$ \\ 
 &  &  $0 \to L_1 \to V_5 \to {\cal W}_3  \to 0$ & $H^1(X,L_1 \otimes L_2^{*})$\\ \hline
 $6$ & $\langle \tilde{C}_3 \rangle,\langle C_2 \rangle \neq 0$ &  $0 \to L_3 \to {\cal W}_3 \to L_2 \to 0$ & $H^1(X,L_3 \otimes L_2^{*})$ \\ 
 &  &  $0 \to {\cal W}_3 \to V_6 \to L_1  \to 0$ & $H^1(X,L_3 \otimes L_1^{*})$\\ \hline
   \end{tabular}
\caption{The possible rank-preserving branches for the split bundle associated with  \eqref{wow}.}
\label{tablebranches}
\end{center}
\end{table}

While all branches in Table \ref{tablebranches} correspond to different vector bundles (with different low-energy particle spectra and other features), they are in the same S-equivalence class as discussed in Subsection \ref{branch_goodness}. That is, although, as in Theorem \ref{HS_fil}, each is ``filtered'' by sub-sheaves differently, they all have the same graded sum \eref{graded_sum} and, hence, share the unique poly-stable representative  $L_1 \oplus L_2 \oplus L_3$ in \eref{wow}. To see this, consider the branch given by
\begin{equation}
\label{nice_eg}
\langle {C}_{1} \rangle  \neq 0 \ , \quad  \langle C_{3} \rangle \neq 0 
\end{equation}
where all other $\langle C_i \rangle=0$. The associated vector bundle is defined by
\bea\label{lara_eg}
& &0 \to L_2 \to {\cal W}_1 \to L_1 \to 0 \ ,\\
& &0 \to {\cal W}_1 \to V_2 \to L_3 \to 0 \ .
\eea
At the stability wall, the D-terms in \eref{two_ds} force the vevs of $C_1$ and $C_3$ to zero and $V_2$ must split as the direct sum of three line bundles in \eref{wow} so as to be poly-stable. However, at this fixed value of K\"ahler moduli, $V_2$ is not supersymmetric for a generic value of its bundle moduli; that is, if $\langle {C}_{1} \rangle\neq 0,  \langle C_{3} \rangle \neq 0$. Instead, it is only semi-stable (rather than poly-stable). For such moduli, the filtration of $V_2$ shown in \eref{filt} is given by
\beq
0 \subset L_2 \subset {\cal W}_1 \subset V_2 \ .
\eeq
In terms of the notation of Theorem \ref{HS_fil} we have the following stable quotients
\beq
{\cal F}_1=L_2,~~\frac{{\cal F}_2}{{\cal F}_1}= \frac{{\cal W}_1}{L_2},~~\frac{{\cal F}_3}{{\cal F}_2}=\frac{V_2}{{\cal W}_1} \ .
\eeq
However, by the definitions of the short exact sequences in \eref{lara_eg} we see that these quotients are simply $\frac{{\cal W}_1}{L_2}=L_1$ and $\frac{V_2}{{\cal W}_1}=L_3$. As a result, the graded sum in \eref{graded_sum} is just
\beq\label{line_graded}
Gr(V_{2})={\cal F}_1\oplus \frac{{\cal F}_2}{{\cal F}_1}\oplus \frac{{\cal F}_3}{{\cal F}_2}=L_1 \oplus L_2 \oplus L_3 \ .
\eeq
That is, the graded sum and the poly-stable decomposition at the stability wall coincide. A similar calculation can be done for each of the $6$ bundles in Table \ref{tablebranches} at the stability wall. In each case, the bundle $V_i$, $i=1,\ldots 6$ is filtered by different sub-bundles. However, each of these will produce the same graded sum \eref{line_graded}. As a result, all six bundles in Table \ref{tablebranches} are in the same S-equivalence class. 
\vspace{5pt}

Finally, let us briefly consider the additional branch structure possible for special values of the complex structure moduli. The six branches listed in Table \ref{tablebranches} were selected by imposing restrictions arising from the superpotential -- specifically, from the tri-linear terms $C_1 \tilde{C}_2 \tilde{C}_3$ and their associated F-terms. Let us analyze these F-term obstructions in more detail. Consider, for example, the branch direction defined by $\langle C_1 \rangle, \langle \tilde{C}_2 \rangle \neq 0$. It is clear by direct substitution that this choice will set both $D^{U(1)}=0$, but it is not one of the six allowed branches listed in Table \ref{tablebranches}. In the effective field theory, there are F-terms of the form
\beq \label{cops1}
\frac{\partial W}{\partial \tilde{C}_3}=\lambda_{1} \langle C_1 \rangle \langle \tilde{C}_2 \rangle \ .
\eeq
As long as the Yukawa parameter is not zero, this F-term is non-trivial and obstructs the $\langle C_1 \rangle, \langle \tilde{C}_2 \rangle$ branch from being a supersymmetric bundle.
However, the coupling parameter $\lambda_{1}$ is generically a function of the complex structure moduli of the CY manifold. That is, $\lambda_{1}=\lambda_{1}(\mathfrak{z}_a)$ with $\mathfrak{z}_a v^a \in H^{2,1}(TX)$. More precisely, $\lambda_{1}(\mathfrak{z}_a)$ is the triple product (Yoneda pairing) in cohomology \cite{Anderson:2009ge} associated with 
$C_1\tilde{C}_2\tilde{C}_3$, 
\beq
\lambda_{1}: H^1(X,L_2 \otimes L_1^{*}) \cup  H^1(X,L_1 \otimes L_3^{*})\cup H^1(X,L_3 \otimes L_2^{*})  \stackrel{\lambda_{1}}{\longrightarrow} \mathbb{C} \ .
\eeq
{\it Generically}, this is non-vanishing. However, there may exist certain higher co-dimensional loci on which $\lambda_{1}(\mathfrak{z}_{0a})=0$ and the F-term in \eqref{cops1} vanishes. Over such loci in complex structure moduli space, one can construct additional stable branches -- in this case, the $\langle C_1 \rangle, \langle \tilde{C}_2 \rangle$ branch -- at the stability wall. The complex structure dependence of such bundles and the geometric interpretation of these obstructions via the Atiyah class is explored in detail in \cite{Anderson:2010mh, us_soon}. However, in this paper we focus on generic points in complex structure space for which coupling parameters, such as $\lambda_{1}$, are non-zero.


\section{Rank-Changing Transitions} \label{rank_change_sec}

In the previous section, {\it rank-preserving} deformations of a poly-stable bundle away from its split locus (moving in both bundle and K\"ahler moduli space) were described in effective field theory through anomalous $U(1)$ D-terms. In that section, we studied the branch-structure that arises when the vevs of all matter fields vanish. 
There the vacuum $U(1)$ D-terms consisted soley of the K\"ahler moduli dependent FI parameters and terms containing $U(1)$ {\it charged  bundle moduli} of the form \eref{ext_types}. In the various vacua, such moduli acquired vevs to preserve $N=1$ supersymmetry, thus guaranteeing that the deformed vector bundles were slope-stable.
 
In this section, we enlarge the analysis to include stable bundle deformations associated with the space of {\it charged matter fields}; that is, fields charged not only under anomalous $U(1)$ symmetries, but also under the ``visible'' $G=E_6, SO(10), SU(5)$ gauge symmetry of the four-dimensional theory. These fields can acquire non-zero vevs consistent with the vanishing of the $U(1)$ and $G$ D-terms, as well as F-flatness. This corresponds to deformations of the vector bundle that change not only the matter spectrum of the four-dimensional theory, but its gauge symmetry as well. We refer to these as {\it rank-changing} deformations.

\subsection{The Geometry of Rank-Changing Deformations}\label{general_rank_change}

The notion of rank-changing deformations of holomorphic bundles is familiar both in the physics and mathematics literature and we briefly review the geometry here. Such deformations have been explored extensively \cite{GSW,Huybrechts_tangent, Li:2004hx}, with the most notable application being to deformations of the tangent bundle of a Calabi-Yau threefold \cite{Huybrechts_tangent, Li:2004hx, Candelas:2008wb,Braun:2009qy}. Generically, however, one is interested in such deformations of any rank $n$ holomorphic vector bundle $V$ over a Calabi-Yau manifold $X$. For specificity (and since it includes the tangent bundle on a threefold), in this section we illustrate these concepts for $n=3$, although the results are applicable to any rank $n$ bundle. In terms of four-dimensional physics, a rank-changing deformation of a bundle corresponds to Higgsing vector-like pairs. For example, by giving vevs to vector-like pairs of the $E_6$ theory associated with a rank $3$ bundle $V$, it is possible to obtain new theories with more realistic $SO(10)$ or $SU(5)$ symmetry. Specifically, a vacuum configuration is chosen in which $E_6$ non-singlets in a ${\bf{27}} \in H^1(X,V)$ and ${\bf{\overline {27}}} \in H^1(X,V^*)$ are given expectation values for which the $E_6$ D-term vanishes. This breaks the gauge group to some subgroup while removing components of the associated ${\bf 27},{\bf \overline{27}}$ pair from the low-energy spectrum. Clearly this deformation changes the rank of the vector bundle -- but what is its geometrical interpretation?

In rank-preserving transitions, we saw that giving a vev to a field associated with the  cohomology group $H^1(X,  U_1\otimes U_{2}^{*})$ corresponds geometrically to turning on a non-trivial extension of the form
\beq
0 \to U_1 \to {\cal{V}} \to U_2 \to 0
\eeq
specified by a non-vanishing element of $Ext^1(U_{2}, U_{1})= H^1(X, U_1\otimes U_{2}^{*})$. In the case of the rank-changing deformations described above, giving a vev to a ${\bf 27}$ field associated with $H^1(X,V)$ can be similarly interpreted. 
Beginning with the rank $3$ bundle $V$, we can define a rank $4$ bundles $V_4$ via the exact sequence
\beq\label{classic_ext}
0 \to V \to V_4 \to \cO_{X} \to 0 
\eeq
specified by 
\begin{equation}
\label{cafe1}
Ext^{1}({\cal{O}}_{X}, V)=H^{1}(X, V \otimes {\cal{O}}^{*}_{X})=H^{1}(X,V).
\end{equation}
It follows that a chosen ${\bf 27}$ vev in $H^1(X,V)$ is simply an extension class of this sequence; that is, a non-trivial deformation of the split rank $4$ bundle
\beq
\label{today}
V\oplus {\cO}_{X} \ .
\eeq
Such a bundle breaks the low energy gauge group from $E_{6}$ to $SO(10)$.
In general, it is possible to iterate this procedure in order to change rank a number of times. This would correspond to deforming $V \oplus \cO_{X}^{\oplus m}$, that is, including $m$ copies of the trivial bundle $\cO_{X}$ via an element of $Ext^{1}( \cO_{X}^{\oplus m} ,V)=H^{1}(X,V)^{\oplus m} $. Taking $m=2$, for example, would correspond to rank 5 extension bundles $V_{5}$ which would break $E_{6}$ to $SU(5)$. Such summands are always available as sub-bundles of $E_8$. Note that the rank of the full bundle is preserved in such processes. Nonetheless, we refer to deformations of this sort as ``rank-changing'', since the bundle changes from a rank $3$ bundle $V$ {\it trivially extended} by $\cO_{X}^{\oplus m}$, to a {\it non-trivial, rank $3+m$ extension}, $V_{3+m}$, of $V$ by $\cO_{X}^{\oplus m}$. Since $c_{1}(V)=c_{1}(\cO_{X})=0$, it follows that $c_{1}({V_{3+m}})=0$. Hence
\begin{equation}
\label{maybe1}
\mu(V_{m+3})=\mu(V)=\mu(\cO_{X})=0
\end{equation}
and any rank $m+3$ bundle $V_{m+3}$ constructed using non-trivial extensions ${\bf {27}} \in H^{1}(X,V)$ of  $V \oplus \cO_{X}^{\oplus m}$ is at best {\it semi-stable}. That is, such bundles do {\it not} preserve $N=1$ supersymmetry.

Similarly, giving vevs to ${\bf{\overline{ 27}}}$ fields associated with $H^1(X,V^{*})$ in the above example can be interpreted as a deformation of the split rank $4$ bundle \eqref{today} via the {\it reverse} extension sequence
\beq\label{classic_ext2}
0 \to \cO_X \to {\tilde{V}}_4 \to V  \to 0 
\eeq
specified by 
\begin{equation}
\label{cafe12}
Ext^{1}(V,{\cal{O}}_{X})=H^{1}(X, {\cal{O}}_{X} \times V^{*} )=H^{1}(X,V^{*}).
\end{equation}
Any such $\tilde{V_{4}}$ bundle breaks the low energy gauge group from $E_{6}$ to $SO(10)$. Again,  it is possible to iterate this procedure so as to change rank a number of times. This would correspond to deforming $V \oplus \cO_{X}^{\oplus m}$; now, however, via the reverse extension $Ext^{1}(V, \cO_{X}^{\oplus m})=H^{1}(X,V^{*})$. It follows from \eqref{maybe1} that any rank $m+3$ bundle ${\tilde{V}}_{m+3}$ constructed as non-trivial extensions ${\bf {\overline{{27}}}} \in H^{1}(X,V^{*})$ of  $V \oplus 
\cO_{X}^{\oplus m}$ is {\it semi-stable}.

Since the bundles associated with turning on either ${\bf {27}}$ or ${\bf {\overline{{27}}}}$ vevs are not properly stable, neither branch contains supersymmetric vacua. From the field theory perspective, this is due to the fact that $G$ D-flatness can only be achieved when {\it both} ${\bf {27}}$ and ${\bf {\overline{{27}}}}$ have non-vanishing vevs. Furthermore, even when $G$ D-flatness is satisfied, other D-terms (for example, anomalous $U(1)$-Dterms associated with stability walls) must also be considered. 
One is led to ask: how do we geometrically describe a deformation of the split poly-stable bundle $V \oplus \cO_{X}^{\oplus m}$ that 1) involves turning on non-zero elements in {\it both}  $H^{1}(X,V)$ and $H^{1}(X,V^{*})$ and 2) is strictly slope-stable? This question was definitively answered by Li and Yau \cite{Li:2004hx}. They demonstrated that a connection satisfying the Hermitian Yang-Mills equations can always be obtained as a small deformation of a connection of a semi-stable bundle ``near'' the poly-stable point $V \oplus \cO_{X}^{\oplus m}$ in moduli space. This stable bundle exists only if the two extensions, \eref{classic_ext} and \eref{classic_ext2}, are non-trivial, and relies heavily on the semi-stability of {\it both} extensions.{\footnote {Note that this differs from the related discussion for rank-preserving transitions in Section 3. There, if one branch was stable the other branch was strictly unstable, containing a sub-sheaf with positive definite slope.}} Although providing a powerful existence theorem, \cite{Li:2004hx} does not present an explicit construction of the stable rank $m+3$ bundle for a general rank $3$ bundle $V$. However, the sought after stable bundle can be described, by differing means, in several important cases. In \cite{Li:2004hx, Huybrechts_tangent}, explicit solutions were found for deformations of the tangent bundle (that is, non-trivial vevs for both ${\bf {27}}$ and ${\bf {\overline{{27}}}}$) of certain simple Calabi-Yau threefolds. In future work \cite{us_tangent}, we will explicitly construct the rank-changing deformations of the tangent bundles of all complete intersection Calabi-Yau threefolds.

In the following subsection, we move beyond tangent bundles and instead use detailed knowledge of K\"ahler cone sub-structure and stability walls to study deformations of more general bundles. Specifically, we show that if the original bundle $V$ contains a stability wall somewhere in it's K\"ahler moduli space, one can explicitly construct the bundle corresponding to the Li-Yau connection. That is, we can {\it solve the Li-Yau problem explicitly and construct stable rank-changing deformations of general holomorphic vector bundles}. The presence of at least one enhanced anomalous $U(1)$ symmetry, and the associated D-terms, gives us the extra insight needed to construct bundles that are automatically slope-stable in some chamber of K\"ahler moduli space. Specifically, we will give vevs to $G$-charged fields in such a way as to satisfy not only $G$ D-flatness, but also D-flatness associated with the anomalous $U(1)$ factors (as well as any new F-terms present). Unlike previous examples of rank-changing deformations in the literature, those associated with stability walls are inherently linked to the K\"ahler moduli and K\"ahler cone sub-structure and, moreover, will be slope-stable by definition.

\subsection{Determining the Li-Yau bundle}\label{li_yau_stab}

To illustrate this, consider an indecomposable $SU(3)$ bundle $V_{3}$ with K\"ahler cone substructure. We will use our knowledge of the effective field theory {\it near the stability wall} to derive a stable rank-changing deformation of the bundle {\it everywhere in a chamber} of K\"ahler moduli space. To begin, suppose that a rank $3$ bundle $V_{3}$ has a stability wall somewhere in the K\"ahler cone for which its poly-stable decomposition is a direct sum of sub-bundles. For example, we can consider the case when a rank 2 bundle ${\cal{F}}$ is the de-stabilizing sub-bundle. In this case one can always construct 
$V_{3}$ from the extension sequence,
\begin{equation}
\label{mid1}
0 \to {\cal{F}} \to V_{3} \to {\cal{K}} \to 0 \ ,
\end{equation}
where ${\cal{K}}$ has rank 1. At the stability wall, defined by the condition $\mu(\cF)=\mu({\cal K})=0$, 
the bundle splits into the two pieces
\beq
V_{3} = \cF \oplus {\cal K} \ .
\eeq 
On this locus, the structure group changes to $S[U(2)\times U(1)] \simeq SU(2)\times U(1)$ and the visible gauge symmetry is enhanced from $E_{6}$ to $E_6 \times U(1)$. The relevant decomposition of the ${\bf 248}$ of $E_8$ is
 \bea
E_8 &\supset& E_6 \times SU(2) \times U(1) \ , \\
{\bf 248} &=& ({\bf 1},{\bf 1})_0 + ({\bf 1},{\bf 2})_3 + ({\bf
  1},{\bf 2})_{-3} + ({\bf 1},{\bf 3})_0 + ({\bf 78},{\bf 1})_0 \\
\nonumber &&+ ({\bf 27},{\bf 1})_{-2} + ({\bf 27},{\bf 2})_1 +
(\overline{{\bf 27}},{\bf 1})_2 + (\overline{{\bf 27}},{\bf 2})_{-1} \ .
\eea
To determine the multiplicity of each multiplet, one must compute the cohomology groups given in Table \ref{table_e6}.

\begin{table}[h]
\begin{center}
\begin{tabular}{|c|c|c|c|}
  \hline
  Representation & Field name & Cohomology & Multiplicity \\ \hline
  $({\bf 1},{\bf 2})_{3}$ & $C_1$ & $H^1(X,{\cal F}^*\otimes {\cal K})$ & 0 \\ \hline
  $({\bf 1},{\bf 2})_{-3}$ & $C_2$ & $H^1(X,{\cal F} \otimes {\cal K}^*)$ & 93  \\ \hline
  $({\bf 1},{\bf 3})_{0}$ & $\phi$ & $H^1(X,{\cal F}^* \otimes {\cal F})$ & 19  \\ \hline
  $({\bf 27},{\bf 1})_{2}$ & $f_1$ & $H^1(X,{\cal K})$ & 0 \\ \hline
  $({\bf 27},{\bf 2})_{-1}$ & $f_2$ & $H^1(X,{\cal F})$ & 12  \\ \hline
  $(\overline{{\bf 27}},{\bf 1})_{-2}$ & $\tilde{f}_1$ & $H^1(X,{\cal K}^*)$ & 9  \\ \hline  $(\overline{{\bf 27}},{\bf 2})_{1}$ & $\tilde{f}_2$ & $H^1(X,{\cal 
F}^*)$ & 1  \\ \hline
  \end{tabular}
\caption{The matter decomposition of an $SU(3)$ bundle at a stability wall into the direct sum of two pieces $\cF\oplus {\cal K}$ with rank $2$ and $1$ respectively. The multiplicities in the last column correspond to the explicit bundles presented in Subsection \ref{explicit_rank_eg}.}
\label{table_e6}
\end{center}
\end{table}
The potential of the effective theory has two D-terms, corresponding
to the anomalous $U(1)$ and the $E_{6}$ factors of the gauge
group. They have the form \bea\label{u1_e6_dterm}
D^{U(1)}&=&\frac{3}{16}\frac{\epsilon_{S}\epsilon_{R}^{2}}{\kappa_4^{2}}\frac{\mu(\cF)}{{\cal V}} - 3G_{L\bar{M}} C_1^L C_1^{\bar M}+3G_{L\bar{M}} C_2^L C_2^{\bar M} \nonumber \\
&&-2{\cal{G}}_{L\bar{M}} f_1^L f_1^{\bar M} +{\cal{G}}_{L\bar{M}}
f_2^L f_2^{\bar M}+2{\cal{G}}_{L\bar{M}} {{\tilde f}_{1}}^L{{\tilde
    f}_1}^{\bar M}-{\cal{G}}_{L\bar{M}} {{\tilde f}_2}^{L}{{\tilde
    f}_{2}}^{\bar M} \eea and \beq\label{e6_dterm}
D^{E_6}=G_{L\bar{M}} f_1^L f_1^{\bar M}+G_{L\bar{M}} f_2^L f_2^{\bar
  M}-G_{L\bar{M}} \tilde{f}_1^L \tilde{f}_1^{\bar M}-G_{L\bar{M}}
\tilde{f}_2^L \tilde{f}_2^{\bar M} \eeq respectively.  Let us explore
the possibility of having D-flat vacua that, in addition to breaking
the anomalous $U(1)$ factor as in the previous section, also break the
$E_6$ symmetry to its $SO(10)$ subgroup. To do this, one must give a
vev to an element of a ${\bf 27}$ and/or ${\bf{\overline{27}}}$
multiplet that is charged under $E_6$ but is an $SO(10)$ singlet. With
respect to $SO(10)\times U(1)' \subset E_6$, the ${\bf 27}$ and ${\bf
  \overline{27}}$ of $E_6$ decompose as \bea \label{night1} E_{6}
&\supset& SO(10) \times U(1)' \ , \\ \label{night2} {\bf {27}} &=&
{\bf 16}_{1} +{\bf 10}_{-2}+{\bf 1}_{-4} \ ,\\ \label{night3} {\bf
  {\overline{27}}} &=& {\bf \overline{16}}_{-1} +{\bf 10}_{2}+{\bf
  1}_{4} \ .  \eea We will consider giving vevs to the pair of
$SO(10)$ singlet fields, ${\bf 1}_{\pm 4}$. As seen below, turning on
such vevs in pairs is required for $E_{6}$ D-flatness.

In terms of the full symmetry $SO(10) \times SU(2) \times U(1)' \times U(1) \subset E_8$, let us focus on singlets transforming as the pair of fields
\beq\label{rank_change_vev}
({\bf{1}},{\bf{2}})_{-4,-1} \subset f_{2} \in H^1(X, \cF)~~~\text{and}~~~({\bf{1}},{\bf{1}})_{4,-2} \subset {\tilde{f}}_{1} \in H^1(X, {\cal K}^{*}) \ ,
\eeq
where the first $U(1)$ charge subscript corresponds to $E_6 \to SO(10) \times U(1)'$ and the second to the anomalous $U(1)$ arising from the stability wall. Note that each such field has negative charge under the anomalous $U(1)$. Choose non-vanishing vevs for the fields in \eref{rank_change_vev}, and set all other vevs to zero. Then, in the region of K\"ahler moduli space in which $\mu({\cal{F}})<0$, we find that one can satisfy D-flatness for {\it both} the original anomalous $U(1)$ D-term in \eref{u1_e6_dterm}, 
\beq
\label{aa}
D^{U(1)} \sim \frac{\mu(\cF)}{{\cal V}}+|f_2|^2+2|\tilde{f}_1|^2 \ ,
\eeq
as well as the $E_6$ D-term \eref{e6_dterm}. Note that, in this case, \eref{e6_dterm} reduces to a simple $U(1)'$ D-term of the form
\beq
\label{bb}
D^{E_6} \to D^{U(1)'} \sim -4|f_2|^2+4|\tilde{f}_1|^2 \ .
\eeq
That is, by choosing vevs of the form \eref{rank_change_vev} we can set both D-terms to zero.
But, what about the F-terms? The superpotential is given by
\beq\label{sup2} 
W \sim C_1f_2\tilde{f}_1 +C_2f_1\tilde{f}_2+f_1f_2f_2 + \tilde{f}_1{\tilde f}_2\tilde{f}_2+ (C_{1}C_{2})^{2}\ldots
\eeq
Clearly all F-terms immediately vanish, with the notable exception of 
\begin{equation}
\label{b12}
F_{C_{1}}=\frac{\partial W}{\partial C_1} \sim \langle f_2 \rangle \langle \tilde{f}_1 \rangle
\end{equation}
which, if present, would be non-zero. It follows that to obtain supersymmetric vacua of this type, for a generic region of moduli space, the charged bundle moduli $C_{1}$ must not appear in the spectrum. 
That such vacua indeed exist is explicitly demonstrated in the following subsection.
We conclude that we have Higgsed both $U(1)$ factors so as to reduce the visible gauge group to $SO(10)$. 

Geometrically, this corresponds to deforming the rank $3$ bundle $V_{3}$ into a stable rank $4$ configuration, $V_{4}$, of the form
\bea\label{ext_by_o}
& &0 \to \cO_{X} \to U \to {\cal K} \to 0 \quad \Leftrightarrow  \quad \ H^1(X, {\cal K}^{*}) \\ \label{ext_by_o2}
& &0 \to \cF \to V_4 \to U \to 0  \quad \ \Leftrightarrow \quad H^1(X, \cF \otimes U^{*}) 
\eea
Clearly, choosing a non-trivial element of the first extension sequence \eqref{ext_by_o} corresponds to turning on a vev $\langle {\tilde{f}}_{1} \rangle \neq 0$. The meaning of the second sequence, however, requires further clarification. Taking the dual of sequence \eqref{ext_by_o} and tensoring with the rank 2 bundle
${\cal{F}}$ gives the short exact sequence
\begin{equation}
\label{home1}
0 \to {\cal{F}}\otimes {\cal{K}}^{*} \to {\cal{F}} \otimes U^{*} \to {\cal{F}} \to 0 \ .
\end{equation}
The associated long exact cohomology sequence is
\begin{equation}
\label{home2}
0 \to H^{1}(X,{\cal{F}}\otimes {\cal{K}}^{*}) \to H^{1}(X,{\cal{F}} \otimes U^{*}) \to H^{1}(X,{\cal{F}}) \stackrel{\delta}{\to}
H^{1}(X,{\cal{F}}^{*} \otimes {\cal{K}})^{*} \to \dots \ ,
\end{equation}
where we have used Serre duality and the fact that the canonical bundle is trivial on a CY threefold to relate
\begin{equation}
\label{home3}
H^{2}(X,{\cal{F}}\otimes {\cal{K}}^{*})=H^{1}(X,{\cal{F}}^{*} \otimes {\cal{K}})^{*} \ .
\end{equation}
It follows from the exactness of sequence \eqref{home2} that
\begin{equation}
\label{home4}
H^{1}(X,{\cal{F}} \otimes U^{*})= H^{1}(X,{\cal{F}}\otimes {\cal{K}}^{*}) + Ker(\delta) \ .
\end{equation}
Note from Table \ref{table_e6} that $C_{2} \in H^{1}(X,{\cal{F}}\otimes {\cal{K}}^{*})$ and $C_{1} \in H^{1}(X,{\cal{F}}^{*} \otimes {\cal{K}})$. If the vev of $C_{2}$ vanishes in vacuum, the only non-trivial elements of $H^1(X,{\cal{F}} \otimes U^{*})$ must come from $Ker(\delta)$ in \eqref{home4}. As a result, as long as $Ker(\delta)\subset H^{1}(X,{\cal{F}})$ is non-trivial, then so is $H^1(X,{\cal{F}} \otimes U^{*})$ and we can define the non-trivial extension in \eref{ext_by_o2}.
Thus, choosing a non-trivial element of $H^1(X,{\cal{F}} \otimes U^{*})$ corresponds to turning on a vev $\langle f_{2} \rangle \neq 0$.

We conclude that the pair of sequences \eqref{ext_by_o},\eqref{ext_by_o2} deform the rank 3 bundle 
$V_{3}$ defined by \eqref{mid1} into the rank 4 bundle $V_{4}$ associated with the vanishing of the $U(1)$ and $E_{6}$ D-terms in \eqref{aa} and \eqref{bb} respectively. It should be emphasized that, although motivated by the decomposition $V_{3} = \cF \oplus {\cal K}$ and the spectrum at the stability wall, the construction of 
$V_{4}$, via the exact sequences \eqref{ext_by_o} and \eqref{ext_by_o2}, depends only on the sub-sheaves ${\cal{F}}$ and ${\cal{K}}$ and their cohomology groups, and is valid everywhere in the K\"ahler cone.  
Note that, as with $V_{3}$, the $SU(4)$ bundle $V_{4}$ is properly slope-stable only in the region of K\"ahler moduli space for which $\mu(\cF)<0$. This $V_{4}$ represents an explicit construction of the Li-Yau bundle. 

As a check on this result, we show that it is possible to recover the original rank $3$ bundle, $V_3$, from $V_4$ in \eref{ext_by_o} and \eref{ext_by_o2} by setting the $E_6$-charged matter vevs in \eref{rank_change_vev} to zero. While keeping the K\"ahler moduli fixed in the stable region, consider tuning $\langle f_2 \rangle, \langle \tilde{f}_1 \rangle \to 0$. In this case, the defining sequences for $V_4$ reduce to
\bea \label{splitU} 
&&~~~~~ \qquad U = \cO_{X} \oplus {\cal K} \\ \label{splitv4}
&&0 \to \cF \to V_4 \to  \cO_{X} \oplus {\cal K} \to 0 
\eea
By definition, the first extension splits. What about the second? It is clear that, in this case, the extension class in $H^1(X, \cF \otimes U^{*})$ associated with \eref{splitv4} is non-trivial {\it only} for non-zero $\langle C_2 \rangle \in H^1(X,\cF \otimes {\cal K}^*)$. Recall that a given $\langle C_2 \rangle \neq 0$ specifies an indecomposable bundle $V_{3}$ through sequence \eref{mid1}.  Choose a non-zero vev of $C_2$ and take $\langle f_2 \rangle =0 \in H^1(X, \cF)$. Then the $\cO_{X}$ term must be a trivial part of the extension (that is, that portion of the sequence splits). As a result, in the presence of  non-vanishing $\langle C_2 \rangle$ but vanishing $f_2,\tilde{f}_1$ vevs, $V_4$ reduces to
\beq
V_4 =V_3 \oplus \cO_{X} \ .
\eeq
Therefore, we have shown that beginning with the Li-Yau solution bundle $V_4$, if we set the defining field vevs  \eref{rank_change_vev} to zero, we recover $V_3 \oplus \cO_{X}$, as expected! Generically then, having fixed $\langle C_2 \rangle$ and, hence, the rank 3 bundle $V_{3}$,
the deformed bundle $V_4$ describes the geometry associated with the
$f_2,\tilde{f}_1$ vevs in \eref{rank_change_vev}. 

A final observation: as with rank-preserving transitions, the extension sequences  \eqref{ext_by_o} and \eqref{ext_by_o2} used to construct $V_{4}$ are not unique. In the Appendix, we use the Snake Lemma to show that $V_{4}$ could equivalently be built using another pair of short exact sequences.
To make the above ideas more concrete, in the next subsection we present an specific example of a rank $3$ bundle with a stability wall which can be deformed into a rank $4$ bundle.
\subsection{An Example}\label{explicit_rank_eg}
Consider the $SU(3)$ monad \cite{okonek, Anderson:2007nc,Anderson:2008uw,Anderson:2008ex} bundle
\beq\label{rank_change_eg}
0 \to V_3 \to \cO_{X}(2,0,0)\oplus \cO_{X}(1,0,1)^{\oplus 2}\oplus \cO_{X}(2,-2,1) \to \cO_{X}(2,2,1) \to 0
\eeq
defined on the CICY threefold
\begin{eqnarray}
\label{eg4cy3}
\left[ \ba{c |c }
\mathbb{P}^1 & 2 \\
\mathbb{P}^1 & 2 \\
\mathbb{P}^2 & 3 \\
\ea \right]^{3,75} \ .
\end{eqnarray}
The bundle $V_3$ in \eref{rank_change_eg} is destabilized by the rank $2$ sub-bundle
\beq\label{rank_mon}
0 \to \cF \to \cO_{X}(2,0,0)\oplus \cO_{X}(1,0,1)^{\oplus 2} \to \cO_{X}(2,2,1)  \to 0
\eeq
in the region of K\"ahler moduli space for which $\mu(\cF) > 0$. On the $\mu(\cF)=0$ plane, there is a unique poly-stable decomposition $\cF \oplus {\cal K}$, where ${\cal K}=\cO_{X}(-2,2,-1)$. As before, the gauge group is enhanced by an anomalous $U(1)$ symmetry (with a D-term as in \eref{u1_e6_dterm}). The zero-mode $E_6 \times U(1)$ spectrum is given in the last column of Table \ref{table_e6}. Note that both  $U(1)$ charged bundle moduli $C_2$ (but {\it not} $C_{1}$) and $E_6$ non-singlets are present in the massless spectrum. As a result, it is, in principle, possible to deform away from the stability wall defined by $\mu(\cF)=0$ into a stable chamber of K\"ahler moduli space by defining either a rank $3$ {\it or} a rank $4$ bundle.

Note that the initial rank $3$ configuration of the bundle, \eref{rank_change_eg}, could equivalently be expressed by the extension sequence
\beq\label{classic}
 0 \to \cF \to V_3 \to {\cal K} \to 0 \quad \Leftrightarrow \quad H^{1}(X,{\cal{F}} \otimes {\cal{K}}^{*}) \ .
\eeq
Choosing a non-trivial extension corresponds to setting the $U(1)$ D-term in \eref{u1_e6_dterm} to zero by giving a vev to a charged bundle modulus $C_2$ in Table \ref{table_e6}, while setting all other vevs, including matter fields charged under $E_{6}$, to zero.  It follows from expression \eqref{e6_dterm} that these vacua are also $D^{E_{6}}$-flat and from \eqref{sup} that all F-terms, as well as W itself, vanish.
Note that $V_{3}$ is stable when $\mu(\cF) <0$.

Let us consider another branch of the low energy theory. Instead of the configuration \eref{classic}, we could, for example, have moved into the same chamber in K\"ahler moduli space by canceling the FI term in \eref{u1_e6_dterm} with the vevs
\beq
\langle f_2 \rangle \in H^1(X,\cF)~~,~~\langle \tilde{f}_1 \rangle \in H^1(X,{\cal K}^{*}) \ .
\eeq
As described above, by giving vevs to elements of such a ${\bf 27},{\bf {\overline {27}}}$ pair one can set to zero both the anomalous $U(1)$ D-term \eref{u1_e6_dterm} as well as the $E_6$ D-term in \eref{e6_dterm}. In addition, one must consider the superpotential in \eref{sup}. By inspection, we see that for the D-flat vevs $\langle f_2 \rangle,\langle \tilde{f}_1 \rangle \neq 0$, and the other vevs set to zero, all F-terms vanish; that is, 
$\partial_{C_{i}} W= \partial_{f_{i}} W=\partial_{{\tilde{f}}_{j}} W=0$, and that $W=0$. As discussed above, this new $SU(4)$ bundle is described geometrically by the pair of sequences.
\bea
& & 0 \to \cO_{X} \to U \to {\cal K} \to 0 \quad  \Leftrightarrow ~~~H^1(X, {\cal K}^{*}) \\
& & 0 \to \cF \to V_4 \to U \to 0  \quad ~ \Leftrightarrow ~~~H^1(X, \cF)~ 
\eea\label{good_v4}
Once again, this bundle is  stable in the region of K\"ahler moduli space for which $\mu(\cF)<0$. This is to be expected since, in this example, both the deformed $SU(3)$ and $SU(4)$ bundles were defined by extension classes which were negatively charged under the anomalous $U(1)$ symmetry.

To conclude, we see that this explicit bundle provides an example of the type described in Subsection \ref{li_yau_stab}. Using the fact that the monad bundle $V_3$ given in \eref{rank_change_eg} gives rise to a stability wall in K\"ahler moduli space, we were able to construct a rank $4$ deformation, $V_4$, of $V_3$ that is stable {\it in the same chamber of K\"ahler moduli space}. $V_{4}$  provides an explicit solution to the Li-Yau construction outlined in Subsection \ref{general_rank_change} and \ref{li_yau_stab} and illustrates the powerful tools that stability walls provide in analyzing such deformations.

\subsection{More General Rank-Changing Deformations}\label{gen_rank_change}

In this subsection, we move beyond the type of rank-changing deformations discussed above and consider more general branches which are associated with {\it different chambers} of K\"ahler moduli space. 

In the previous section, we analyzed a rank $3$ bundle $V_{3}$ that was stable in the chamber of K\"ahler moduli space for which $\mu(\cF)<0$ for some sub-sheaf ${\cal F}$. Using the effective theory near the stability wall, we were able to derive a rank $4$ deformation of $V_{3}$ which was stable in {\it the same chamber} of K\"ahler moduli space and realize the Li-Yau solution.
What happens in the chamber where $\mu(\cF)>0$? To answer this question, we note that it in this region there might exist a reverse extension rank $3$ bundle defined by 
\begin{equation}
\label{pen2}
0 \to {\cal{K}} \to {\tilde{V}}_{3} \to {\cal{F}} \to 0 \ .
 \end{equation}
Recall that this shares the same stability wall as $V_{3}$, where it again splits as
\begin{equation}
\label{pen3}
{\tilde{V}}_{3} = {\cal{F}} \oplus {\cal{K}}  \ .
\end{equation}
However, even in the case when \eref{pen2} does not exist (that is, $Ext^1(\cF,{\cal K})=0$), there may still exist important branch structure in the region where $\mu(\cF)>0$.  To see this, we once again begin with the theory on the wall defined by the split bundle in \eref{pen3}.

Similarly to the above discussion, here we may give vevs to the $U(1)^{'}$-singlets in the matter fields \beq\label{rank_change_vev2}
({\bf{1}},{\bf{2}})_{4,1}\subset {\tilde{f}}_{2} \in H^1(X, \cF^{*})~~~\text{and}~~~({\bf{1}},{\bf{1}})_{-4,2}\subset f_{1} \in H^1(X, {\cal K}) 
\eeq
that are positively charged under the anomalous $U(1)$. All other vevs are chosen to vanish. Geometrically, these vevs correspond to defining a stable rank $4$ configuration ${\tilde{V}}_{4}$ of the form
\bea\label{good_bun}
& & 0 \to \cO_{X} \to \tilde{U} \to \cF \to 0  \quad \Leftrightarrow \quad H^1(X, {\cal F}^{*}) \\ \label{good_bun2}
& & 0 \to {\cal K} \to \tilde{V}_4 \to {\tilde{U}} \to 0  \quad  \ \Leftrightarrow \quad H^1(X, {\cal K} \otimes {\tilde{U}}^{*}) 
\eea
A non-trivial extension of sequence \eqref{good_bun} clearly corresponds to a vev $\langle {\tilde{f}}_{2} \rangle \neq 0$. Furthermore, an argument similar to the one above shows that for vanishing $C_{1}$ and $C_{2}$ vevs, $H^1(X, {\cal K} \otimes {\tilde{U}}^{*})$ is determined by
\begin{equation}
\label{pen1}
H^{1}(X,{\cal{K}}) \subset H^{1}(X,{\cal{K}}\otimes {\tilde{U}}^{*}) \ .
\end{equation}
Hence, choosing a non-zero element of this cohomology corresponds to a vev $\langle f_{1} \rangle \neq 0$. We conclude that the pair of sequences \eqref{good_bun},\eqref{good_bun2} deform the rank 3 bundle ${\tilde{V}}$ defined by \eqref{pen2} into the rank 4 bundle ${\tilde{V}}_{4}$ associated with the vanishing of the $U(1)$ and $E_{6}$ D-terms respectively. 
The bundle $\tilde{V}_{4}$ is clearly stable when $\mu({\cal K})<0$. This corresponds to the ``reverse'' chamber of K\"ahler moduli space to the one defined by \eref{ext_by_o}, \eref{ext_by_o2}.

The vacua in this subsection are D-flat. What about the F-terms? Recall from \eqref{sup2} that
\beq\label{sup} 
W \sim C_1f_2\tilde{f}_1 +C_2f_1\tilde{f}_2+f_1f_2f_2 + \tilde{f}_1{\tilde f}_2\tilde{f}_2+ (C_{1}C_{2})^{2}\ldots
\eeq
For the vacuum defined by \eref{good_bun}, \eqref{good_bun2} with $\langle f_1 \rangle,\langle \tilde{f}_2 \rangle \neq 0$ and the other vevs zero, all F-terms vanish with the exception of
\beq
F_{C_{2}}=\frac{\partial W}{\partial C_2} \sim \langle f_1 \rangle \langle \tilde{f}_2 \rangle \ .
\eeq
If there are no $C_2$ fields present in the massless spectrum, then this F-term doesn't appear. Examples of such vacua are easily constructed.  Otherwise, generically, such vacua break supersymmetry. We note that there is an important caveat to this. As discussed in previous sections and in \cite{us_soon},  it may be possible to move to a special locus in complex structure moduli space (or a special locus in bundle moduli space) for which the coupling parameter for $C_2f_1\tilde{f}_2$ vanishes. This is similar to the moduli-specializing structure that we saw in the rank-preserving deformations of Subsection \ref{more_comp}. On this higher-codimensional locus of moduli space, such vacua remain supersymmetric.

In summary, we see that given a decomposition $V_1 \oplus V_2 \oplus \ldots$ at a stability wall, one must consider all possible rank-preserving and rank-changing branches generated not only by the vevs of charged bundle moduli $C_{ij} \in Ext^1(V_i,V_j)$ (as in \eref{ext_types}), but also by the $G$-charged matter fields
\beq
f_i \in H^1(X, V_i) \ .
\eeq
Considering the possible choices of vevs for both the $C_{ij}$ and $f_i$ fields, it is clear the branch structure associated with a stability wall is very rich indeed! However, for any given example, the possible web of deformations can be straightforwardly enumerated by considering the ``wall theory" and choosing vevs for any combination of fields in such a way that D- and F- flatness are preserved. 

Considering the simplest case of a K\"ahler cone with two chambers separated by a stability wall, one would generically expect there to be {\it both} rank-changing and rank-preserving deformations in {\it each} chamber. However, it can be shown that the existence of rank-preserving and rank-changing branches can, in some cases, be correlated by certain topological obstructions .
As an example, one might ask whether there exist stability walls with the property that 1) on one side of the stability wall only rank-changing deformations are possible while 2) on the other, only 
rank-preserving deformations exist. In the case of a single anomalous $U(1)$ this would, for example, occur for field content with only positively charged bundle moduli $C_1$ and negatively  charged $E_6$ non-singlets $f_2,\tilde{f}_1$. This corresponds geometrically to
\bea\label{hypo}
&& H^1(X, \cF^{*} \otimes {\cal K}) \neq 0 \\ \label{hypotoo}
&& H^1(X,\cF) \neq 0~~,~~H^1(X,{\cal K}^{*}) \neq 0
\eea
with all other such cohomology groups vanishing.
However, a simple index argument tells us that such configurations cannot occur. First, note that
\beq
\rm{Ind}(\cF \otimes {\cal K}^{*})=\rm{Ind}(\cF) + 2 \rm{Ind}({\cal K}^{*}) \ .
\eeq
Then, if
\beq
\rm{Ind}({\cal{F}})<0 \ , \quad \rm{Ind}({\cal K}^{*}) <0 \ ,
\eeq
as assumed in \eref{hypotoo}, then $\rm{Ind}(\cF \otimes {\cal K}^{*})<0$ as well. As a result, the field content postulated above cannot occur. There must be additional negatively charged bundle moduli $C_2$ {\it or} additional positively charged $E_6$ matter $f_1,\tilde{f}_2$ (and, hence, rank-preserving branches) in the low energy spectrum. From this index argument, one can infer that at stability walls either 1) only one type of either rank-preserving or rank-changing branches is possible, or 2) at least two types (that is, {\it both} rank-changing/preserving) of branches are available in some chamber of the K\"ahler cone.

\section{Donaldson-Thomas Invariants and Wall-Crossing \\Formulae}\label{DT_section}
In this section, we turn to a different application of stability walls in heterotic compactifications -- counting the number of possible string vacua that lead to realistic low energy particle physics. 
More precisely, given a Calabi-Yau threefold $X$ how many holomorphic vector bundles $V$ exist that  satisfy the required phenomenological conditions on their Chern classes \cite{GSW,Anderson:2008ex} listed in Table \ref{Chern_bounds}? We note that the number of distinct Chern classes of physical interest \cite{Braun:2005ux,Braun:2005nv,Bouchard:2005ag,Braun:2005zv,Ambroso:2009jd,Ambroso:2009sc,burt1,Anderson:2009mh} is bounded and, generally, quite small.

\begin{table}[!ht]
\begin{center}
\begin{tabular}{|l|l|}
  \hline
 Constraint on a Chern class & Physical Implication \\ \hline \hline
 $c_1(V)=0$ (mod $2$) & Spinor representations\\ \hline
  $c_2(TX)-c_2(V)=[W]$~~($[W]$ an effective curve class) &  anomaly cancellation \\ \hline
  $c_3(V)=3|{\cal G}|$~~(${\cal G}$ a discrete automorphism of $X$) &  three chiral generations \\ \hline
$V$ slope-stable & $N$=1 supersymmetric vacua\\ \hline
\end{tabular}
\caption{Necessary conditions on the Chern classes of stable holomorphic bundles associated with realistic four-dimensional theories.}
\label{Chern_bounds}
\end{center}
\end{table}

To proceed, we must be able to count how many stable bundles there are with 
fixed ``total'' Chern class $c=({\rm rank}, c_1,c_2,c_3)$ and K\"ahler ``polarization'' $\omega$\footnote{By polarization $\omega$ we mean the entire line in K\"ahler moduli space passing through  a point $\omega$. In the remainder of this paper, we use the same notation for a point and its one-dimensional polarization.}. 
Fortunately, this question has been addressed in the mathematics literature in recent years.
In particular, it is known that the number of such bundles is always finite \cite{langer1,langer2}. Furthermore, the exact number is counted by the absolute value{\footnote{ The Donaldson-Thomas invariant can be negative in some cases. Hence, one must take its absolute value to compute the number of stable bundles.}}
of a geometric invariant of the Calabi-Yau threefold -- known as a Donaldson-Thomas (DT) invariant \cite{DT,Thomas} -- defined by
\beq\label{dt_def}
\lambda_{\omega}(X,c)=c_{top}(T{\cal{M}}^{*}_{\omega}(c)) \ .
\eeq
Here, ${\cal M}_{\omega}(c)$ is the moduli space of stable bundles at $\omega$ with total Chern class $c$ and $c_{top}$ is the integer value of the highest Chern class of this space. Note that $\lambda_{\omega}(X,c)$ is not to be confused with the Yukawa couplings discussed earlier.
The mathematics literature on this topic is vast, and we will not attempt to review it here (see \cite{Li_Qin,gottsche_notes}, for example). We note, however, that Donaldson-Thomas invariants are notoriously difficult to compute since few examples of total moduli spaces ${\cal M}_{\omega}(c)$ are known. Areas in which some examples are understood include the spectral cover construction \cite{FMW} for elliptically fibered spaces, the work of \cite{BM} on Fourier-Mukai
transforms for elliptic and K3-fibrations, and Thomas' original examples \cite{Thomas}, constructed using $K3$-fibrations and the Serre construction. 

In recent years, stability walls have appeared in the mathematics literature as a powerful technique for determining moduli spaces and, hence, DT-invariants. They have been used in a variety of contexts \cite{ellingsrud,Qin,Friedman_Qin,Li_Qin,Li_Qin2}. These include the computation of DT-invariants for rank $1$ and $2$ bundles defined on surfaces \cite{ellingsrud,Qin,Friedman_Qin} and for bundles defined on certain Calabi-Yau threefolds \cite{Li_Qin,Li_Qin2}. In particular, there has been considerable interest in both the mathematics and physics literature \cite{Nekrasov1,Nekrasov2,Joyce} in understanding ``wall-crossing" formulae involving DT-invariants defined on distinct polarizations; that is, formulae relating $\lambda_{\omega_1}(X,c)$ and $\lambda_{\omega_2}(X,c)$ for two different polarizations $\omega_{1}$ and $\omega_{2}$ respectively. 
In this section, we review the relationship between K\"ahler cone substructure, stability walls and DT-invariants and extend the results. Specifically, we relate the computation of these geometric invariants to the field theoretical picture developed in the preceding sections. We also extend the mathematical results to Calabi-Yau threefolds and certain higher rank vector bundles.

In broad outline, our computations will proceed as follows. First, fix a total Chern class $c$ and consider all bundles with this topological data. Next, we note that the slope stability of any such bundle can be related to its topological data via a criterion known as the Bogomolov-L\"ubke inequality \cite{Huybrechts}: \\

\noindent {\it Bogomolov-L\"ubke Criterion}: If a rank $n$ bundle $V$ is slope-stable with respect to a 
chosen K\"ahler form $\omega=t^{k}\omega_k$, then 

\beq\label{bogomolov}
\int_{X} \left(2nc_{2}(V)-(n-1)c_{1}^{2}(V)\right)\wedge \omega \geq 0 \ .
\eeq
\\
\noindent This condition is necessary, but not sufficient, for stability. 
Now, consider a total Chern class for which $c_{1}$ and $c_2$ violate this condition for some polarization $\omega$. Then there exists a 
co-dimension one wall in the K\"ahler cone, call it the $BL$-wall, on one side of which any bundle with this topological data cannot be slope-stable. Let $V$ be such a bundle.
Then, in this region there must exist a sub-sheaf $\cF$ which de-stabilizes $V$. It follows that, as discussed previously, the bundle can be described {\it everywhere in K\"ahler moduli space} by the extension sequence
\beq
\label{someday1}
0 \to \cF \to V \to \frac{V}{\cF} \to 0 \ .
\eeq
Thus, the number of different bundles $V$ can be counted by determining all non-isomorphic sub-sheaves
$\cF$, and locally free quotients $V/\cF$, with the property that $V$ has total Chern class $c$. 
In special cases, 
all such $\cF$ and $V/\cF$ sheaves can indeed be determined.

To continue, note that computing 
$\lambda_{\omega}(X,c)$ requires one to specify all such bundles which are {\it slope-stable} for  polarization $\omega$. To do this, we proceed as follows. Consider one such bundle $V$, specified by extension \eqref{someday1}, and the slopes of both $\cF$ and $V$. We already know $\mu(\cF)>\mu(V)$ in the BL unstable region. If $\mu(\cF)>\mu(V)$ {\it everywhere} in the K\"ahler cone, then $V$ is {\it nowhere} stable and, hence, never contributes to a DT-invariant. If, however, $\mu(\cF)$ becomes equal to $\mu(V)$ somewhere in the K\"ahler cone, then there will exist a co-dimension one {\it stability wall} -- not to be confused with the BL-wall -- on which $V$ can split as $\cF \oplus \frac{V}{\cF}$. Then, using the formalism developed in this paper, one can compute the stable chambers associated with this bundle as well as the dimension of the bundle moduli space in each chamber. This dimension will be the contribution of this specific bundle $V$ to the DT-invariant for a polarization in that chamber. The total DT-invariant $\lambda_{\omega}(X,c)$ will be the sum over all such bundles which are stable for this polarization. That is, {\it the stability wall formalism we have developed in this paper can be used to study DT-invariants}.

To make these concepts more concrete, we now turn to a specific example.
\subsection{An $SU(2)$ Example}\label{DT_su2}

Once again consider the CICY threefold $X$ with $h^{(1,1)}=2$ defined by a degree $\{2,4\}$ hypersurface in $\mathbb{P}^{1}\times \mathbb{P}^3$. Following \cite{Li_Qin,Li_Qin2}, let us analyze $SU(2)$ bundles 
on $X$ with the fixed total Chern class
\beq\label{egc}
c=(2,0,c_{2}^{rs}d_{rst}=(-4,6),0) \ ,
\eeq
where the second Chern class, $c_{2}^{rs}$ is expressed as a vector $c_{2}^{rs}d_{rst}=(-4,6)$ using the intersection numbers $d_{rst}$ in \eqref{someday2}.
Then, the Bogomolov-L\"ubke inequality \eref{bogomolov} gives
\bea
& & \int_{X} \left(2nc_{2}(V)-(n-1)c_{1}^{2}(V)\right)\wedge \omega=4d_{rst}c_2(V)^{rs}t^t \\
& &=4(2d_{122}c_2^{12}t^2+d_{221}c_{2}^{22}t^1+d_{222}c_2^{22}t^2)=4(6t^2-4t^1)\geq 0 \ .
\eea
This inequality is saturated  at the one-dimensional $BL$-wall given by $\frac{t^2}{t^1}=\frac{2}{3}$.
It follows that in the region of K\"ahler moduli space defined by
\begin{equation}
\label{red1}
\frac{t^2}{t^1}<\frac{2}{3} \ ,
\end{equation}
{\it any} bundle with total Chern class \eqref{egc} must be unstable. 
If $V$ is one such bundle, there must exist a rank $1$ sub-sheaf, $\cF$, which de-stabilizes $V$ everywhere in region \eqref{red1}. Typically, $\cF$ is a rank $1$ torsion-free sheaf. However,
by taking the double-dual of both sides of $\cF \subset V$, we find that this induces a line bundle ${\cal L}=\cF^{{*}{*}} \subset V^{{*}{*}} \simeq V$ (since $V$ is a vector bundle, it is isomorphic to its double-dual \cite{AG}). As a result, each such $V$ can be defined by the sequence
\beq
0 \to {\cal L} \to V \to {\cal{L}}^{*}\otimes {\cal I}_{{\cal Z}} \to 0 \ ,
\eeq
where ${\cal Z}$ is a (possibly empty) co-dimension $2$ sub-scheme. In the following discussion we shall consider the case when ${{\cal Z}}$ is empty (and hence ${\cal I}_{{\cal Z}}$ is not present) since it was shown in \cite{Li_Qin2} that the results are unchanged even when it is non-trivial{\footnote{Note that in general, such ideal sheaves {\it do} contribute to DT-invariants and must be carefully analyzed.}}.
The defining sequence then simplifies to
\beq
\label{bird1}
0 \to {\cal L} \to V \to {\cal{L}}^{*} \to 0 \ .
\eeq

We now want to find all line bundles $\cL$ with the property that extension bundle $V$ has total Chern class \eqref{egc}.  To do this,  start with the general line bundle $\cL=\cO_{X}(m,n)$, compute the total Chern class of $V$ using \eqref{bird1} and set it equal \eqref{egc}. It is trivially true that $c_{1}=c_{3}=0$. Using the additivity of the Chern character
\begin{equation}
\label{bird2}
Ch(V)=Ch(\cO_{X}(m,n))+Ch(\cO_{X}(-m,-n)) \ , 
\end{equation}
it follows that $c_{2}(V)_{t}=c_{1}(\cL)^{r}c_{1}(\cL)^{s}d_{rst}$
with $c_{1}(\cL)^{r}=(m,n)$ and $d_{rst}$ given in  \eqref{someday2}. Setting this equal to $(-4,6)$, we find
\begin{equation}
\label{bird3}
-2(m^{2}+n^{2})=-4~, \quad 3(m^{2}+n^{2})=6
\end{equation}
and, hence, 
\begin{equation}
\label{bird4}
m^{2}+n^{2}=2 \ .
\end{equation}
Therefore, there are four possible line bundles, specified by 
\begin{equation}
\label{bird5}
c_{1}(\cL)=\{(1,1), (1,-1), (-1,1), (-1,-1)\} \ ,
\end{equation}
for which $V$ has total Chern class \eqref{egc}.

To compute DT-invariants, one must decide which of these four bundles $V$ is slope-stable {\it somewhere} in K\"ahler moduli space. To do this, let us compute the slope of a general line bundle 
$\cL=\cO_{X}(m,n)$. We find that
\begin{equation}
\label{bird6}
\mu(\cL)=4m(t^{2})^{2}+8nt^{1}t^{2}+2n(t^{2})^{2} \ .
\end{equation}
Note that each of the four line bundles $\cL$ in \eqref{bird5} satisfies $\mu(\cL)>\mu(V)=0$ everywhere in the unstable region specified by the $BL$ criterion, as it must. Furthermore, for three of these line bundles, namely (1,1), (1,-1) and (-1,-1), the associated bundle, $V$, is unstable {\it everywhere} in the K\"ahler cone (in fact, the extension \eref{bird1} splits in these three cases). Hence, these three possibilities do not contribute to the DT-invariant. On the other hand, using \eqref{bird6} it is straightforward to see that $\mu(\cO_{X}(-1,1))=\mu(V)=0$ on the line defined by
\begin{equation}
\label{bird8}
\frac{t_{2}}{t_{1}}=4 \ .
\end{equation}
Note that this is above the $BL$-wall given in \eqref{red1}. For any polarization with $t_{2}/t_{1}>4$, the associated bundle $V$ {\it is} slope-stable and contributes to the DT-invariant. Hence, \eqref{bird8} defines the stability wall of $V$. It follows that the entire DT-invariant arises from one bundle $V$ constructed from $\cL=\cO_{X}(-1,1)$ via the extension
\beq\label{eg_line}
0 \to \cO_{X}(-1,1) \to V \to \cO_{X}(1,-1) \to 0 \ .
\eeq

Given this result, the DT-invariant for any polarization $\omega$ in the stable chamber of $V$, that is, where $t_{2}/t_{1}>4$, can be determined geometrically as follows. It is well known \cite{Li_Qin,Sharpe:1998zu} that for any rank 2 bundle $V$ defined by
\beq\label{lines}
0 \to L_1 \to V \to L_2 \to 0 \ ,
\eeq
its moduli space is
\begin{equation}
\label{white1}
{\cal M}_\omega=\mathbb{P}^m
\end{equation}
where
\begin{equation}
\label{white2}
m=h^1(X,V\otimes V^{*})={\rm dim}(Ext^1(L_2,L_1))-1 \ .
\end{equation}
The Chern classes of this space can be computed using the defining sequence for the cotangent bundle \cite{AG, AG2} given by 
\beq\label{pn}
0 \to (T\mathbb{P}^{m})^{*} \to \cO_{\mathbb{P}^{m}}(-1)^{m+1} \to \cO_{\mathbb{P}^{m}} \to 0 \ ,
\eeq
where $\cO_{\mathbb{P}^{m}}(1)$ denotes the hyperplane bundle in $\mathbb{P}^m$. In particular, we find that the top Chern class of $T{\cal{M}}_{\omega}^{*}$ is given by
\beq
\label{white3}
c_{m}((T\mathbb{P}^{m})^{*})=-\binom{m+1}{m}=-(m+1) \ .
\eeq
It then follows from \eqref{dt_def} that
\beq
\label{white4}
\lambda_{\omega}(X,c)= -(m+1) \ .
\eeq
Applying these results to the specific example above, with extension sequence \eqref{eg_line}, we conclude that the total moduli space of stable bundles with fixed Chern class \eref{egc} and polarization $\omega$ in the stable chamber $t^2/t^1>4$ is
\beq
\label{white5}
{\cal M}_{\omega}=\mathbb{P}^9
\eeq
and, hence,
\beq
\label{white6}
\lambda_{\omega}(X,c)=c_{9}((T\mathbb{P}^{9})^{*})=-10 \ .
\eeq

In summary:  we find that for fixed Chern class \eref{egc} the K\"ahler cone is divided into two chambers by a stability wall with slope $t^2/t^1=4$. The DT-invariant for any polarization is defined simply by the chamber containing that polarization. Let $\omega_{r}=\omega_1+s\omega_2$ be the K\"ahler form. Then the wall-crossing formula is simply
\beq\label{wall_cross}
\left\{
\begin{array}
[c]{l}%
\lambda_{\omega_{s}}(X,c)=c_{top}(T{\cal M}_{\omega}^{*})=-10~~4<s < \infty \\
\lambda_{\omega_{s}}(X,c)=0~~~~~~~~~~~~~~~~~~~~0<s \leq 4\\
\end{array}
\right.
\eeq
Having provided this direct computation of a Donaldson-Thomas invariant, we can now ask -- how does this relate to the effective field theory description of stability walls and branch structure developed in previous sections of this paper?

In Section \ref{stab_wall_review}, we have already seen the low-energy $E_7 \times U(1)$ theory associated with the bundle $0 \to \cO_{X}(-1,1) \to V \to \cO_{X}(1,-1) \to 0$ (that is, \eref{E7_eg}) and its stability wall at $t^{2}/t^{1}=4$. We found that, at the stability wall, the bundle decomposes as $\cO_{X}(-1,1)\oplus \cO_{X}(1,-1)$ and an additional $U(1)$ gauge factor appears in the effective theory. Furthermore, we showed that the {\it only} fields charged under this $U(1)$ in the spectrum are the bundle moduli,
\beq
C_2 \in H^1(X, \cL^2)=H^1(X, \cO_{X}(-2,2)) \ ,
\eeq
in Table \ref{SU2_table}.
As a result, the $FI$ parameter in the $U(1)$-Dterm, \eref{dterm_e7}, can be canceled in the vacuum, and, hence, supersymmetry preserved, by a single $\langle C_2 \rangle \neq 0$ in the region for which $\mu(\cO_{X}(-1,1))<0$. However, when $\mu(\cO_{X}(-1,1)) >0$, there are no positively $U(1)$-charged fields available to cancel the FI term. Hence, a non-zero $D^{U(1)}$-term potential forbids any vacua in this region of K\"ahler moduli space. 
Moreover, in Section \ref{stab_wall_review}, the number of heterotic vacua in the stable region of K\"ahler moduli space was simply counted by the $h^1(X,\cO(-2,2))=10$ independent choices of $\langle C_2 \rangle$ vevs available to cancel the FI-term. Further, the flat directions in the potential in the stable region corresponded to \eref{moduli_count} which gave us
\beq
h^1(X, V\otimes V^{*})=h^1(X, \cO_{X}(-2,2)) -1 =9 \ .
\eeq
which exactly matches the dimension of the local tangent space $T\mathbb{P}^9$. Thus, the results of Section \ref{stab_wall_review} are simply the DT wall-crossing formula in \eref{wall_cross}, phrased in terms of the effective field theory!

As a final comment, note that the effective field theory and geometrical methods introduced in this paper can only be used to compute the branch structure and DT-invariants of a set of bundles if they have Chern classes such that they are guaranteed to be unstable somewhere in K\"ahler moduli space. One cannot, in general, guarantee that {\it every} bundle on a Calabi-Yau threefold will have a stability wall or, if it does, that it is described by the same effective theory. In the above example, we saw that the presence of a stability wall in certain cases of rank $2$ bundles can be used to compute the Donaldson-Thomas invariants. But what about higher-rank bundles? For realistic heterotic compactifications, we are interested in rank $n$ bundles with $n=3,4,5$. However, to date, no DT-invariants have ever been computed for rank $n>2$ bundles over a compact Calabi-Yau threefold. While this problem remains a difficult one, in the following subsection we outline how, in principle, the branch structure of stability walls can be used to recursively construct higher-rank Donaldson-Thomas invariants. We illustrate this by computing the DT-invariants of a specific rank $3$ bundle on a CY threefold.

\subsection{Donaldson-Thomas Invariants By Recursion}\label{DT_recursion}
In this section, we show how the structure of heterotic stability walls can, in principle, be used to recursively construct Donaldson-Thomas invariants. 
To illustrate the idea, suppose that we attempt to apply the same reasoning as the previous section to a rank three bundle over the same threefold? Consider for example the following total Chern class,
\beq\label{top_data_principle}
c=(3,0,c_2,c_3)~.
\eeq
We will refer to any bundle with this topological data as $V$. 

Once again, we shall apply the Bogomolov-L\"ubke inequality to a bundle of this form. Let us suppose that as before, the Bogomolov inequality indicates that this rank three bundle must be de-stabilized somewhere in the K\"ahler cone. Applying \eref{bogomolov}, it follows that
\beq\label{bog_again}
\int_X c_2(V)\wedge \omega \geq 0~~~\Rightarrow~~\frac{t^2}{t^1} \geq \rho 
\eeq
for some positive rational number $\rho$, is a necessary condition for stability of $V$. As a result, there must exist a sub-sheaf, $\cF \subset V$ which de-stabilizes $V$ in the region $\frac{t^2}{t^1} < \rho$ . Hence, each $V$ is either stable no-where or has a stability wall at some fixed ray, $t^2/t^1=r$ in K\"ahler moduli space, where $r \geq \rho$ (recall that the Bogomolov bound is necessary but not sufficient and the stable region could in fact be smaller than that indicated by \eref{bog_again} alone). However, unlike in Section \ref{DT_su2}, where we could be assured that the sub-sheaf was rank $1$, here we could have {\it either}  $rk(\cF)=1$ or $2$.

As we saw in the previous subsection, in order to find the DT-invariant, we must determine all possible sub-sheaves, $\cF$, and quotients, $V/\cF$. If an $\cF$ is found that de-stabilizes its associated bundle $V$ everywhere in K\"ahler moduli space, we discard it, since it will not contribute to the DT-invariant. For the remaining sub-sheaves, $\cF$, we will proceed further using the techniques developed in this work.

Suppose that upon determining all possible sub-sheaves, some sub-sheaf $\cF$ exists which does not destabilize $V$ everywhere, but instead leaves some stable region ($t^2/t^1>r$). Then, as we saw in Section \ref{rank_pres_sec}, we can infer the full structure of the theory in the stable region from the Harder-Narasimhan filtration of $V$ at the stability wall ($t^2/t^1=r$) generated by $\cF$. At the wall defined by $\mu(\cF)=0$, the graded sum in \eref{graded_sum} could take one of two possible forms:
\bea\label{two_wall_poss1}
1)~~V \to \bigoplus_{i=1}^{3} L_i~~~~\text{with}~~~~rk(L_i)=1~,~\mu(L_i)=0~\forall~i~~~~~~~~~~~~~~~~~~~~ \\ \label{two_wall_poss2}
2)~~V\to {\cal Q} \oplus L~~~~\text{with}~~~~rk({\cal Q})=2,~rk(L)=1~,~\mu({\cal Q})=\mu(L)=0\eea
where all the sheaves are torsion-free and stable for $\frac{t^2}{t^1} \geq r$. These two possibilities each constitute a ``wall theory" of the kind we have studied throughout this work. If it were possible to classify all possible vacua corresponding to these two Harder-Narasimhan graded sums (That is,  ``wall decompositions") in \eref{two_wall_poss1} and \eref{two_wall_poss2} then in principle the DT-invariant associated with \eref{top_data_principle} could be determined. In particular for the stable chamber of K\"ahler moduli space the DT-invariant would be determined by the sum $no.~of~vacua_{1}+no.~of~vacua_{2}$ where $1,2$ refer to the branches of the wall theories described above. That is,
\beq\label{mods}
{\cal M}^{total}_{\omega}(c)= {\cal M}^{1}_{\omega}(c) \times {\cal M}^{2}_{\omega}(c)
\eeq
where ${\cal M}^{1}_{\omega}(c)$ corresponds to the stable branches (that is, moduli space) determined by \eref{two_wall_poss1} and ${\cal M}^{2}_{\omega}(c)$ to the branches associated with \eref{two_wall_poss2} (one or both of which could be trivial or contain multiple components of its general form). Thus, the DT-invariant associated with the polarization $\omega_s=t\omega_1 +s t\omega_2$ and the total Chern class in \eref{top_data} would be
\beq\label{wall_cross2}
\left\{
\begin{array}
[c]{l}%
\lambda_{\omega_s}(X,c)=c_{top}(T{{\cal M}^{1}_{\omega}}^{*})+c_{top}(T{{\cal M}^{2}_{\omega}}^{*})~~r<s < \infty \\
\lambda_{\omega_{s}}(X,c)=0~~~~~~~~~~~~~~~~~~~~0<s \leq r\\
\end{array}
\right.
\eeq
for some integer $r$.

The difficulty in explicitly analyzing this system comes in the fact that there could in principle be many different rank $1$ sub-sheaves, $L$ and many rank $2$ sub-sheaves, ${\cal Q}$, consistent with the structure described above (which would lead to many components to \eref{mods} and integers $r$ in \eref{wall_cross2}). We must first ask, can the problem be bounded? That is, can we reduce the possible wall components, $L_i,L,{\cal Q}$ to a set of sheaves with bounded Chern classes? If so, can we then find all possible sub-sheaves in \eref{two_wall_poss1} and \eref{two_wall_poss2}? In this section, it is important to observe that {\it if the moduli spaces (and DT-invariants) associated with all possible sub-sheaves $L,{\cal Q},L_i$ were already known}, then in principle we could use this information to recursively compute the moduli spaces \eref{mods} via the techniques outlined in Section \ref{rank_pres_sec}. 

Whatever technique one employs to determine the sub-DT-invariants, $\lambda_{\omega}(X,c_{L,{\cal Q},L_i})$ (characterizing $L,{\cal Q},L_i$, respectively) once they are obtained, they can be used with the decompositions in \eref{two_wall_poss1} and \eref{two_wall_poss2} to determine the DT-invariant associated with \eref{top_data_principle}. In this work however, we do not have a ``black box" for determining these $\lambda_{\omega}(X,c_{L,{\cal Q},L_i})$. Instead, we have only the Bogomolov-L\"ubke bound and we will consider cases where it is possible to obtain information by applying this condition recursively on possible rank $2$ sub-sheaves. This will be illustrated below in the following sections. However, because the analysis involves the details of integer systems and is necessarily quite convoluted, we shall outline the basic approach here before we begin.

We will employ the following algorithm:
\begin{enumerate}
\item Consider all possible rank $2$ and rank $1$ bundles which do not de-stabilize $V$ everywhere in the K\"ahler cone.
\item For each, enumerate the possible stability wall decompositions (that is, graded sums) in \eref{two_wall_poss1} and \eref{two_wall_poss2}
\item In the case of \eref{two_wall_poss1}, solve for all possible rank $1$ sheaves $L_i$. Enumerate all possible $V$ using the extension bundle techniques described in this work.
\item In the case of \eref{two_wall_poss2}, recursively apply the Bogomolov-L\"ubke bound, \eref{bogomolov}, to the Chern class of all possible rank $2$ sheaves ${\cal Q}$.
\item If the Bogomolov-L\"ubke bound is violated somewhere in K\"ahler moduli space, classify ${\cal Q}$ using all possible rank $1$ sub-sheaves as in the previous section.
\item Solve for all possible $V$ of each type to determine the moduli space in \eref{mods}.
\end{enumerate}

To see how this could be possible, and to illustrate the algorithm in a concrete case, we will investigate the given total Chern class
\beq\label{top_data}
c=(3,0,c_{2}^{rs}d_{rst}=(-12,18),-20)~,
\eeq
We will consider in turn each the possibilities in  \eref{two_wall_poss1} and \eref{two_wall_poss2} and recursively apply the Bogomolov-L\"ubke bound to show that in fact, only one of them is possible. In this simple case, we will find that it is possible to {\it  fully determine the moduli space of stable, rank $3$ bundles with total Chern class given by \eref{top_data}}.
\subsection*{Case 1}
First, we consider the case shown in \eref{two_wall_poss1}, where the rank $3$ bundle $V$ decomposes as a sum of three rank $1$ sheaves. What constraints can we place on the three rank $1$ sheaves $L_i$? Let $c_1(L_i)=(n^1_{i},n^{2}_{i})$ for some integers $n^{1}_{i},n^{2}_{i}$, $i=1,2,3$. Given the topological data in \eref{top_data} we can constrain these integers as follows. 

Recall the that the Chern character is additive. That is,
\beq
ch(V)=\sum_{i=1}^{3}ch(L_i)
\eeq
and hence we can use the topological data in \eref{top_data} to constrain the integers $n^{1}_{i},n^{2}_{i}$ via
\bea
c_1(V)=0=\sum_{i=1}^{3} c_1(L_i) \Rightarrow \sum_i n^1_{i}=0~~\forall i \\
d_{rst}ch_2(V)^{st}=-d_{rst}c_2(V)^{st}=(12,-18)=\sum_i ch_2(L_i)=\sum_i (2(n^{2}_{i})^2,\frac{1}{2}((n^{2}_{i})^2+8n^{2}_{i}n^{1}_{i}) \\
ch_3(V)=\frac{c_3(V)}{2}=-10=\sum_ich_3(L_i)=\sum_i \frac{1}{3}((n^{2}_{i})^3+6(n^{2}_{i})^2(n^{1}_{i}))
\eea
As it turns out this integer system is highly constrained and in fact not only bounds the integers $n^{1}_{i},n^{2}_{i}$, but produces a unique solution! This system can be straightforwardly solved to find that\beq
c_1(L_1)=(-2,2)~,~c_1(L_2)=(1,-1)~,~c_1(L_3)=(1,-1)
\eeq
In general then, the rank $1$ sheaves take the form
\beq
L_1=\cO(-2,2)\otimes {\cal I}_{Z_{1}}~,~L_2=\cO(1,-1)\otimes {\cal I}_{Z_{2}}~,~L_3=\cO(1,-1)\otimes {\cal I}_{Z_{3}}
\eeq
where ${\cal I}_{Z_{i}}$ are ideal sheaves associated with co-dimension (at least) $2$ sub-schemes $Z_i$. In  \cite{Li_Qin2}, a detailed analysis was given for the conditions under which the $Z_i$ can be taken trivial without effecting the final structure of the moduli space. The present example is a special case of this analysis and as a result, we shall take $Z_i =\emptyset$ in the following discussion. In this case, the filtration in \eref{two_wall_poss1} is given by
\beq\label{case1_goodness}
V \to L_1 \oplus L_2 \oplus L_3= \cO_{X}(-2,2) \oplus \cO_{X}(1,-1) \oplus \cO_{X}(1,-1)
\eeq

Direct computation of $\mu(\cO_{X}(-2,2))=0$ shows that in fact the position of this stability wall is at $r=4$ as in the previous subsection (once again the Bogomolov bound gives a weaker region of stability than the real K\"ahler cone sub-structure). In terms of the effective field theory, at this wall, two enhanced anomalous $U(1)$ symmetries appear in the low energy theory as described in Section \ref{more_comp}. Moreover, we have $U(1)$-charged bundle moduli of the form
\beq\label{case1cs}
(\tilde{C}_1)_{(-1/2,-3)} \in H^1(X,L_1 \otimes L_{2}^*)~~,~~(\tilde{C}_2)_{(1/2,-3)} \in H^1(X,L_1 \otimes L_{3}^*)
\eeq
where $h^1(X,L_1 \otimes L_{2}^*)=h^1(L_1 \otimes L_{3}^*)=40$ and all other charged bundle moduli $C_i$ vanish.

Since there are two types of $U(1)$ charged bundle moduli available to satisfy the anomalous $U(1)$ D-terms, we see that we can construct a stable bundle in the region $t^2/t^1 >4$ (where $\mu(L_1)<0$), as expected. Explicitly, we construct a stable $SU(3)$ bundle with total Chern class \eref{top_data} as
\bea
0 \to L_1 \to {\cal W} \to L_2 \to 0 \\
0 \to {\cal W} \to V \to L_3 \to 0
\eea
Note that this bundle is equivalent to the bundle referred to as `branch 3' in Table \ref{tablebranches}. The bundle moduli space of $V$ has dimension
\beq\label{case1_bund_mod}
h^1(X,V\otimes V^*)=h^1(X,L_1 \otimes L_{2}^*)+h^1(X,L_1 \otimes L_{3}^*)-2=78
\eeq
As a result, by the correspondence established in the previous section between the effective field theory and the DT-Invariants we would expect $c_{top}(T{\cal M}^{1}_{\omega})=80$ for $\omega$ in the region $t^2/t^1 >4$. If there were no other ways in which the bundle $V$ could decompose at its stability wall, we would have the complete answer here. But what about the second possibility given in \eref{two_wall_poss2}? 

\subsection*{Case 2}
Next we must consider the decomposition $V \to {\cal Q} \oplus L$ at the stability wall. We must be careful at this point, because a priori, the stability wall associated with this decomposition need not be the same as that in Case 1. All that we can infer from the topological data is that the the bundle is certainly unstable for $t^2/t^1 <2/3$. Let us assume that the real stability wall lies at $t^2/t^1=r \geq 2/3$. As in the previous case, we will try to determine $r$ explicitly. At this stability wall, we require that $\mu({\cal Q})=\mu(L)=0$, however, only one of ${\cal Q}, L$ can destabilize $V$ in the unstable region. We must treat each of these cases in turn

\subsubsection*{Case 2a}
We will begin by considering ${\cal Q}$ to be the de-stabilizing sub-sheaf and parameterize its unknown  first Chern class as $c_1({\cal Q})=(-n,m)$ for some positive integers, $n,m$ ($c_1(L)=(n,-m)$. The slope of ${\cal Q}$ is then
\beq
\mu({\cal Q})=t^2(-2nt^2+4mt^1+mt^2)
\eeq
From the condition that $\mu({\cal Q})=0$ at some polarization $t^2/t^1=r \geq 2/3$ we infer that
\bea\label{stab_int}
m<2n \leq 7m~~,~ n,m>0
\eea

Next, we must attempt to bound the full topological data (that is, Chern classes) of $L$ and ${\cal Q}$. From the topological data in \eref{top_data} we have first a constraint on the first Chern class
\beq\label{ch1}
c_1(V)=0 \Rightarrow c_1(L)=-c_1({\cal Q})
\eeq
and next a condition that is most easily phrased in terms of the second Chern character
\bea\label{ch2}
d_{rst}ch_2(V)^{st}=d_{rst}(ch_2(L) +ch_2({\cal Q}))^{st}=(12,-18) \Rightarrow \\\nn
ch_2(L)=\frac{1}{2}c_1(L)^2 ~~\text{and}~~~d_{rst}ch_{2}({\cal Q})^{st}=(12-2m^2,-18-(m^2-4mn))
\eea
and finally, the condition on the third Chern character (or $\rm{Ind}(V)$)
\bea\label{ch3}
ch_3(V)=ch_3({\cal Q})+ch_3(L)=-10 \\\nn
 \Rightarrow \frac{1}{2}c_1({\cal Q})c_2({\cal Q})=-6m^2n+6n+9m+m^3=10
\eea
Once again, the integer system given by \eref{stab_int}-\eref{ch3} proves remarkably restrictive. Simply combining the condition in \eref{stab_int} with the final condition, \eref{ch3}, we find that there are only three   positive integer solutions given by 
\beq
n=m=1 ~~~ ,~~~n=2, m=1~~~,~~~n=3, m=1
\eeq
and we find that in fact the stability walls (defined by $\mu({\cal Q})=0$) in these three cases would be give by given by $r=4,4/3,4/5$ respectively.

As a result, we have the possibility of three different $SU(3)$ bundles of the form $0 \to {\cal Q} \to V \to L \to 0$,
\bea\label{possibilities}
0 \to {\cal Q}_1 \to V_1 \to \cO_{X}(1,-1) \to 0 \\
 0 \to {\cal Q}_2 \to V_2 \to \cO_{X}(2,-1) \to 0 \\
0 \to {\cal Q}_3 \to V_3 \to \cO_{X}(3,-1) \to 0
\eea
where ${\cal Q}$ is a still-unknown rank $2$ sheaf. However, to decide if such bundles can really occur, we must verify that in each case it is possible for 
{\it ${\cal Q}$ to be slope-stable} at the given stability wall defined by $\mu({\cal Q})=0$. If ${\cal Q}$ is not stable, then we do not have a graded sum of the form assumed in \eref{two_wall_poss2} and instead we could further decompose ${\cal Q}$ into its own graded sum, returning us to the case of three rank $1$ sheaves in \eref{two_wall_poss2}. To decide whether each ${\cal Q}_i$ in  is stable, we must apply to the Bogomolov bound, \eref{bogomolov}, to ${\cal Q}_i$ directly.

For each of the three possibilities, we can read off the second Chern class of ${\cal Q}_i$ via \eref{ch2} to obtain
\bea\label{c2qs}
c_2({\cal Q}_1)_r=d_{rst}c_2({\cal Q}_1)^{st}=(-18,27)\\ \nn
c_2({\cal Q}_2)_r=d_{rst}c_2({\cal Q}_2)^{st}=(-18,15)\\ \nn
c_2({\cal Q}_3)_r=d_{rst}c_2({\cal Q}_3)^{st}=(-18,3)~~
\eea
Directly substituting into \eref{bogomolov}, the Bogomolov bound tells us that necessary though not sufficient conditions for the stability of each ${\cal Q}_i$ are given by
\bea\label{qstabs}
{\cal Q}_1:~~~t^2/t^1 \geq 2/3~~~~~r=4~~~~ \\ \nn
{\cal Q}_2:~~~t^2/t^1 \geq 6/5~~~~~r=4/3~ \\ \nn
{\cal Q}_3:~~~~t^2/t^1 \geq 6~~~~~~~r=4/5~
\eea
where we have listed the position of the stability wall associated with $V_i$ immediately after the stable region for ${\cal Q}_i$. From the above, we immediately see that since ${\cal Q}_3$ is only stable for $t^2/t^1 \geq 6$ it cannot be stable at the stability wall of $V_3$ located at $t^2/t^1=4/5$. As a result, we can immediately disregard this possibility! Similarly, we can look more closely at the possible rank $1$ sub-sheaves which could de-stabilize ${\cal Q}_1$ and ${\cal Q}_2$ to determine if these two possibilities are valid. 

Applying the same arguments to ${\cal Q}_1,{\cal Q}_2$ that we did to the rank $2$ bundle in Section \ref{DT_su2} and solve for de-stabilizing rank $1$ sheaves
\beq
0 \to L^{i}_{1} \to {\cal Q}_{i} \to L^{i}_2 \to 0~~~i=1,2
\eeq
where $c_1(L^{i}_1)+c_1(L^{i}_2)=c_1({\cal Q}_i)$, consistent with the second Chern classes in \eref{c2qs} the and the stable regions in \eref{qstabs}. It is straightforward to check that for ${\cal Q}_2$ no such rank $1$ objects exist. Once again, the stability of ${\cal Q}_2$ has eliminated it from consideration. Finally, for ${\cal Q}_1$, we can solve for $L^{1}_1$ and $L^{1}_2$ explicitly and that the two solutions are
\beq
c_1(L^{1}_1)=(-2,2),~c_1(L^{1}_2)=(1,-1)~~~. ~~~c_1(L^{1}_1)=(1,-1),~c_1(L^{1}_2)=(-2,2).
\eeq
That is, ${\cal Q}_1$ is not stable at the stability wall given by $r=4$ and in fact must decompose itself. Thus, the possibility of ${\cal Q}_1$ has reduced exactly to Case 1 in \eref{case1_goodness} above! 

Having dramatically reduced the possible structure of the moduli space in \eref{mods} we turn now to our final possibility.

\subsubsection*{Case 2b}
To finish our analysis, we must at last consider the possibility that $L$ in \eref{two_wall_poss2} is the de-stabilizing sub-sheaf. As before, we parameterize its unknown first Chern class as $c_1(L)=(-n,m)$ for some positive integers, $n,m$. As before the slope of $L$ is
\beq
\mu(L)=2t^2(-2nt^2+4mt^1+mt^2)
\eeq
and as in Case 2a), the condition that $\mu(L)=0$ at some polarization $t^2/t^1=r \geq 2/3$ leads to
\bea\label{stab_int_again}
m<2n<7m~~,~ n,m>0
\eea

Next, we must attempt to bound the total Chern characters of $L$ and ${\cal Q}$. From the topological data in \eref{top_data} we have the following conditions on the Chern classes. The first Chern class gives us
\beq\label{ch1_again}
c_1(V)=0 \Rightarrow c_1(L)=-c_1({\cal Q})
\eeq
and next the second Chern character
\bea\label{ch2_again}
d_{rst}ch_2(V)^{st}=d_{rst}(ch_2(L) +ch_2({\cal Q}))^{st}=(12,-18) \Rightarrow \\\nn
ch_2(L)=\frac{1}{2}c_1(L)^2 ~~\text{and}~~~d_{rst}ch_{2}({\cal Q})^{st}=(12-2m^2,-18+(m^2-4mn))
\eea
and finally, the condition on the third Chern character 
\bea\label{ch3_again}
ch_3(V)=ch_3({\cal Q})+ch_3(L)=-10 \\\nn
 \Rightarrow \frac{1}{2}c_1({\cal Q})c_2({\cal Q})=6m^2n-6n-9m-m^3=10
\eea
In this final scenario, the integer system given by \eref{stab_int_again}-\eref{ch3_again} once again gives us a firm bound on the integers in question. From the condition in \eref{stab_int_again} and the final condition, \eref{ch3_again}, we find that the unique positive integer solution is given by 
\beq
n=m=2
\eeq
and unsurprisingly, the stability wall is located at $r=4$ as in the previous sections.

As a result, in principle we have the stable bundle
\beq
0 \to L \to V \to {\cal Q} \to 0
\eeq
However, once again we must ask whether the rank $2$ quotient sheaf ${\cal Q}$ is slope-stable at $r=4$? Checking the Bogomolov bound for this ${\cal Q}$ with $c_1({\cal Q})=(2,-2)$ leads us to our final result. At the stability wall this rank $2$ sheaf is unstable and must decompose into the poly-stable sum of rank $1$ sheaves
\beq 
{\cal Q} \to L_1 \oplus L_2~~\text{with}~~ c_1(L_1)=(1,-1), c_1(L_2)=(1,-1)
\eeq
Thus, this possibility too reduces to the case of the graded sum in \eref{two_wall_poss1}. Since both Case 2a) and Case 2b) reduce in this way, we conclude that no wall decompositions of the form \eref{two_wall_poss2} are possible, and hence the component ${\cal M}^{2}_{\omega}$ of the moduli space is trivial.

\vspace{5pt}

To conclude then, we have found something significant. As we hoped, the Bogomolov bound, \eref{bogomolov}, and the topological data in \eref{top_data} provided bounds on the possible sub-sheaves appearing in the wall theories in \eref{two_wall_poss1} and \eref{two_wall_poss2}. Moreover, the recursive nature of our stability analysis provided a unique solution for the stability wall structure of any bundle with the the total Chern class given in \eref{top_data}. {\it Any such bundle} must possess a stability wall, located at $t^2/t^1=4$ in K\"ahler moduli space and at this wall the bundle takes the poly-stable form
\beq
V \to \cO_{X}(-2,2) \oplus \cO_{X}(1,-1) \oplus \cO_{X}(1,-1)
\eeq
Furthermore, in the stable region of Kahler moduli space, $V$ can be described by the extension sequences
\bea
0 \to \cO_{X}(-2,2) \to {\cal W} \to \cO_{X}(1,-1) \to 0 \\
0 \to {\cal W} \to V \to \cO_{X}(1,-1) \to 0
\eea
As we saw in Case 1), this set of extension sequences, formed by giving vevs to the $U(1)$ charged matter in \eref{case1cs} is the {\it only stable rank $3$ branch} to theory. Thus, the dimension of the total moduli space is simply counted by \eref{case1_bund_mod}, that is, $h^1(X,V \otimes V^*)=2h^1(X,{\cal W} \otimes {\cal W}^*)=2\cdot 39=78$. More precisely, the moduli space in this case is $\mathbb{P}^{39}\times \mathbb{P}^{39}$.

To conclude then, recursively applying the techniques outlined in this section we have computed the Donaldson-Thomas invariant associated with the topological data in \eref{top_data}. For the polarization $\omega_{s}=\omega_1 +s \omega_2$ the complete formula is
\beq\label{wall_cross_rank3}
\left\{
\begin{array}
[c]{l}%
\lambda_{\omega_{s}}(X,c)=c_{top}(T{{\cal M}^{1}_{\omega_{s}}}^{*})=80~~4<s < \infty \\
\lambda_{\omega_{s}}(X,c)=0~~~~~~~~~~~~~~~~~~~~0<s \leq 4\\
\end{array}
\right.
\eeq
Thus, we have computed a DT-invariant associated with a rank $3$ $SU(3)$ bundle on a compact Calabi-Yau threefold. 

The above example illustrates that stability walls are a versatile tool for the study of Donaldson-Thomas invariants. Even in cases of higher rank bundles, a recursive analysis of the structure possible at the stability wall can yield information about the total moduli space of stable bundles and shed light on the structure of realistic heterotic vacua.

\section*{Acknowledgments}
The authors would like to thank A. Lukas, E. Sharpe, T. Pantev and R. Donagi for useful discussions. J.~Gray
would like to thank the University of Pennsylvania for hospitality
while part of this work was completed. L.A. and B.O. are supported in part by
the DOE under contract No.  DE-AC02-76-ER-03071and by NSF RTG Grant
DMS-0636606. 
\appendix

\section{Appendix: Extension Bundles, Isomorphisms and the Snake Lemma}\label{snake_lemma}
In this section, we clarify the correspondence between giving vevs to a fixed set of $C$-fields in the vacuum and the local construction of the bundle in that branch using extension sequences. In particular, we prove that while there may appear to be many different bundles corresponding to the same {\it fixed} choices of $C_i$ vevs, these naively different extension bundles are, in fact, isomorphic. 

\subsection{An Isomorphism Theorem}
As an example, consider the system described in Subsection \ref{more_comp}. There, at a stability wall we began with the direct sum of three line bundles $L_1 \oplus L_2 \oplus L_3$ and the two anomalous $U(1)$ D-terms associated with them 
\bea\label{two_ds2}
D_{1}^{U(1)} \sim \frac{\mu(L_1)}{{\cal V}} -q_1 G_{IJ}C^{I}{\bar C}^{J} \ , \quad
D_{2}^{U(1)} \sim \frac{\mu(L_2)}{{\cal V}} -q_2 G_{IJ}C^{I}{\bar C}^{J} \ .
\eea
As discussed in Subsection \ref{more_comp}, one can preserve vacuum supersymmetry by giving a vev to two 
$C$-fields in Table \ref{tabledoubled} to cancel the non-trivial FI terms (in some chamber of K\"ahler moduli space) while setting all other vevs to zero. For example, as in \eref{bad1}, we can choose
\beq\label{vev_choice}
\langle \tilde{C}_2 \rangle \neq0 \in H^1(X, L_1 \otimes L_3^{*})~~\ , ~~\langle C_3 \rangle \neq 0 \in H^1(X, L_2 \otimes L_3^*)
\eeq
with all other $\langle C_i \rangle=0$. The vevs of fields $\tilde{C}_2$ and $C_3$ are adjusted so that the two D-terms in \eqref{two_ds2} vanish. Following the arguments of Subsection \ref{more_comp}, we can describe this vacuum configuration geometrically. For, example, it was shown that this choice of field vevs corresponds to defining a stable indecomposable rank $3$ bundle $V$ via the two sequences
\bea \label{firstone2}
& &0 \to L_1 \to {\cal{W}} \to L_3 \to 0  \ ,\\ \label{secondone2}
& &0 \to L_2 \to V \to {\cal{W}} \to 0 \ . 
\eea
However, one might ask: is this the only consistent possibility? Could we not, for example, have written down
\bea \label{firstone3}
& &0 \to L_2 \to {\cal{W}}^{'} \to L_3 \to 0  \ ,\\  \label{secondone3}
& &0 \to L_1 \to V^{'} \to  {\cal{W}}^{'} \to 0 \ .
\eea
The first of these sequences is clearly defined by the extension group $Ext^1(L_3, L_2)=H^1(X,L_2 \otimes L_3^*)$. Likewise, the second sequence is defined by $Ext^1({\cal W}^{'},L_1)=H^1(X,L_1\otimes {{{\cal W}^{'}}}^{*})$. A simple analysis -- similar to \eref{tensored} and \eref{tensored_cohom} -- shows that this second extension group is determined by $H^1(X,L_1\otimes L_{3}^{*})$. That is, these sequences correspond to non-zero values of $\langle C_{3} \rangle$ and  $\langle \tilde{C}_2 \rangle$ respectively, as in \eqref{vev_choice}.
To compare the two apparently different bundles $V$ in \eref{secondone2} and $V'$ in\eref{secondone3}, we will make use of the following well-known result \cite{okonek} and its corollary.
\begin{theorem}[Morphism Lemma]
  \label{OSS_lemma_thm}
  Let $\phi: V_1 \to V_2$ be a non-trivial sheaf homomorphism between semi-stable bundles $V_1,V_2$. If at least one of the bundles is properly stable and $\mu(V_1)=\mu(V_2)$, then $\phi$ is a monomorphism or generically, an epimorphism.
   \end{theorem}
   
\begin{theorem}[Corollary]
  \label{OSS_corollary_thm}
Let $\phi: V_1 \to V_2$ be a nontrivial sheaf homomorphism between two semistable vector bundles $V_1,V_2$ with $rk(V_1)=rk(V_2)$ and $c_1(V_1)=c_1(V_2)$. Let at least one of the bundles be properly stable. Then $\phi$ is an isomorphism.
 \end{theorem}
We will use these results to establish an isomorphism between the two bundles in \eref{secondone2} and \eref{secondone3}. In particular, since $V$ and $V^{'}$ are properly stable in a given region of K\"ahler moduli space, we have only to establish that there exists {\it some} non-vanishing homomorphism, $\phi: V \to V^{'}$ (that is, $\rm{dim}(Hom(V,V^{'}))=h^0(X,V^{'}\otimes V^{*}) \neq 0$) in order to show that the two descriptions are indeed equivalent.

To begin then, we must compute $h^0(X,V^{'}\otimes V^{*})$. Combining the sequences in \eref{secondone2} and \eref{secondone3} we find that
\beq\label{bigarray1}
{\small
\ba{cccccccccl}
&&0&&0&&0&&& \\
&&\downarrow&&\downarrow&&\downarrow&&& \\
0&\to& L_{1}\otimes {\cal W}^{*} &\to& L_{1} \otimes V^{*} &\to& L_1 \otimes L_{2}^{*} &\to&0
 \\
&&\downarrow&&\downarrow&&\downarrow&&& \\
0&\to& {\cal W}^* \otimes V^{'} &\to& V^* \otimes V^{'} &\to& L_{2}^* \otimes V^{'} &\to&0
 \\
&&\downarrow&&\downarrow&&\downarrow&&& \\
0&\to& {\cal W}^{'}\otimes {\cal W}^* &\to& {\cal W}^{'} \otimes V^{*} &\to& L_{2}^* \otimes {\cal W}^{'} &\to&0
 \\
&&\downarrow&&\downarrow&&\downarrow&&& \\
&&0&&0&&0&&& \\
\ea
}
\eeq
where ${\cal W}^{'}\otimes {\cal W}^*$ is determined by \eref{firstone2} and \eref{firstone3}, which lead to the array
\beq\label{bigarray2}
{\small
\ba{cccccccccl}
&&0&&0&&0&&& \\
&&\downarrow&&\downarrow&&\downarrow&&& \\
0&\to& L_{2}\otimes L_{3}^{*} &\to& L_{2} \otimes W^{*} &\to& L_2 \otimes L_{1}^{*} &\to&0
 \\
&&\downarrow&&\downarrow&&\downarrow&&& \\
0&\to& L_{3}^{*} \otimes {\cal W}^{'} &\to& {\cal W}^* \otimes {\cal W}^{'} &\to& L_{1}^* \otimes {\cal W}^{'} &\to&0
 \\
&&\downarrow&&\downarrow&&\downarrow&&& \\
0&\to& \cO_{X}&\to& L_3 \otimes {\cal W}^{*} &\to& L_{1}^* \otimes L_{3} &\to&0
 \\
&&\downarrow&&\downarrow&&\downarrow&&& \\
&&0&&0&&0&&& \\
\ea
}
\eeq
We will consider here the generic case in which $Hom(L_i,L_j)=0$ for $i,j=1,2,3$ (that is, all three line bundles are distinct with mixed sign $c_1$)\footnote{In the case that $Hom(L_i,L_j) \neq 0$ for some $i,j$, the isomorphism result is straightforward to prove. As a result, we consider the more difficult, non-trivial case above.}. Now, from the last column of \eref{bigarray1}, we have the following long exact sequence in cohomology
\beq
0  \to H^0(X,L_2^{*} \otimes V^{'}) \to H^0(X,L_2^{*} \otimes W^{'}) \stackrel{\delta}{\to} H^1(X,L_1 \otimes L_{2}^{*}) \to \ldots
\eeq
By tensoring \eref{firstone3} by $L_{2}^{*}$ it is clear that $H^0(X,L_2^{*} \otimes W^{'})=H^0(X, \cO_{X})=\mathbb{C}$. Now, we must analyze the co-boundary map $\delta: H^0(X, \cO_{X}) \to H^1(X,L_1 \otimes L_{2}^{*})$ in the chosen vacuum. The map, $\delta$, is determined by the vacuum value of the fields, in this case those corresponding to $H^1(X,L_1 \otimes L_{2}^{*})$. However, since we have chosen the vacuum configuration given in \eref{vev_choice}, which has only $\langle\tilde{C}_{2} \rangle$ and $\langle C_3 \rangle$ non-zero, it is clear that {\it $\delta$ is the zero element of $H^1(X,L_1 \otimes L_{2}^{*})$} (that is, $\delta=\langle \tilde{C}_{1}\rangle=0$). As a result, $Ker(\delta)=\mathbb{C}$ and 
\beq\label{firstbadger}
H^0(X,L_2^{*} \otimes V^{'})=\mathbb{C}~.
\eeq

A similar analysis can be applied to the first column in \eref{bigarray1} (using the defining sequences in \eref{bigarray2}). The long exact sequence in cohomology associated with this first column gives us $H^0(X,{\cal W}^{*}\otimes V^{'})=0$ and 
\beq\label{middlebadger}
0 \to H^1(L_1\otimes {\cal W}^{*}) \to H^1(X, {\cal W}^{*} \otimes V^{'}) \to H^1(X,{\cal W}^{'}\otimes {\cal W}^{*}) \to \ldots
\eeq
where
\beq
H^1(X,L_1\otimes {\cal W}^{*})=\frac{H^1(X, L_1 \otimes L_{3}^{*})}{\mathbb{C}}~~~,~~~H^1(X,{\cal W}^{'}\otimes {\cal W}^{*})=\frac{H^1(X, L_2 \otimes L_{3}^{*})}{\mathbb{C}} \ .
\eeq
Substituting back into \eref{middlebadger}, we then find
\bea\label{nextbadger}
H^1(X, {\cal W}^{*} \otimes V^{'})=\frac{H^1(X, L_1 \otimes L_{3}^{*})}{\mathbb{C}}\oplus Ker(\beta) \ ,\\
\beta: \frac{H^1(X, L_2 \otimes L_{3}^{*})}{\mathbb{C}} \to H^2(X,L_1\otimes {\cal W}^{*}) \ .
\eea

With these results in hand, we are in a position to analyze $H^0(X,V^{*}\otimes V^{'})$ directly. From the middle row of \eref{bigarray1} we take the short exact sequence
\beq\label{thebigkahuna}
0 \to {\cal W}^{*} \otimes V^{'} \to V^{*}\otimes V^{'} \to L_{2}^{*} \otimes V^{'} \to 0 \ .
\eeq 
Next, using the results above, the long exact sequence in cohomology associated with \eref{thebigkahuna} gives us
\beq
0 \to H^0(X, V^{*}\otimes V^{'}) \to H^0(X,L_{2}^{*} \otimes V^{'}) \stackrel{\gamma}{\to} H^1(X,{\cal W}^{*} \otimes V^{'} ) \to \ldots
\eeq
which, combing the results from \eref{firstbadger} and \eref{nextbadger}, reduces to
\bea
H^0(X, V^{*}\otimes V^{'})=Ker(\gamma) \ ,\\
\gamma: \mathbb{C} \to \frac{H^1(X, L_1 \otimes L_{3}^{*})}{\mathbb{C}} \oplus Ker(\beta) \ , \\
Ker(\beta) \in \frac{H^1(X, L_2 \otimes L_{3}^{*})}{\mathbb{C}} \ .
\eea
However, the coboundary map $\gamma=(\gamma_1, \gamma_2)$ is precisely given by the vevs
\beq
\gamma_1=\langle \tilde{C}_2 \rangle \in H^1(X, L_1 \otimes L_3^{*})~~\ , ~~\gamma_2=\langle C_3 \rangle \in H^1(X, L_2 \otimes L_3^*) \ .
\eeq
However, as we have seen in previous sections, by construction of the extension sequences\footnote{See  Appendix A of \cite{Anderson:2009nt} for a detailed discussion.} these two vev choices (that is,  maps) correspond exactly to the zero elements in the quotients $H^1(X, L_1 \otimes L_{3}^{*})/\mathbb{C}$ and $H^1(X, L_2 \otimes L_{3}^{*})/\mathbb{C}$, respectively! In terms of field theory, these are the two degrees of freedom that are removed from the zero-mode spectrum by the $U(1)$ D-terms in \eref{two_ds2}.

As a result, the map $\gamma$ automatically has a zero image and the kernel of $\gamma$ is simply $\mathbb{C}$. Hence,
\beq
H^0(X, V^{*}\otimes V^{'})=Ker(\gamma)=\mathbb{C}
\eeq
and we have reached a conclusion. Since there exists a non-trivial homomorphism between $V$ and $V^{'}$, by the Corollary, Theorem \ref{OSS_corollary_thm}, the two bundles are isomorphic.

\subsection{Isomorphism via the Snake Lemma}
We illustrate below another method that can be used to analyze some of the redundant descriptions of the extension bundles used in this paper. As a concrete example, we consider the system described in Subsection \ref{li_yau_stab}. There, we began with both $U(1)$ and $E_6$ D-terms and explored flat directions corresponding to Li-Yau bundles for rank-changing deformations. Specifically, we choose as an example non-vanishing vevs of the matter fields
\beq\label{rank_change_vev_again}
\langle f_{2} \rangle \in H^1(X, \cF)~~~\text{and}~~~\langle {\tilde{f}}_{1} \rangle \in H^1(X, {\cal K}^{*}) \ ,
\eeq\
with all other vevs set to zero. These satisfy D-flatness for 
\bea\label{u1_e6_dterm2}
D^{U(1)}&=&\frac{3}{16}\frac{\epsilon_{S}\epsilon_{R}^{2}}{\kappa_4^{2}}\frac{\mu(\cF)}{{\cal V}} - 3G_{L\bar{M}} C_1^L C_1^{\bar M}+3G_{L\bar{M}} C_2^L C_2^{\bar M} \nonumber \\
&&-2{\cal{G}}_{L\bar{M}} f_1^L f_1^{\bar M} +{\cal{G}}_{L\bar{M}} f_2^L f_2^{\bar M}+2{\cal{G}}_{L\bar{M}} {{\tilde f}_{1}}^L{{\tilde f}_1}^{\bar M}-{\cal{G}}_{L\bar{M}} {{\tilde f}_2}^{L}{{\tilde f}_{2}}^{\bar M}
\eea
and
\beq\label{e6_dterm2}
D^{E_6}=G_{L\bar{M}} f_1^L f_1^{\bar M}+G_{L\bar{M}} f_2^L f_2^{\bar M}-G_{L\bar{M}} \tilde{f}_1^L \tilde{f}_1^{\bar M}-G_{L\bar{M}} \tilde{f}_2^L \tilde{f}_2^{\bar M} 
\eeq
in the region of moduli space for which $\mu(\cF)<0$.
We showed in Subsection \ref{li_yau_stab} that geometrically this solution corresponds to 
\bea\label{oext}
& &0 \to \cO_{X} \to U \to {\cal K} \to 0 \quad \Leftrightarrow  \quad \ H^1(X, {\cal K}^{*}) \\ \label{oext2}
& &0 \to \cF \to V_4 \to U \to 0  \quad \ \Leftrightarrow \quad H^1(X, \cF \otimes U^{*})
\eea
However, as in the previous subsection, there is more than one way that we could imagine defining $V_4$. For example, the bundle $V_4$ could also be described by another pair of sequences
\bea\label{oextagain}
& &0 \to \cF \to U^{'} \to \cO_{X} \to 0 \quad \Leftrightarrow  \quad \ H^1(X, {\cal F}) \\ \label{oextagain2}
& &0 \to U^{'} \to V_{4}^{'} \to {\cal K} \to 0  \quad \ \Leftrightarrow \quad H^1(X, U^{'}\otimes {\cal K}^{*}) 
\eea
The first of these sequences is clearly defined by the extension group $Ext^1(\cO_X, \cF)=H^1(X,\cF)$. Likewise, the second sequence is defined by $Ext^1({\cal K},U^{'})=H^1(X, U^{'}\otimes {\cal K}^{*})$. As we have seen previously, this second extension group is determined by $Ker(\delta) \in H^1(X,U^{'}\otimes {\cal K}^{*})$ where $\delta: H^1(X,{\cal K}) \to H^2(X,\cF \otimes{\cal K}^*)$.

We now demonstrate explicitly that these two descriptions are equivalent. To prove that the bundles $V_{4}$ in \eref{oext2} and $V_{4}'$ in \eref{oextagain2} are isomorphic, we can make use of the following well-known Lemma\cite{AG,AG2}.

\begin{theorem}[Snake Lemma]
  \label{snake_lemma_thm}
  Given a commutative diagram involving two exact sequences of vector bundles and morphisms $\phi_{A},\phi_{B},\phi_{C}$
  \begin{equation}
 \begin{array}{ccccccccc}
  0&\rightarrow&A&\stackrel{f_1}{\longrightarrow}&B&\stackrel{f_2}{\longrightarrow}&C&\rightarrow&0\\
    &                    &\quad \downarrow\phi_{A}  & \quad    & \quad \downarrow\phi_{B}&& \quad\downarrow\phi_{C}&&\\
  0&\rightarrow&A^{'}&\stackrel{g_1}{\longrightarrow}&B^{'}&\stackrel{g_2}{\longrightarrow}&C^{'}&\rightarrow&0 
\end{array}\label{snake_lemma1}
\end{equation}  
there exists a long exact sequence
\beq\label{long_snake}
0 \to Ker(\phi_{A}) \to Ker(\phi_{B}) \to Ker(\phi_{C}) \to Coker(\phi_{A}) \to Coker(\phi_{B}) \to Coker(\phi_{C}) \to 0 \ .
\eeq
 \end{theorem}
We can use the Snake lemma to verify the equivalence of \eref{oext2} and \eref{oextagain2}. To establish the isomorphism between $V_4$ and $V_{4}^{'}$, consider the following commutative diagram

 \begin{equation}
 \begin{array}{ccccccccc}
  0&\rightarrow&\cF &\stackrel{f_1}{\longrightarrow}&V_{4}&\stackrel{f_2}{\longrightarrow}&U &\rightarrow&0\\
    &                    &\quad \downarrow\phi_{\cF}  & \quad    & \quad \downarrow\phi_{V_{4}}&& \quad\downarrow\phi_{U}&&\\
  0&\rightarrow&U^{'}&\stackrel{g_1}{\longrightarrow}&V_{4}^{'}&\stackrel{g_2}{\longrightarrow}&{\cal K}&\rightarrow&0
\end{array}\label{snake_lemma_app}
\end{equation}  
where we have {\it chosen the morphisms} $\phi_{\cF}$ and $\phi_{U}$ to be those induced from the extension sequences \eref{oext} and \eref{oextagain} respectively. The map $\phi_{V_{4}}$ we will be taken to be any non-trivial morphism defined to be consistent with the commutative diagram. Using \eref{long_snake} this leads to
\beq\label{snake_iso}
0 \to Ker(\phi_{\cF}) \to Ker(\phi_{V_{4}}) \to Ker(\phi_{U}) \to Coker(\phi_{\cF}) \to Coker(\phi_{V_{4}}) \to Coker(\phi_{U}) \to 0 \ .
\eeq
To proceed, we note that from \eref{oextagain}, that 
\beq
Ker(\phi_{\cF})=0~~~,~~~Coker(\phi_{\cF})=\cO_{X}
\eeq
while from \eref{oext} we have 
\beq
Ker(\phi_{U})=\cO_{X}~~~,~~~Coker(\phi_{U})=0 \ .
\eeq
Substituting this into \eref{snake_iso}, we get
\beq
0 \to 0 \to Ker(\phi_{V_{4}}) \to \cO_{X} \stackrel{g}{\to} \cO_{X} \to Coker(\phi_{V_{4}}) \to 0 \to 0 \ .
\eeq
A simple analysis verifies that the induced map $g$ is non-trivial, an isomorphism and that 
\beq
Ker(\phi_{V_{4}})=Coker(\phi_{V_{4}})=0 \ .
\eeq
As a result, the map $\phi_{V_{4}}$ is an isomorphism, $V_{4} \simeq V_{4}^{'}$. A similar analysis can be performed for many of the extension bundles we define throughout this paper. 

The results of this Appendix demonstrate that the choice of vevs uniquely determines a vector bundle and, as a result, we are free to choose whichever description is most convenient.

\newpage




\begin{thebibliography}{99}

\bibitem{Candelas:1985en}
  P.~Candelas, G.~T.~Horowitz, A.~Strominger and E.~Witten,
  ``Vacuum Configurations For Superstrings,''
  Nucl.\ Phys.\ B {\bf 258}, 46 (1985).

\bibitem{Witten:1985bz}
  E.~Witten,
  ``New Issues In Manifolds Of SU(3) Holonomy,''
  Nucl.\ Phys.\  B {\bf 268} (1986) 79.

\bibitem{GSW}
M.~B.~Green, J.~H.~Schwarz and E.~Witten,
``Superstring Theory, Vol II,'' CUP, 1987.

 \bibitem{Lukas:1997fg}
  A.~Lukas, B.~A.~Ovrut and D.~Waldram,
  ``On the four-dimensional effective action of strongly coupled heterotic string theory,''
  Nucl.\ Phys.\  B {\bf 532}, 43 (1998)
  [arXiv:hep-th/9710208].

\bibitem{Lukas:1998yy}
  A.~Lukas, B.~A.~Ovrut, K.~S.~Stelle and D.~Waldram,
  ``The universe as a domain wall,''
  Phys.\ Rev.\  D {\bf 59}, 086001 (1999)
  [arXiv:hep-th/9803235].
  
  \bibitem{Lukas:1998tt}
  A.~Lukas, B.~A.~Ovrut, K.~S.~Stelle and D.~Waldram,
  ``Heterotic M-theory in five dimensions,''
  Nucl.\ Phys.\  B {\bf 552}, 246 (1999)
  [arXiv:hep-th/9806051].


\bibitem{Lukas:1998hk}
  A.~Lukas, B.~A.~Ovrut and D.~Waldram,
  ``Non-standard embedding and five-branes in heterotic M-theory,''
  Phys.\ Rev.\  D {\bf 59}, 106005 (1999)
  [arXiv:hep-th/9808101].
  
  \bibitem{Donagi:2004ia}
  R.~Donagi, Y.~H.~He, B.~A.~Ovrut and R.~Reinbacher,
  ``The particle spectrum of heterotic compactifications,''
  JHEP {\bf 0412}, 054 (2004)
  [arXiv:hep-th/0405014].\\
  Y.~H.~He,
  ``GUT particle spectrum from heterotic compactification,''
  Mod.\ Phys.\ Lett.\  A {\bf 20} (2005) 1483.

\bibitem{Donagi:2004qk}
  R.~Donagi, Y.~H.~He, B.~A.~Ovrut and R.~Reinbacher,
  ``Moduli dependent spectra of heterotic compactifications,''
  Phys.\ Lett.\  B {\bf 598}, 279 (2004)
  [arXiv:hep-th/0403291].
  
  \bibitem{Donagi:2004su}
  R.~Donagi, Y.~H.~He, B.~A.~Ovrut and R.~Reinbacher,
  ``Higgs doublets, split multiplets and heterotic SU(3)C x SU(2)L x U(1)Y
  spectra,''
  Phys.\ Lett.\ B {\bf 618}, 259 (2005) [arXiv:hep-th/0409291].
   R.~Donagi, Y.~H.~He, B.~A.~Ovrut and R.~Reinbacher,
  ``The spectra of heterotic standard model vacua,''
  JHEP {\bf 0506}, 070 (2005)
  [arXiv:hep-th/0411156].
  
  \bibitem{Donagi:2003tb}
  R.~Donagi, B.~A.~Ovrut, T.~Pantev and R.~Reinbacher,
  ``SU(4) Instantons on Calabi-Yau Threefolds with $Z_2 x Z_2$ Fundamental
  Group,''
  JHEP {\bf 0401}, 022 (2004)
  [arXiv:hep-th/0307273].
   
\bibitem{burt4}
V.~Braun, Y.-H.~He, B.A.~Ovrut, T.~Pantev, 
``Moduli dependent mu-terms in a heterotic standard model''
JHEP 0603:006,2006,
hep-th/0510142

\bibitem{burt5}
V.~Braun, Y.-H.~He, B.A.~Ovrut, T.~Pantev, 
 ``Heterotic standard model moduli''
JHEP 0601:025,2006, 
hep-th/0509051

\bibitem{burt6}
 V.~Braun, B.A.~Ovrut, T.~Pantev, R.~Reinbacher,
``Elliptic Calabi-Yau threefolds with Z(3) x Z(3) Wilson lines''
JHEP 0412:062,2004, 
hep-th/0410055

\bibitem{burt7}
 R.~Donagi, Y.-H.~He, B.A.~Ovrut, R.~Reinbacher, 
``The Spectra of heterotic standard model vacua''
JHEP 0506:070,2005, 
hep-th/0411156

\bibitem{Blaszczyk:2010db}
  M.~Blaszczyk, S.~G.~Nibbelink, F.~Ruehle, M.~Trapletti and P.~K.~S.~Vaudrevange,
  ``Heterotic MSSM on a Resolved Orbifold,''
  JHEP {\bf 1009}, 065 (2010)
  [arXiv:1007.0203 [hep-th]].
  
\bibitem{Nibbelink:2009sp}
  S.~G.~Nibbelink, J.~Held, F.~Ruehle, M.~Trapletti and P.~K.~S.~Vaudrevange,
  ``Heterotic Z6-II MSSM Orbifolds in Blowup,''
  JHEP {\bf 0903}, 005 (2009)
  [arXiv:0901.3059 [hep-th]].
  
\bibitem{Greene:1986bm}
  B.~R.~Greene, K.~H.~Kirklin, P.~J.~Miron and G.~G.~Ross,
  ``A Three Generation Superstring Model. 1. Compactification And Discrete
  Symmetries,''
  Nucl.\ Phys.\  B {\bf 278} (1986) 667.
  
\bibitem{Greene:1986ar}
  B.~R.~Greene, K.~H.~Kirklin, P.~J.~Miron and G.~G.~Ross,
  ``A Superstring Inspired Standard Model,''
  Phys.\ Lett.\  B {\bf 180} (1986) 69.

\bibitem{Greene:1986jb}
  B.~R.~Greene, K.~H.~Kirklin, P.~J.~Miron and G.~G.~Ross,
  ``A Three Generation Superstring Model. 2. Symmetry Breaking and the
  Low-Energy Theory,''
  Nucl.\ Phys.\  B {\bf 292} (1987) 606.

\bibitem{Candelas:2008wb}
  P.~Candelas and R.~Davies,
  ``New Calabi-Yau Manifolds with Small Hodge Numbers,''
  Fortsch.\ Phys.\  {\bf 58} (2010) 383
  [arXiv:0809.4681 [hep-th]].
  
\bibitem{Braun:2009qy}
  V.~Braun, P.~Candelas and R.~Davies,
  ``A Three-Generation Calabi-Yau Manifold with Small Hodge Numbers,''
  Fortsch.\ Phys.\  {\bf 58}, 467 (2010)
  [arXiv:0910.5464 [hep-th]].
  
   \bibitem{Sharpe:1998zu}
  E.~R.~Sharpe,
  ``Kaehler cone substructure,''
  Adv.\ Theor.\ Math.\ Phys.\  {\bf 2}, 1441 (1999)
  [arXiv:hep-th/9810064].  

\bibitem{Anderson:2009nt}
  L.~B.~Anderson, J.~Gray, A.~Lukas and B.~Ovrut,
  ``Stability Walls in Heterotic Theories,''
  JHEP {\bf 0909}, 026 (2009)
  [arXiv:0905.1748 [hep-th]].
  
\bibitem{Anderson:2009sw}
  L.~B.~Anderson, J.~Gray, A.~Lukas and B.~Ovrut,
  ``The Edge Of Supersymmetry: Stability Walls in Heterotic Theory,''
  Phys.\ Lett.\  B {\bf 677}, 190 (2009)
  [arXiv:0903.5088 [hep-th]].
  
\bibitem{Anderson:2010tc}
  L.~B.~Anderson, J.~Gray and B.~Ovrut,
  ``Yukawa Textures From Heterotic Stability Walls,''
  JHEP {\bf 1005} (2010) 086
  [arXiv:1001.2317 [hep-th]].
  

\bibitem{DT}
S.~K.~ Donaldson and R.~P.~ Thomas, 
``Gauge theory in higher dimensions, In: The geometric Universe", (Oxford, 1996), Oxford University Press, Oxford, 1998, 31Ð47.

\bibitem{Thomas}
R.~Thomas,
``A holomorphic Casson invariant for Calabi-Yau 3-folds, and bundles on $K3$ fibrations,"
J.\ Differential\ Geom.\ {\bf 54}, Number 2 (2000), 367-438. 

 \bibitem{ellingsrud}
G.~ Ellingsrud and L. Gottsche,
``Variation of moduli spaces and Donaldson invariants under change of polarization," 
J.\ reine\ angew.\ Math.\ {\bf 467} (1995), 1-49.
[arXiv:alg-geom/9410005v2].


\bibitem{Qin}
Z.~Qin,
``Birational Properties of Moduli spaces of stable locally free rank-2 sheaves on algebraic
surfaces,"
Manuscripta\ Mathematica {\bf 72}, (1991), 163-180.

\bibitem{Li_Qin}
W.~P.~Li and Z.~Qin,
``Donaldson-Thomas invariants of certain Calabi-Yau 3-folds,"
arXiv:1002.4080v1 [math.AG].

\bibitem{Li_Qin2}
W.~P.~Li and Z.~Qin,
``Stable rank-2 bundles on Calabi-Yau manifolds,"
Internat.\ J.\ Math. {\bf 14}, (2003), 1097-1120.

\bibitem{Friedman_Qin}
R.~Friedman and Z.~Qin,
``Flips of moduli spaces and transition formulas for Donaldson polynomial invariants of rational surfaces,"
Commu.\ in Anal.\ and \ Geom.\  {\bf 3} (1995), 11-83. 
[arXiv:alg-geom/9410007v1].

\bibitem{Nekrasov1}
D.~ Maulik, N.~Nekrasov, A.~Okounkov and R.~Pandharipande, 
``GromovÐWitten theory and DonaldsonÐThomas theory, I,"
Compositio Mathematica {\bf 142}, (2006) 1263-1285.

\bibitem{Nekrasov2}
D.~Maulik, N.~Nekrasov, A.~Okounkov and R.~Pandharipande, 
``GromovÐWitten theory and DonaldsonÐThomas theory, II,". 
Compositio Mathematica, {\bf 142}, (2006) 1286-1304.

\bibitem{Lukas:1999nh}
  A.~Lukas and K.~S.~Stelle,
  ``Heterotic anomaly cancellation in five dimensions,''
  JHEP {\bf 0001}, 010 (2000)
  [arXiv:hep-th/9911156].

\bibitem{Blumenhagen:2005ga}
  R.~Blumenhagen, G.~Honecker and T.~Weigand,
  ``Loop-corrected compactifications of the heterotic string with line bundles,''
  JHEP {\bf 0506}, 020 (2005)
  [arXiv:hep-th/0504232].
  
  \bibitem{Blumenhagen:2006wj}
R.~Blumenhagen, S.~Moster and T.~Weigand,
 ``Heterotic GUT and standard model vacua from simply connected Calabi-Yau
  manifolds,''
  Nucl.\ Phys.\ B {\bf 751}, 186 (2006)
  [arXiv:hep-th/0603015].\\
  R.~Blumenhagen, S.~Moster, R.~Reinbacher and T.~Weigand,
  ``Massless spectra of three generation U(N) heterotic string vacua,''
  [arXiv:hep-th/0612039].
  
\bibitem{Kuriyama:2008pv}
  M.~Kuriyama, H.~Nakajima and T.~Watari,
  ``A Theoretical Framework for R-parity Violation,''
  Phys.\ Rev.\  D {\bf 79}, 075002 (2009)
  [arXiv:0802.2584 [hep-ph]].

  
  \bibitem{hartshorne_deformation}
R.~Hartshorne,  ``Deformation Theory, Springer,'' GTM 257,
Springer-Verlag, 2010. 

  \bibitem{Huybrechts}
 D.~Huybrechts and M.~Lehn, ``The geometry of the moduli space of stable of sheaves". Aspects of Mathematics, E 31 (1997).
 
  \bibitem{Huybrechts_tangent}
 D.~Huybrechts,
 ``The tangent bundle of a Calabi-Yau manifold---deformations and restriction to rational curves,"
Commun.\ Math.\ Phys.\  {\bf 171} (1995) 139.
   
\bibitem{Li:2004hx}
  J.~Li and S.~T.~Yau,
  ``The existence of supersymmetric string theory with torsion,''
  arXiv:hep-th/0411136.
  
  \bibitem{duy}
K. Uhlenbeck and S.-T. Yau,  ``On the existence of Hermitian Yang-Mills 
connections in stable bundles,''  Comm. Pure App. Math. 39 257 (1986), 
42 703 (1986). 
S. Donaldson,  ``Anti Self-Dual Yang-Mills Connections over 
Complex Algebraic Surfaces and Stable Vector Bundles,''  
Proc. London Math. Soc. 3 1 (1985).

  \bibitem{AG}
R.~Hartshorne,  ``Algebraic Geometry, Springer,'' GTM 52,
Springer-Verlag, 1977. 

\bibitem{AG2}
P. Griffith, J. Harris, ``Principles of algebraic geometry,'' 1978.

\bibitem{Braun:2005ux}
  V.~Braun, Y.~H.~He, B.~A.~Ovrut and T.~Pantev,
  ``A heterotic standard model,''
  Phys.\ Lett.\ B {\bf 618}, 252 (2005)
  [arXiv:hep-th/0501070].
  V.~Braun, Y.~H.~He, B.~A.~Ovrut and T.~Pantev,
  ``A standard model from the E(8) x E(8) heterotic superstring,''
  JHEP {\bf 0506}, 039 (2005)
  [arXiv:hep-th/0502155].
  
\bibitem{Anderson:2010ke}
  L.~B.~Anderson, V.~Braun, R.~L.~Karp and B.~A.~Ovrut,
  ``Numerical Hermitian Yang-Mills Connections and Vector Bundle Stability in
  Heterotic Theories,''
  JHEP {\bf 1006} (2010) 107
  [arXiv:1004.4399 [hep-th]].
  
  \bibitem{burt2}
V.~Braun, T.~ Brelidze, M.R.~Douglas, B.A.~Ovrut, 
 ``Eigenvalues and Eigenfunctions of the Scalar Laplace Operator on Calabi-Yau Manifolds'',
JHEP 0807:120,2008
arXiv:0805.3689 [hep-th]

\bibitem{burt3}
V.~Braun, T. ~Brelidze, M.R.~Douglas, B.A.~Ovrut,  
``Calabi-Yau Metrics for Quotients and Complete Intersections'',
JHEP 0805:080,2008, 
arXiv:0712.3563 [hep-th]

\bibitem{Anderson:2009ge}
  L.~B.~Anderson, J.~Gray, D.~Grayson, Y.~H.~He and A.~Lukas,
  ``Yukawa Couplings in Heterotic Compactification,''
  Commun.\ Math.\ Phys.\  {\bf 297} (2010) 95
  [arXiv:0904.2186 [hep-th]].
  
\bibitem{Anderson:2010mh}
  L.~B.~Anderson, J.~Gray, A.~Lukas and B.~Ovrut,
  ``Stabilizing the Complex Structure in Heterotic Calabi-Yau Vacua,''
  arXiv:1010.0255 [hep-th].
  
   \bibitem{us_soon}
 L.~B.~Anderson, J.~Gray, A.~Lukas and B.~Ovrut. To appear.
 
    \bibitem{us_tangent}
 L.~B.~Anderson, J.~Gray, and B.~Ovrut. To appear.
 
  
\bibitem{Anderson:2007nc}
L.~B.~Anderson, Y.~H.~He and A.~Lukas,
``Heterotic compactification, an algorithmic approach,''
JHEP {\bf 0707}, 049 (2007) [arXiv:hep-th/0702210].

\bibitem{Anderson:2008uw}
  L.~B.~Anderson, Y.~H.~He and A.~Lukas,
  ``Monad Bundles in Heterotic String Compactifications,''
  JHEP {\bf 0807}, 104 (2008)
  [arXiv:0805.2875 [hep-th]].
  
  \bibitem{Anderson:2008ex}
  L.~B.~Anderson,
  ``Heterotic and M-theory Compactifications for String Phenomenology,''
  arXiv:0808.3621 [hep-th].
  
   \bibitem{okonek}
  C.~Okonek, M.~Schneider, and H.~Spindler, ``Vector bundles on complex projective spaces". Birkhuser Boston, Mass., 1980. 
 
\bibitem{Braun:2005nv}
  V.~Braun, Y.~H.~He, B.~A.~Ovrut and T.~Pantev,
  ``The exact MSSM spectrum from string theory,''
  JHEP {\bf 0605}, 043 (2006)
  [arXiv:hep-th/0512177].

\bibitem{Bouchard:2005ag}
  V.~Bouchard and R.~Donagi,
  ``An SU(5) heterotic standard model,''
  Phys.\ Lett.\ B {\bf 633}, 783 (2006)
  [arXiv:hep-th/0512149].

\bibitem{Braun:2005zv}
  V.~Braun, Y.~H.~He, B.~A.~Ovrut and T.~Pantev,
  ``Vector Bundle Extensions, Sheaf Cohomology, and the Heterotic Standard
  Model,''
  Adv.\ Theor.\ Math.\ Phys.\  {\bf 10}, 4 (2006)
  [arXiv:hep-th/0505041].

\bibitem{Ambroso:2009jd}
  M.~Ambroso and B.~Ovrut,
  ``The B-L/Electroweak Hierarchy in Heterotic String and M-Theory,''
  JHEP {\bf 0910}, 011 (2009)
  [arXiv:0904.4509 [hep-th]].
  
\bibitem{Ambroso:2009sc}
  M.~Ambroso and B.~Ovrut,
  ``The B-L/Electroweak Hierarchy in Smooth Heterotic Compactifications,''
 Int.J.Mod.Phys. A25:2631-2677,2010  
 [ arXiv:0910.1129 [hep-th]].
  
\bibitem{burt1}
  M.~ Ambroso and B.A.~Ovrut,
``The Mass Spectra, Hierarchy and Cosmology of B-L MSSM Heterotic Compactifications'',
e-Print: arXiv:1005.5392 [hep-th]

  
\bibitem{Anderson:2009mh}
  L.~B.~Anderson, J.~Gray, Y.~H.~He and A.~Lukas,
  ``Exploring Positive Monad Bundles And A New Heterotic Standard Model,''
  JHEP {\bf 1002} (2010) 054
  [arXiv:0911.1569 [hep-th]].
  
  \bibitem{langer1}
A.~Langer,
``Semistable Sheaves in Positive Characteristic,"
Ann.\ Math.\ {\bf 159}, No.1 (2004), 251-276.

\bibitem{langer2}
A.~Langer,
``Moduli Spaces of Sheaves in Mixed Characteristic,"
Duke.\ Math.\  J.\ {\bf 124}, No.3 (2004), 571-586.

  \bibitem{gottsche_notes}
L.~Gottsche, ``Donaldson Invariants in Algebraic Geometry", Abdus Salam ICTP Lecture notes, 1999. 


\bibitem{Joyce}
D.~Joyce,
``Generalized Donaldson-Thomas invariants," 
arXiv:0910.0105v2 [math.AG].

\bibitem{FMW}
R.~Friedman, J.~Morgan and E.~Witten, 
``Vector bundles over elliptic fibrations,"
 J.\ Alg.\ Geom. \ {\bf 8} (1999), 279Ð401.

\bibitem{BM}
T.~ Bridgeland and A.~Maciocia, 
``Fourier-Mukai transforms for K3 and elliptic fibrations,"
J.\ Alg.\ Geom.\  {\bf 11} (2002), 629Ð657.

  
 

 \end{thebibliography}
\end{document}